\documentclass[12pt]{iopart}
\usepackage{iopams,amsthm} 
\usepackage{graphicx}

\begin{document}
 
\title[Brick walls and AdS/CFT]{Brick walls and AdS/CFT}

\author{Bernard S. Kay and L. Ort\'iz\footnote{\textit{present address}: Institute for Nuclear Sciences, UNAM, Nexico City, Mexico.}}

\address{Department of Mathematics, University of York, York YO10 5DD, UK}
 
\ead{({\rm address for correspondence}) bernard.kay@york.ac.uk}

\begin{abstract} 
We discuss the relationship between the bulk-boundary correspondence in Rehren's algebraic holography (and in other `fixed-background', QFT-based, approaches to holography) and in mainstream string-theoretic `Maldacena AdS/CFT'. Especially, we contrast the understanding of black-hole entropy from the point of view of QFT in curved spacetime -- in the framework of 't Hooft's `brick wall' model  -- with the understanding based on Maldacena AdS/CFT.   We show that the brick-wall modification of a Klein Gordon field in the Hartle-Hawking-Israel state on 1+2 dimensional Schwarzschild AdS (BTZ) has a well-defined boundary limit with the same temperature and entropy as the brick-wall-modified bulk theory.   One of our main purposes is to point out a close connection, for general AdS/CFT situations, between the puzzle raised by Arnsdorf and Smolin regarding the relationship between Rehren's algebraic holography and mainstream AdS/CFT and the puzzle embodied in the `complementarity principle' proposed by Mukohyama and Israel in their work on the brick-wall approach to black hole entropy.  Working on the assumption that similar results will hold for bulk QFT other than the Klein Gordon field and for Schwarzschild AdS in other dimensions, and recalling the first author's proposed resolution to the Mukohyama-Israel puzzle based on his `matter-gravity entanglement hypothesis', we argue that, in Maldacena AdS/CFT, the algebra of the boundary CFT is isomorphic only to a proper subalgebra of the bulk algebra, albeit (at non-zero temperature) the (GNS) Hilbert spaces of bulk and boundary theories are still the `same' -- the total bulk state being pure, while the boundary state is mixed (thermal).  We also argue from the finiteness of its boundary (and hence, on our assumptions, also bulk) entropy at finite temperature, that the Rehren dual of the Maldacena boundary CFT cannot itself be a QFT and must, instead, presumably be something like a string theory.
\end{abstract}

\section{Introduction}
\label{Sect:intro}

\subsection{Maldacena holography, fixed-background holography and the Arnsdorf-Smolin puzzle}

The remarkable conjectured `holographic' correspondence \cite{Aharony:1999ti}, due originally to Maldacena
\cite{Maldacena:1997re}, between certain superstring theories in the bulk of Anti de Sitter space (AdS) (in suitable dimensions and times suitable-dimensional spheres and certain other compact spaces
$<$\ref{Note:AdSCFT}$>$)\footnote{The small Roman numerals in angle-brackets refer to the end section, Section
\ref{Sect:notes}, entitled `Notes'.} and certain supersymmetric conformal field theories (CFT) on the AdS conformal
boundary has inspired some authors to study simpler, but more general, types of correspondence between quantum
theories defined on a (fixed) AdS background and associated (again, conformal) quantum theories on the AdS conformal
boundary. In particular, in Rehren's `algebraic holography' \cite{Rehren:1999jn, Rehren:2000tp} a geometrically
natural bijection between bulk wedges and boundary double-cones is seen to induce a bijection between
nets of local *-algebras in the bulk and corresponding nets of local *-algebras on the conformal boundary
$<$\ref{Note:vNvsCstar}$>$ which bijects between bulk sub*-algebras for wedges and boundary sub*-algebras for
double cones, in such a way that the action of the AdS isometry group on the bulk algebra goes over to the action of
the conformal group on the boundary algebra. The nets of local *-algebras involved in this correspondence are of the
sort introduced earlier by Haag \cite{Haag} and others in order to give a general mathematical formulation of quantum
field theory (on a fixed background); the main axioms being those of isotony (i.e.\ that the *-algebra for a
subregion of a given region is a sub*-algebra of the *-algebra for the region) and commutativity at spacelike
separation (see Note $<$\ref{Note:wrap}$>$). And, indeed, it is clear that the Rehren correspondence amounts, at least in some known simple examples
involving bulk `ordinary' linear field theories, to a correspondence between theories which, in both bulk and
boundary, are quantum field theories. In particular, for a bulk theory consisting of the (real) covariant Klein-Gordon
field of mass $m$ satisfying
\begin{equation}
\label{KG}
(g^{ab}\nabla_a\nabla_b+m^2)\phi=0
\end{equation}
on AdS $<$\ref{Note:wrap}$>$ of any spacetime dimension 1+$d$ satisfying vanishing boundary conditions at the 
$d$-dimensional conformal boundary,  Kay and Larkin \cite{Kay:2007rf} have made use of a classical counterpart to
 holography, which they introduced and called `pre-holography', to obtain by second quantization (for integer or half-integer $\Delta$ in Equation (\ref{Delta}) below) concrete examples of Rehren's algebraic holography in which the bulk net of local *-algebras is the usual (C$^*$) algebra for the Klein Gordon theory and the boundary net of local *-algebras 
is a certain subalgebra $<$\ref{Note:thinloc}$>$ of the algebra of a real scalar conformal generalized free field theory \cite{Greenberg:1961mr, Duetsch:2002hc} on its conformal boundary with anomalous scaling dimension $\Delta$, where $\Delta$ is related to the mass, $m$, of the bulk field and the bulk spatial dimension ($=$ boundary spacetime dimension) $d$, by
\begin{equation}  
\label{Delta}
\Delta =\frac{d}{2} + \frac{1}{2}(d^2+4m^2\ell^2)^{\frac{1}{2}}
\end{equation}
where $\ell$ is the AdS radius (see Equations (\ref{E:a12}) and (\ref{E:104})).
(I.e.\ the algebra on the boundary is determined by having a commutator function equal to $W_b(x,x') - W_b(x',x)$ where $W_b$ is as in Equation (\ref{Wanom}) below.)
 
Moreover, Bertola et al.\ have proven \cite{Bertola:2000pp} that a Wightman theory (as they define it) on
the bulk of AdS, again in any spacetime dimension, $d+1$, will have, as its boundary limit, a (conformally invariant) Wightman theory on its ($d$-dimensional) conformal boundary $<$\ref{Note:Wightman}$>$ provided the boundary limit of the theory's $n$-point functions is defined so as to correspond to a field mapping:
\begin{equation}
\label{Blim}
\phi^{\mathrm{bulk}} \mapsto \phi^{\mathrm{boundary}}\ \ \hbox{where} \ \ \phi^{\mathrm{boundary}}({x^i})=\lim_{q\rightarrow q_0}
\Xi(q)^{\delta} \phi^{\mathrm{bulk}}(q,x^i)
\end{equation}
where $\Xi(q)$ is a suitable conformal factor $<$\ref{Note:confboundary}$>$ and $\delta$ an appropriate power.   

In the case of the covariant Klein Gordon equation for mass $m$ in the bulk with vanishing boundary conditions, this boundary-limit holography links the bulk theory in the geometrically natural global ground state (i.e.\ the ground state with respect to AdS global time) on AdS to the (as made clear in \cite{Kay:2007rf}) same conformal generalized free field theory with anomalous scaling dimension $\Delta$ which figures in Rehren's algebraic holography correspondence and, in this case, the $\delta$ in Equation (\ref{Blim}) above is the same as the $\Delta$ in Equation (\ref{Delta}).  We remark that, because the relevant quantum field theories in this case are free (respectively generalized-free) the latter boundary theory is fully determined by the boundary limit of the bulk two point function and, when we conformally map the boundary of a Poincar\'e chart to $d$-dimensional Minkowski space, this takes the form, 
\begin{equation}
\label{Wanom}
W_b(x,x')=\frac{1}{2\pi^{\frac{d}{2}}}\frac{\Gamma(\Delta)} 
{\Gamma(\Delta-d/2+1)}\frac{1} {[-(x^0-x'^0-i\epsilon)^{2}+({\bf x}-{\bf x'})^2]^\Delta} 
\end{equation}
which, by the way, specifies exactly what we mean by (the two point function of) a real scalar conformally invariant generalized free field with anomalous scaling dimension $\Delta$.

Clearly there are important differences between the Bertola et al.\ and Rehren versions of holography on the one hand
(below, we shall refer to either/both of these as {\it fixed-background holography}) and the mainstream work on the
Maldacena conjecture (which we shall refer to as {\it Maldacena holography}). In particular, in Maldacena holography,
the bulk theory (i.e.\ a certain superstring theory) involves a dynamical gravitational field rather than a fixed
background. Arnsdorf and Smolin \cite{Arnsdorf:2001qb} have pointed out a seeming paradox, resulting from this
difference.  Their point is that, if the quantum field theory on the conformal boundary of AdS$_5$ in the Maldacena work (see Note $<$\ref{Note:AdSCFT}$>$)  is both equivalent -- via Rehren's algebraic holography -- with a bulk quantum theory formulable in terms of a net of local *-algebras on a fixed AdS background and also equivalent -- via Maldacena-holography -- with a bulk theory which incorporates quantum gravity, then, it would logically follow that the latter, quantum gravity, theory must be formulable in terms of a net of local *-algebras on a fixed AdS background (with a notion of commutativity at spacelike separation built into its axioms -- see \cite{Rehren:1999jn} $<$\ref{Note:Rehgeo}$>$) -- which would seem to be in contradiction with evidence (based on the bending of light etc.) that Einstein gravity is a background-independent theory. 

On the other hand, some of the objections which have been made to fixed-background holography $<$\ref{Note:blog}$>$
appear to be unfounded.  These objections have to do with the very different-looking prescriptions for obtaining the bulk-boundary correspondence in Maldacena holography and in fixed-background holography.  In a quantum field theoretic version of Maldacena holography, which is believed to capture much of the essence of the Maldacena correspondence in suitable limiting cases, the bulk and boundary theories are thought of as being {\it dual} to one another according to the
schematic prescription (say in a Euclidean formulation) \cite{Witten:1998qj, Gubser:1998bc}
\begin{equation}
\label{dualstring}
<\exp(-\int {\cal O}j)>=\int \exp(-I[\phi(j)])
\end{equation}
where the integral on the right hand side is a path integral and $I(\phi(j))$ represents the action of an appropriate supergravity theory for fields, schematically represented by $\phi(j)$, with boundary limit (defined according to appropriate generalizations of (\ref{Blim}) above) equal to $j$, while, on the left hand side, $\cal O$ is the `dual' field on the boundary of AdS and is integrated over this boundary with $j$.
In a further limiting case which (see also the discussion in \cite{Rehren:2004yu}) involves  (say, in the case of the AdS$_5\times S^5$ version of the Maldacena correspondence) a classical supergravity theory in the bulk and a certain large $N$ limit of the corresponding CFT on the conformal boundary, the above schematic formula gets replaced \cite{Witten:1998qj, Gubser:1998bc}  by
\begin{equation}
\label{dual}
<\exp(-\int {\cal O}j)>=\exp(-I_{\mathrm{class}}[\phi(j)]) 
\end{equation}
where $I_{\mathrm{class}}[\phi(j)]$ now represents the (Euclidean) classical action of supergravity for fields (again, schematically represented by $\phi(j)$) with boundary limits (defined according to appropriate generalizations of (\ref{Blim})) equal to $j$.  Although, following custom, the correspondences in (\ref{dualstring}) and (\ref{dual}) are referred to as {\it dualities}, we note, in passing, that, as emphasized e.g.\ in \cite{Arnsdorf:2001qb}, which refers to the relation (\ref{dualstring}) as `conformal induction', [see also Note $<$\ref{Note:unfortunate}$>$] 
the mapping defined by (\ref{dualstring}) is one way: from bulk to boundary, and, at least not obviously, reversible. We shall return to this point in Section \ref{Sect:discuss}.

In apparent contrast, in, say, the Bertola et al.\ boundary limit holography \cite{Bertola:2000pp}, the relation between bulk and boundary theories is, as we saw above, that the boundary theory is the {\it direct} boundary limit, (\ref{Blim}), of the bulk theory and this, at first sight, appears rather different from the `duality' in e.g.\ the schema (\ref{dual}) $<$\ref{Note:unfortunate}$>$.  However, as pointed out by Rehren \cite{Rehren:2004yu} (and as Rehren notes there, as was also realized in some of the string-theory literature itself) mainstream work on Maldacena holography, based on the duality schema (\ref{dual}) e.g.\ by Witten \cite{Witten:1998qj} for a simple model with the bulk covariant Klein Gordon equation (\ref{KG}) (on a fixed bulk background) leads to exactly the same conformal generalized free field theory (specified by (\ref{Wanom})) with anomalous scaling dimension $\Delta$ (\ref{Delta}) on the conformal boundary as that given by Bertola et al.'s boundary-limit prescription.  And in fact Duetsch and Rehren \cite{Duetsch:2002wy} (see also \cite{Rehren:2004yu}) have given general arguments, valid for an interacting bulk field theory, as to why the direct boundary limit prescription and the duality prescription based on (\ref{dual}) will give the same boundary theory at least for certain field theories and to all finite orders in perturbation theory -- the reasons being not at all trivial.

So in spite of the differences, there would still seem to be at least a valid and interesting `family resemblance'
between the two types of holography.  As we have discussed, the situations considered in Maldacena holography involve
theories with very special properties (involving superstrings in the bulk and very special supersymmetric CFTs on the boundary)  and with deep features -- in particular a bulk geometry which participates in the dynamics -- which are not captured in the Bertola et al.\ framework and seemingly also not in the Rehren framework.   Nevertheless, 
fixed-background holography does appear to capture {\it some} of the more basic features of Maldacena holography while having the virtue of being simple enough to admit of fully mathematically controllable formulations.   Even if its study were to lead to the conclusion that a fully quantum field theoretic fixed-background holography must necessarily have certain bad properties, it could still be fruitful to try to understand how Maldacena holography manages to circumvent these bad properties.   In particular such a study could perhaps contribute to a better understanding of the relationship between full quantum gravity and quantum field theory in curved spacetime, and indicate how, in general (i.e.\ not just in AdS/CFT situations) quantum gravity (as realized by string theory) manages to resolve some of the difficulties inherent in an approximate semi-classical treatment of quantum gravity based on quantum field theory on fixed backgrounds.

In fact, Rehren has already pointed out in \cite{Rehren:1999jn} that a fully quantum field theoretic (fixed-background) holography will, on his algebraic holography approach, indeed necessarily have (for 1+$d$ dimensional AdS when $d$ is greater than 1) {\it potentially} bad properties in the sense that, while the bulk and boundary theories can both be Wightman theories, they cannot both be Wightman theories which satisfy the time-slice condition $<$\ref{Note:slice}$>$.  In the simple scalar field model discussed above, the bulk theory (i.e.\ the covariant Klein Gordon equation) is a Wightman theory satisfying the time-slice axiom but the boundary theory (i.e.\ the conformally invariant generalized free field with anomalous scaling dimension $\Delta$ as in Equation (\ref{Wanom})) doesn't satisfy the time-slice axiom \cite{Duetsch:2002hc} and is also known to certainly have bad properties.  In particular, we know, thanks e.g.\ to the general results of \cite{BuchJung86}, that it will have anomalous thermodynamical properties.  In general, however, it seems that failing to satisfy the time-slice axiom may not necessarily be bad and there seems to be evidence it could fail (or at least a lack of evidence that it holds) for certain physically realistic theories such as non-abelian gauge theories, cf. \cite{Rehren:1999jn, Rehren:2000tp}.  This is relevant to the discussion at the end of Section \ref{Sect:discuss}.

\subsection{Fixed-background holography on Schwarzschild AdS and BTZ}

In the present article, we wish to report on an investigation into the fixed-background version of holography due to Bertola et al.\ (i.e.\ {\it direct-boundary-limit holography}) where, instead of a plain AdS bulk spacetime, one considers an asymptotically AdS black hole spacetime.  In particular we shall consider the Bertola et al.\ boundary limit holography $<$\ref{Note:bhalgholo}$>$ for the covariant Klein Gordon equation with (for simplicity) zero mass  (and again with vanishing boundary conditions on the conformal boundary) on the (zero angular momentum) BTZ spacetime \cite{Banados:1992gq} which, as is well known, can be regarded as a black hole solution to Einstein's equations with negative cosmological constant in 1+2 dimensions.  Mathematically, BTZ is simpler than its counterparts in higher dimensions in that it is not only globally asymptotically AdS but also locally exactly AdS -- arising in fact from AdS by a certain quotient construction \cite{Banados:1992gq} and this construction can be exploited to obtain \cite{Lifschytz:1993eb} the bulk two-point function  for the BTZ-analogue of the Hartle-Hawking-Israel state (see below)  by the use of an image sum method from the two-point function for the AdS global ground state \cite{Avis:1977yn}  (for the case of vanishing boundary conditions on the conformal boundary).

Here, we recall that the Hartle-Hawking-Israel state (HHI state) \cite{Hartle:1976tp, Israel:1976ur} is the unique \cite{Kay:1988mu, Kay:1992gr} quantum state for the covariant Klein Gordon equation on the Kruskal spacetime which is invariant with respect to its one-parameter group of Schwarzschild isometries and for which the two-point function has the Hadamard form \cite{Kay:1988mu} required to have an everywhere finite renormalized stress-energy tensor.  Restricted to the exterior Schwarzschild spacetime (i.e.\ say, the right wedge of Kruskal) and regarded as a stationary state with respect to Schwarzschild time-evolution, it is a thermal equilibrium (i.e.\ KMS -- see Section \ref{Sect:prelim}) state at the Hawking temperature.  The BTZ analogue of the HHI state obtained by the image sum mentioned above will have appropriately analogous properties, being, again, a Hadamard state and stationary and thermal on the (say) right BTZ wedge at the appropriate Hawking temperature with respect to the (right-BTZ-wedge preserving) BTZ time translations.  From now on, we shall also call this an HHI state, relying on the context to make it clear when we are referring to the BTZ analogue.  (Similarly we shall refer to the HHI state on Schwarzschild AdS in higher dimensions.)

We shall also consider a simple 1+1 dimensional analogue to this bulk theory in which (cf. again \cite{Lifschytz:1993eb})   what takes the place (see Figure \ref{Fig:ads2}) of the 1+2 dimensional BTZ spacetime is a certain four-wedge
region of the 1+1 dimensional AdS spacetime; here the HHI state is just the restriction to the four-wedge region of the AdS ground state.  But of course the one-parameter group of AdS-isometries which gets identified, under the analogy, with `BTZ time-evolution' in (say) the right wedge differs from (global) `AdS-time'.  In fact, it stands in relation to global AdS-time in much the same way as Rindler time stands in relation to Minkowski time -- and thus the thermal property can be understood as analogous to the Unruh effect \cite{Unruh:1976db}.   

\smallskip

For ease of exposition, we shall actually treat this latter, simpler, 1+1 dimensional model first and in detail (in Section \ref{Sect:1+1}) postponing a concise treatment of the 1+2 dimensional case to Section \ref{Sect:1+2}.

\subsection{The contrasting accounts of the origin of thermality and entropy in the two versions of holography}

Our main purpose will be to compare and contrast what happens in this quantum-field-theory-in-curved-spacetime model with what happens in Maldacena holography when one considers appropriate superstring/supergravity theories on a Schwarzschild-AdS bulk.  In particular we shall compare and contrast the different understandings, suggested by each version of holography, of how and why the bulk theory goes over to a thermal theory on the conformal boundary, and we shall compare and contrast the different understandings of the origin of black hole entropy suggested by each version.

As is well-known (see e.g.\ \cite{Maldacena:2001kr})  such a bulk geometry will, when maximally extended, have a conformal boundary which consists of two cylinders (see Figure \ref{Fig:BTZcylinders}) -- each cylinder being the conformal boundary of one of the two spacelike exterior-Schwarzschild-like `wedges' of the maximal extension.  One of the notable results \cite{Maldacena:2001kr} of the Lorentzian signature version of Maldacena holography is that the state of the relevant CFT on this disconnected boundary is believed to have a restriction to each single cylinder which is a thermal state at the appropriate Hawking temperature -- the pair of restricted states being mutually entangled in just such a way that the total state on the pair of cylinders is pure.  ({\it Added note}: In \cite{Kay:enclosed} one of us nevertheless argues against the correctness of the picture in \cite{Maldacena:2001kr} in the context of quantum gravity -- see also Note $<$\ref{Note:double}$>$ and also the two paragraphs following our italicized statement in Section \ref{subsect:matgrav}.)

  On observing that (using a Euclidean approach and then Wick-rotating) the appropriate replacement for (\ref{dualstring}) formally implies that the bulk and boundary partition functions coincide, and assuming that this remains true in full string-theoretic Maldacena holography, one expects the entropy of the thermal state on each single cylinder to equal the entropy of the bulk which, in turn, one expects (in a suitable limit of weak string coupling and a large ratio $\ell/\ell_s$ where $\ell_s$ is the string length scale and for temperatures above the Witten phase transition \cite{Witten:1998zw} -- see Note $<$\ref{Note:AdSCFT}$>$) to equal 
the Hawking value (one quarter of the area of the event horizon $<$\ref{Note:onequarter}$>$) of the Bekenstein-Hawking entropy for the bulk asymptotically AdS black hole.  Most remarkably, this is indeed the value of the entropy obtained for the thermal state of the CFT on a single cylinder.  (In the case of AdS$_5\times S^5$ with a 1+3 dimensional conformal boundary, this agreement cannot be checked exactly because one can only calculate the boundary entropy in a different regime -- see Note $<$\ref{Note:AdSCFT}$>$ and the references we cite there.   But the agreement is exact in the AdS$_3$ versions with a BTZ bulk and a 1+1 dimensional conformal boundary -- see \cite{Ross:2005sc} and refs. therein.) 

One might summarize the situation by saying:

\smallskip

\noindent
\textit{In Maldacena holography, the (bulk) entropy of the black hole (which equals one quarter of the area of its event horizon) is equal to the entropy of the dual field on (say) the right boundary cylinder.  In the bulk entropy calculation in Maldacena holography, the entropy is regarded (as it is in Euclidean quantum gravity) as an attribute mainly of the (dynamical) quantum gravitational field.}

\smallskip 

We remark, concerning the last sentence of this statement and in preparation for our later discussion (about the Mukohyama-Israel `complementarity' hypothesis -- see below) that, in the case of asymptotically flat black holes, the Hawking value for the entropy (i.e.\ one quarter of the area of the event horizon)  is deriveable directly from the bulk partition function if one calculates this within the framework of  Euclidean quantum gravity \cite{Gibbons:1976ue, Hawking:1980gf}.  This defines it \cite{Gibbons:1976ue} as the path integral of the exponential of minus the appropriate Euclidean quantum gravity action for Einstein gravity with appropriate matter fields and with a suitable boundary geometry periodic in imaginary time with period $\beta=1/{\mathrm{temperature}}$. In fact, Gibbons and Hawking \cite{Gibbons:1976ue} approximate this by the classical action which they point out is entirely the gravitational action of the relevant classical Euclidean black hole solution and they show that this leads to the Hawking value for the entropy.  Thus, on the assumption (which Gibbons and Hawking make) that the one and higher-loop terms also contributing to the total action (which represent physically the [gravity and matter] `thermal atmosphere' of the black hole) are small, one re-obtains (approximately) the Hawking value for the entropy.   All these developments were shown by Hawking and Page to generalize to asymptotically AdS black holes in \cite{Hawking:1982dh} and indeed, as these authors point out, this case is more satisfactory in that there are stable equilibria involving a black hole surrounded by its thermal atmosphere whereas for Schwarzschild black holes, there is the well-known difficulty of negative specific heat \cite{Gibbons:1976ue, Hawking:1982dh}.

We want to draw particular attention to the importance, in this explanation of the origin of the Hawking value for the entropy, of the assumption that the entropy of the thermal atmosphere is small (or perhaps, alternatively, for some other reason, neglectable).   In \cite{Hawking:1980gf} Hawking argues for its smallness by proposing that this term will be well-approximated by the entropy of an equal metrical volume of thermal radiation in flat spacetime and points out that, in relevant situations, this will indeed be small compared to the Hawking value of the black hole entropy.  In the work in \cite{Hawking:1982dh} on the thermodynamics of Schwarzschild AdS, it appears to be tacitly assumed that the entropy of the corresponding thermal atmosphere can similarly be neglected, and/or estimated as the (again, small) entropy of thermal radiation at the same temperature in plain AdS for the same cosmological constant (but without a black hole). 

As we shall discuss further below, this raises a puzzle (related to the Mukohyama-Israel `complementarity' principle which we discuss below) because, as we shall see, the (Lorentzian) brick wall approach (which we advocate) leads us to conclude that, in addition to the small `volume' piece, the thermal atmosphere entropy also has a piece proportional to the area of the event horizon which is comparable in size to the Hawking value of the black hole entropy.  In fact, as we shall see in Section \ref{Sect:brick}, for our scalar Klein Gordon equation on the (1+2 dimensional) BTZ background, the brick wall approach entails that the area piece is substantial while (with the approximation we use) the volume piece vanishes!

Turning next from Maldacena holography back to fixed-background holography,  if we assume our bulk theory (i.e.\ the covariant  Klein Gordon equation on the maximally extended BTZ spacetime with vanishing boundary conditions on the conformal boundary) to be in the HHI state, then, one of us (LO) \cite{Ortiz:2011mi, lOrt11} has obtained, by taking a direct boundary limit as in (\ref{Blim}), a similar result involving two mutually entangled thermal states (i.e.\ for the conformal generalized free field with anomalous scaling dimension $\Delta$ (\ref{Delta})) on the same pair of boundary cylinders -- again at the relevant Hawking temperature $<$\ref{Note:bhalgholo}$>$.

This result had previously been obtained, with reference to the duality schema (\ref{dual}), for the same covariant Klein Gordon model, by Keski-Vakkuri \cite{KeskiVakkuri:1998nw} (and quoted in the paper by Maldacena \cite{Maldacena:2001kr} to which we referred above).   However, the method by which they are obtained in \cite{Ortiz:2011mi, lOrt11}, summarized here in Sections \ref{Sect:1+1} and \ref{Sect:1+2}, may be of interest as an alternative derivation, closer in spirit to the boundary-limit holography work of \cite{Bertola:2000pp}.

 One can, in fact, argue on general grounds (cf. \cite{Kay:1985zs}), using the algebraic approach to quantum field theory (see Section \ref{Sect:prelim}) that the direct boundary limit on the pair of cylinders of a bulk theory defined on BTZ would be expected $<$\ref{Note:expect}$>$ to inherit the entanglement and thermal properties of the bulk theory in this way.  To argue this, we note first that we would expect that the quantum dynamical system $({\cal A}_{\mathrm{DW}}, \alpha_{\mathrm{DW}}(t))$ consisting of the *-algebra for the BTZ double wedge (i.e.\ in the 1+1 dimensional analogue, the union of the triangles ACF and CED in Figure \ref{Fig:ads2}) together with the one-parameter group of automorphisms induced by the one-parameter group of wedge-preserving BTZ isometries to be equivalent to the quantum dynamical system $({\cal A}_{\mathrm{DC}}, \alpha_{\mathrm{DC}}(t))$ consisting of the *-algebra for the appropriate boundary theory on the `double cylinder' -- i.e.\ the union of the right and left boundary cylinders -- together with the one-parameter group of automorphisms which time-translates towards the future on the right cylinder and toward the past on the left cylinder (this arising from the one-parameter group of conformal isometries -- actually isometries -- on the boundary inherited from the one-parameter group of wedge-preserving isometries in the bulk).  Thus, to every bulk state (i.e.\ positive normalized linear functional on ${\cal A}_{\mathrm{DW}}$ -- see Section \ref{Sect:prelim}) -- with given entanglement properties and given thermal properties with respect to $\alpha_{\mathrm{DW}}(t)$, one expects an equivalent boundary state with equivalent entanglement properties and equivalent thermal properties with respect to $\alpha_{\mathrm{DC}}(t)$  $<$\ref{Note:equiv}$>$.  Moreover, we expect  that suitably smeared  boundary limits (in the sense of (\ref{Blim})) of bulk fields will be identifiable with elements of the *-algebra for the appropriate boundary theory on the union of the  right and left boundary cylinders in such a way that the action of bulk time-translations goes over to the action of the above-mentioned cylinder time-translations.  (Further, we expect all these expectations to generalize from BTZ to Schwarzschild AdS in other dimenions.)

However, a formal calculation now results in an {\it infinite} value for the entropy of each of these (bulk and boundary) thermal states. 

This infinite entropy should not have been a surprise for two rather different reasons which we point out below.  First, we remark that the bulk and boundary entropies are expected, on general grounds, necessarily to be equal and thus, if one is infinite, then so must the other be.  In fact, $<$\ref{Note:Ham}$>$, we expect the Hamiltonian generating the BTZ-analogue of Schwarzschild time-evolution in the BTZ-analogue of the Boulware vacuum representation \cite{Boulware:1974dm} for the field *-algebra in the right BTZ wedge to be unitarily equivalent to the Hamiltonian generating time-evolution in the ground-state representation for the field algebra on the right cylinder.  Thus, since the entropy of a thermal state (at any inverse temperature, $\beta$) is given by the standard formulae 
$S=-\tr\rho\ln\rho$ where $\rho=e^{-\beta H}/Z$ where $Z=\tr(e^{-\beta H})$, and since the trace is a unitary invariant, if the Hamiltonians are unitarily equivalent, then the entropies must necessarily be equal.  

The first reason why we expect this entropy to be infinite, is, thinking of the entropy as the boundary entropy, because of the `bad thermodynamics' of the conformal generalized free field theory with anomalous scaling dimension $\Delta$ (see (\ref{Delta}) and (\ref{Wanom})) on the conformal boundary which we mentioned above.

The second reason why we expect this entropy to be infinite, now thinking of the entropy as bulk entropy, is because we expect the bulk Hamiltonian (and thus also, by the unitary equivalence, the boundary Hamiltonian) to have continuous spectrum whereupon, of course, the trace in the definitions of $Z$ and $\rho$ above will be infinite.  We next turn to discuss the reason why we expect the bulk Hamiltonian to have continuous spectrum and explain the relevance of the brick wall model.

The reason for the continuous spectrum is essentially the same as the reason that the spectrum of the `Boulware Hamiltonian' \cite{Boulware:1974dm} on the right Schwarzschild wedge of the Kruskal spacetime (i.e.\ the Hamiltonian generating Schwarzschild time-evolution in the ground-state representation of the *-algebra for the right Schwarzschild wedge)  is continuous, even when one encloses the field in a box at some large Schwarzschild radius (the $L$ of Equation (\ref{Sapprox})) and imposes (say) vanishing boundary conditions at the wall of this box.   This continuous spectrum, in turn, is what is responsible for the fact that the entropy of  the HHI state for the same system is infinite even when enclosed in a large box.  This sort of infinity is in fact well-known in the context of quantum field theory in curved spacetime; it also occurs, e.g., for the Minkowski vacuum state restricted to a Rindler wedge (again, even when enclosed in a box at large Rindler spatial coordinate) in Minkowski space.  Its origin may regarded as due to the fact that (in the Kruskal example) as far as Schwarzschild wave-modes of our covariant Klein-Gordon equation are concerned, the relevant radial coordinate is not the metrical distance from the horizon $<$\ref{Note:metrical}$>$ but rather the appropriate Regge-Wheeler or `tortoise' coordinate $r^*\in (-\infty, \infty)$, related to the Schwarzschild $r\in (2M,\infty)$ by
\begin{equation}
\label{Schwtortoise}
r^*=r+2M\ln(r/2M-1)
\end{equation}
which, if taken as a sort of `distance' measure places the horizon infinitely far away from any exterior point.  (In Minkowski space it is the Rindler coordinate, $x$, related, at Minkowski time $T=0$, to Minkowski $X$ coordinate by $x=\ln X$ -- see e.g.\ \cite{Kay:1985zs}.)  Alternatively, it may be regarded as due to the fact that the local temperature (relative to Schwarzschild time or Rindler time -- see Section \ref{Sect:prelim}) tends to infinity as one approaches the horizon -- like inverse metrical distance.

\subsection{'t Hooft's brick wall model and the Mukohyama-Israel puzzle}

Following 't Hooft \cite{'tHooft:1984re} and subsequent authors (see especially the important correction and clarification to the brick wall approach due to Mukohyama and Israel \cite{Mukohyama:1998rf}) it is well known -- for say the Schwarzschild metric -- that, for any linear model field theory, one will obtain a Hamiltonian with discrete spectrum if one encloses the system in a large box and also separates it from the horizon by another wall (called a `brick wall' in \cite{'tHooft:1984re}) just outside the horizon where, say, vanishing boundary conditions are again imposed.  If one does this, and takes the system between the brick wall and the box wall to be at the appropriate Hawking temperature, the entropy of the resulting system turns out to be finite.   It  turns out (in the case of the massless Klein Gordon field) to consist of the sum of a `volume' piece which is close in magnitude to the entropy of the same Klein Gordon field in an equal metrical volume of flat space and an `area piece' which is proportional to the area of the event horizon.  

Also following 't Hooft and Mukohyama and Israel, we shall assume that the fields which actually exist in nature can be modelled, for thermodynamic purposes, by a number,  $\mathsf{N}$ $<$\ref{Note:MIcorrection}$>$, of `effective (real massless) Klein Gordon fields' in the sense that their combined entropy will be $\mathsf{N}$ times the entropy of a single (real massless) Klein Gordon field.  

By adjusting, `by hand',  the metrical distance, $\alpha$, from the brick wall to the horizon to be (approximately) the Planck length (i.e.\ by taking $\alpha\approx 1$ in natural units) one then finds that the area piece of the entropy is significantly large.  In fact, following the treatment of \cite{Mukohyama:1998rf}, one obtains the Hawking value for the proportionality constant, i.e.\ one quarter, provided one takes ${\mathsf{N}}$ to be of the order of (only) 300 or so.  (In fact, for the area piece to take the Hawking value, one requires ${\mathsf{N}}/\alpha^2$ to have the value
$90\pi\simeq 300$. See Note $<$\ref{Note:300}$>$ for details.)  This of course does not prove that it is physically correct to take $\alpha\approx 1$  and ${\mathsf{N}} \approx 300$ and, in particular,  it does not prove that all the entropy of a black hole arises in this way.  But it does suggest that at least a substantial part of it does.  And it is then tempting to argue, as 't Hooft and Mukohyama and Israel do, that the entire entropy of a black hole can be seen as arising in this way and that this way of arriving at the entropy should be thought of as a {\it complementary description} of black hole entropy to the description of it as arising from the Gibbons-Hawking Euclidean classical gravitational action $<$\ref{Note:complementarity}$>$ which we discussed above.  

Thus with Mukohyama and Israel's proposed complementarity principle, it seems that such a brick-wall modification of a quantum field theory near a horizon is able to mimic what we believe (see Note $<$\ref{Note:onequarter}$>$) would be an exact and correct result for the entropy of a black hole in a full quantum gravitational treatment.  Moreover, by the way,  if one accepts their argument, then the number 300 may be regarded (see Note $<$\ref{Note:300}$>$) as an order of magnitude prediction for the effective number of matter fields needed in any consistent quantum theory of gravity describing physics in 1+3 dimensions. 

The Mukohyama and Israel complementarity principle actually embodies, what is, for us, a second puzzle (our first being that of Arnsdorf and Smolin).  After all, as we just remarked,  the `volume' piece of the entropy can be estimated as the entropy of thermal radiation at the Hawking temperature in flat space in a box of the same metrical volume.  Thus it may, presumably, be identified with the `small' volume piece which Hawking proposed for the entropy of the thermal atmosphere in the Euclidean quantum gravity approach to black hole entropy which we discussed earlier.  However, as we anticipated in that discussion, and as we have just seen, with the brick wall approach, the main part of the entropy of the thermal atmosphere is expected to be the piece proportional to the area of the event horizon which, following Mukohyama and Israel \cite{Mukohyama:1998rf} can, as we just saw, plausibly already account for the full Hawking entropy of the black hole.  Thus it seems one is forced to conclude, with Mukohyama and Israel, that, for reasons we perhaps do not yet understand, the thermal atmosphere (perhaps both the [small] volume and the [large] area pieces) should not be added to the gravitational entropy of a black hole (which already accounts for the full Hawking value for its entropy) but rather should be regarded as an alternative calculation (or `complementary description') of the same physical quantity.  (See Section \ref{Sect:discuss} for further discussion.)

Returning to the  case of interest here of Schwarzschild-AdS black holes, a similar brick-wall scenario is applicable (see e.g.\  \cite{hep-th/0011176} -- but see Note $<$\ref{Note:complementarity}$>$) and we note that, in this case, a
large box is no longer needed and gets replaced by imposing vanishing boundary conditions on our quantum field(s) on the conformal boundary (i.e.\ at spacelike infinity).  Again one expects the entropy to consist of the sum of a `volume' piece and a piece proportional to horizon area (i.e.\ circumference in the case of BTZ).   As we have already remarked above,  it turns out \cite{Kim:1996eg} that in the case of 1+2 dimensional BTZ (at least to a good approximation) the entire entropy (see equation (\ref{1plus2bricksecondresult}) in Section \ref{Sect:brick})  is proportional to the circumference of the horizon (i.e.\ the appropriate notion of `horizon area' in 1+2 dimensions)  and, strikingly, the `volume piece' vanishes.  (Again in line with the Mukohyama-Israel complementarity principle/puzzle, this is in sharp contrast to the assumptions of Hawking and Page in their work \cite{Hawking:1982dh} on the thermodynamics of Schwarzschild AdS, where it is assumed that the only term which goes like horizon area is the term arising from the zero-loop gravitational action.)   We shall re-derive this result here, by applying the method and ideas of Mukohyama and Israel \cite{Mukohyama:1998rf} in Section \ref{Sect:brick}.  We find, by the way, that, for a brick wall of metrical distance 1 natural unit from the horizon, the entropy will equal the usual Hawking value (one quarter of the circumference) provided the matter fields behave as an effective number of around 34 Klein Gordon fields.  (In fact, by the results of Section \ref{Sect:brick}, the entropy will take the Hawking value if ${\mathsf{N}}/\alpha=4\pi^3/(3\zeta(3))\simeq 34$. Cf. the discussion of the 1+3 dimensional Schwarzschild case above, where the corresponding quantity, ${\mathsf{N}}\alpha^2$, needed to take a value around 300.)  (As far as we are aware this has not been pointed out before.)

At a computational level, our main new development in the present paper is that we show, in Section \ref{Sect:brick},  in our 1+1 and 1+2 dimensional BTZ models with a single real massless Klein Gordon field, that the introduction of a suitable brick wall into the bulk alters the state induced from the appropriate bulk thermal state by the direct boundary limit (\ref{Blim}) on our (say) right cylinder boundary (in the case of 1+1 dimensional BTZ, right boundary line) in such a way that the altered boundary state remains a thermal state at the Hawking temperature, but in such a way that the boundary Hamiltonian has a discrete spectrum and the state induced on the boundary has a finite entropy.   In fact, in both 1+1 and 1+2 dimensional cases, the finite boundary entropy is equal to the finite brick-wall-modified bulk entropy. 

We may summarize much of what we have said above with the statement:

\smallskip

\noindent
\textit{We assume we can mimic a matter atmosphere realistically enough by considering a multiplet of $\mathsf{N}$
(massless) Klein Gordon fields on BTZ. By introducing a suitable brick wall, the entropy of this model thermal atmosphere
becomes finite and is equal to the entropy of the CFT on the conformal boundary obtained from the brick-wall modified bulk
theory with fixed-background (specifically direct-boundary-limit) holography. Moreover, by adjusting the metrical
distance, $\alpha$, from the brick wall to the horizon to be around 1 in natural units (i.e.\ around the `Planck
length' -- see Section \ref{Sect:prelim}), the entropy can be made to equal the Hawking value of one quarter of the
area (i.e.\ circumference) of the event horizon by taking $\mathsf{N}/\alpha$ to be around 34.  In the bulk entropy calculation in the brick wall approach, the entropy is regarded as an attribute of a collection of (mainly) matter fields propagating on a fixed gravitational background.}
 
\subsection{The connection between the Arnsdorf-Smolin and Mukohyama-Israel puzzles}

A comparison of our two italicised displayed statements above seems to suggest a new and interesting connection between, on the one hand, the puzzling contrast between Maldacena holography and fixed-background holography -- as highlighted e.g.\ by Arnsdorf and Smolin's paradox -- and, on the other hand, the puzzle embodied in the 
Mukohyama-Israel complementarity principle, according to which the (entire) entropy of a black hole should describable either as the entropy derived from the zero-loop quantum gravitational partition function, or, in a complementary description, as the entropy of its thermal (mostly matter) atmosphere.  In fact our first italicised statement above indicates that Maldacena holography goes together with the first description and our second italicised statement indicates that fixed-background holography goes together with the second description.

Thus, we have pointed out an apparently interesting and new connection between the two puzzles.  

\subsection{Outline of paper and brief preview of conclusions}

After a section (Section \ref{Sect:prelim}) entitled Preliminaries, we recall the relevant facts about direct boundary-limit holography for BTZ, treating the 1+1 dimensional case in Section \ref{Sect:1+1} and the 1+2 dimensional case in Section
\ref{Sect:1+2}.  For each case, we construct the HHI state for a single massless Klein Gordon field and show that its direct boundary limit on (say) the right boundary cylinder (= boundary line in the 1+1 case) is a thermal state at the relevant Hawking temperature.   Then, in Section \ref{Sect:brick} we introduce the brick wall and show that the brick-wall modified HHI state for a collection of $\mathsf{N}$ massless Klein Gordon fields remains a thermal state on the right boundary cylinder (/line) at the Hawking temperature, but now has a finite entropy given by equation (\ref{1plus2bricksecondresult}) and that this is equal to the entropy of the state induced on the boundary from the brick-wall modified HHI state.

In the first part of our Discussion Section -- Subsection \ref{subsect:summary} of Section \ref{Sect:discuss} -- we summarize our main results.  In the second part (Subsections \ref{subsect:working} to \ref{subsect:dual}) we discuss further our two puzzles, i.e.\ those of Arnsdorf and Smolin and of Mukohyama and Israel, in the light of our results and with reference to the AdS$_5$ version of Maldacena holography -- and we discuss what conclusions can be drawn, in the light of our results, regarding these puzzles.  We shall work on the assumption that our main results and, in particular, the equality of bulk and boundary entropies (in the presence of suitable brick walls) 
will hold for bulk QFT other than the Klein Gordon theory and for Schwarzschild AdS in dimensions other than 1+2.  We shall also recall the proposed resolution to the Mukohyama-Israel puzzle due to one of us (BSK)  based on the `matter-gravity entanglement hypothesis' of references \cite{Kay:1998vv, Kay:1998cj, Abyaneh:2005tc, Kay:2007rx}.   The main tentative conclusions we arrive at, and our arguments for them, may be very briefly summarized as follows:   First we shall argue from this latter hypothesis that, in Maldacena AdS/CFT, the algebra of the boundary CFT is isomorphic only to a proper subalgebra of the bulk algebra, albeit (at non-zero temperature) the (GNS) Hilbert spaces of bulk and boundary theories are still the `same' -- the total bulk state being pure, while the boundary state is mixed (thermal).   Secondly, we shall argue from the finiteness of its boundary (and hence, on our assumptions, also bulk) entropy at finite temperature, that the Rehren dual of the Maldacena boundary CFT cannot itself be a QFT and must, instead, presumably be something like a string theory.   As we say again, at the end of Section \ref{Sect:discuss}, while these tentative conclusions still do not resolve the Arndorf-Smolin puzzle, they seem to go at least some of the way towards a possible resolution.

\section{Preliminaries}
\label{Sect:prelim}

We adopt signature (+,-,-,-).  Concerning units:  In the main part of the paper, we omit all factors of $c$, $\hbar$, $G$, and $k$ (Boltzmann's constant) from our formulae and so regard all physical quantities as pure numbers.  In 1+3 dimensions, where theory can be compared with experiment, this of course amounts to using Planck units in which lengths are multiples of the Planck length ($\approx 10^{-33}$ cm), times are multiples of the Planck time ($\approx 10^{-43}$ sec) and masses are multiples of the Planck mass ($\approx 10^{-5}$ g).  (In Note $<$\ref{Note:AdSCFT}$>$ and Section \ref{Sect:discuss} we shall restore $G$.)  We use the symbol `$\approx$' to denote `of the order of' and `$\simeq$' to denote approximate equality.

The appropriate mathematical framework for quantum theory in curved spacetime is the algebraic approach to quantum field theory -- see \cite{Haag:1967sg} for the general framework and  \cite{Kay:2006jn}  and \cite{Kay:1988mu} for the specific application to linear quantum field theories in curved spacetime such as that of our Klein Gordon equation (\ref{KG}).  We briefly recall some of the salient points although many of the details will only be needed in the Notes section (Section \ref{Sect:notes}).  Observables are regarded as self-adjoint elements of a *-algebra, for which we shall typically use the symbol $\cal A$, and a state, $\omega$, on $\cal A$ is a positive normalized linear functional and, in the case of a quantum field theory, is specified in practice by listing all of its (smeared) $n$-point functions. 

The GNS representation of a *-algebra, $\cal A$, for a choice of state, $\omega$, is a representation, $\rho$, of $\cal A$ as operators on a Hilbert space $\cal H$, together with a vector, $\Omega\in {\cal H}$ such that $\Omega$ is {\it cyclic} for $\rho$ (meaning the set $\rho({\cal A})\Omega$ is dense in $\cal H$) and satisfies
$\omega(A)=\langle\Omega|\rho(A)\Omega\rangle \ \forall A\in {\cal A}$.  Moreover, if we have an algebraic quantum dynamical system $({\cal A}, \alpha(t))$ consisting of a *-algebra, ${\cal A}$, together with a one-parameter group, $\alpha(t)$, of `time-translation' automorphisms, $\alpha(t)$, then there will be a one-parameter unitary group $U(t)$  (which we shall assume to be strongly continuous and which will, hence, arise in the form $U(t)=e^{-iHt}$) for a self-adjoint `Hamiltonian generator', $H$, on the GNS Hilbert space $\cal H$ which {\it implements} $\alpha(t)$ in the sense that $\rho(\alpha(t)(A))=U(-t)\rho(A)U(t) \ \forall A\in {\cal H}$.  It is well-known and easy to show that $U(t)$ is fixed uniquely if we demand $U(t)\Omega=\Omega$ (i.e.\ $H\Omega=0$) for all $t\in {\cal H}$.  When such a condition holds for each of two equivalent  quantum dynamical systems
$({\cal A}_1, \alpha_1(t))$ and $({\cal A}_2, \alpha_2(t))$, then one easily sees that the Hamiltonian generators,  $H_1$ and $H_2$, say, must then be equivalent in the sense that there exists an isomorphim $V: {\cal H}_1\rightarrow {\cal H}_2$ such that
\begin{equation}
\label{equivHam}
VH_1=H_2V. 
\end{equation}

Whenever we refer to a `ground state' or to a `thermal state' at some given `inverse temperature', there is always, at least implicitly, a notion of `time' being assumed and so these notions are defined formally relative to a given quantum dynamical system $({\cal A}, \alpha(t))$.  A state, $\omega$ on the $\cal A$ of such a dynamical system is said to be
a ground state if its Hamiltonian generator, defined, and made unique as in the above paragraph, is a positive operator. 
Ground states are characterised by having two-point functions which are translationally invariant and, regarded as functions of $t=t_1-t_2$, say, are the boundary limits of holomorphic functions in the lower half $t$ plane (and an appropriate statement for other $n$-point functions).  The two-point function of a thermal state, at inverse temperature, $\beta$, will satisfy the KMS condition (see again \cite{Haag} for the general theory and especially \cite{Haag:1967sg}; see also  \cite{Kay:2006jn} and \cite{Kay:1988mu} for the case of linear field theories in curved spacetimes and see also \cite{saFullsnRui87})  which means that it is, again, translationally invariant in time and extends to a function, $G$ of $t=t_1-t_2$ which is holomorphic in the strip, $-\beta <\mathrm{Im} t < 0$ such that the boundary value, $G(-t)$ equals the boundary value $G(t-i\beta)$. 

We note that, if $H$ is a Hamiltonian on a Hilbert space $\cal H$ and if $Z={\rm tr}(e^{-\beta H})$ is finite, then the state, $\omega_\beta$, defined on operators, $A$, on $\cal H$ by $\omega_\beta(A)=
{\rm tr}(e^{-\beta H} A)/Z$  will be a KMS state at inverse temperature $\beta$ and its GNS Hilbert space will, in a certain sense explained in \cite{Kay:1985yw} (cf. also the `thermo-field dynamics' of Takahashi and Umezawa \cite{TakUme, Takahashi:1996zn}) be larger than $\cal  H$ (in fact ${\cal H}\otimes {\cal H}$) and may be thought of as carrying a GNS representation associated to a pure state (the `purification' of $\omega_\beta$) on a larger system whose reduced density operator on $\cal H$ is equal to $e^{-\beta H}/Z$ .  This remark will be relevant in Note $<$\ref{Note:double}$>$ and in Subsection \ref{subsect:matgrav} of Section \ref{Sect:discuss}.

In a spherically symmetric spacetime of dimension 1+$d$ ($d>1$) with a metric of form
\begin{equation}
\label{generalsphmetric}
ds^2=f(r)dt^2-f(r)^{-1}dr^2-r^2ds_{{\mathrm S}^{d-1}}^2,
\end{equation}
where $ds_{{\mathrm S}^{d-1}}^2$  represents the metric on a sphere of dimension $d-1$,  then, for a given temperature, $\cal T$, with respect to $t$, one can also refer to a `local temperature' (see (\ref{Tlocal})) equal to $f^{-1/2}{\cal T}$. (This also applies in $d=1$ with the last term in the above equation absent.)  In the case of the Schwarzschild metric, $\cal T$ then coincides with the temperature at infinity since the relevant $f$ tends to $1$ at infinity.  In the case of the Schwarzschild-AdS metric (such as our BTZ metrics, given by (\ref{E:bb3}) and (\ref{E:16}) in 1+1 dimensions and (\ref{E:109}) and (\ref{E:110}) in 1+2 dimensions) the local temperature is infinity on the horizon and zero at infinity.  However, because Schwarzschild-AdS time (in 1+1 and 1+2 dimensions, what we call here BTZ-time) in the bulk goes over to `cylinder-time' on the conformal boundary, temperature with respect to cylinder time on the conformal boundary will coincide with temperature with respect to BTZ-time in the bulk.

Finally, we note that, for metrics of the above spherically symmetric form,
if a horizon is located at $r_+$, indicated by the vanishing of $f(r_+)$, then its surface gravity, $\kappa$ is equal to $f'(r_+)/2$.

\section{The boundary limit of the massless Klein Gordon QFT on 1+1 dimensional AdS and 1+1 dimensional BTZ}
\label{Sect:1+1}

We first briefly recall some well-known facts about coordinate systems for 1+1 dimensional AdS (we'll just call it AdS for the rest of this section) and then, following Lifschytz and Ortiz \cite{Lifschytz:1993eb}, define what we mean by 1+1 dimensional BTZ.

1+1 dimensional AdS with radius $\ell$ can be defined to be the surface
\begin{equation}
\label{E:a12}
u^{2}+v^{2}-x^{2}=\ell^{2}
\end{equation}
embedded in $\mathbb{R}^{3}$ with metric
\begin{equation}
\label{E:2}
ds^{2}=du^{2}+dv^{2}-dx^{2}.
\end{equation}

The metric is given by the pullback of
(\ref{E:2}) to (\ref{E:a12}) under the inclusion map. We are interested in three parametrizations of (\ref{E:a12}). The first is with
\begin{equation}
\label{E:a13}
u=\ell\sec\rho\sin\lambda
\hspace{0.7cm}v=\ell\sec\rho\cos\lambda\hspace{0.7cm}x=\ell\tan\rho,
\end{equation}
where $\lambda\in[-\pi,\pi)$ and $\rho\in(-\pi/2, \pi/2)$ which gives the metric in global coordinates:
\begin{equation}
\label{E:a14}
ds^{2}=\ell^{2}\sec^{2}\rho\left(d\lambda^{2}-d\rho^{2}\right).
\end{equation}

The second is 
\begin{equation}
\label{E:77}
u=\frac{\ell T}{z}\quad
v=\frac{1}{2z}\left(z^{2}+\ell^{2}-T^{2}\right)\quad
x=\frac{1}{2z}\left(-z^{2}+\ell^{2}+T^{2}\right),
\end{equation}
where $(T,z)\in(-\infty,\infty)\times(0,\infty)$
which gives the metric in Poincar\'e coordinates: 
\begin{equation}
\label{E:79}
ds^{2}=\frac{\ell^{2}}{z^2}\left(dT^{2}-dz^{2}\right).
\end{equation}

A third possibility (cf. \cite{Lifschytz:1993eb}) is to choose a positive number, $M$ (which will get interpreted as the `BTZ mass') and to parametrize (\ref{E:a12}) with
\[
v=\ell\left(\frac{r^{2}-r_{+}^{2}}{r_{+}^{2}}\right)^{1/2}\sinh\kappa
t\hspace{0.5cm}x=\ell\left(\frac{r^{2}-r_{+}^{2}}{r_{+}^{2}}\right)^{1/2}\cosh\kappa t
\]
\begin{equation}
\label{E:bb2}
u=-\ell\frac{r}{r_{+}},
\end{equation}
where $r_+=\ell\sqrt{M}$ and $\kappa=\frac{r_{+}}{\ell^{2}}$ $(=\sqrt M/\ell)$. This gives the
metric as
\begin{equation}
\label{E:bb3}
ds^{2}=f(r)dt^{2}-f(r)^{-1}dr^{2},
\end{equation}
($t\in (-\infty,\infty), r=(r_+,\infty)$) where
\begin{equation}
\label{E:16}
f(r)=\frac{r^{2}}{\ell^{2}}-M.
\end{equation}
We call $(t,r)$ BTZ coordinates.  We note that
$\kappa=f'(r_+)/2$ and so (see Section \ref{Sect:prelim}) it is equal to the surface gravity of the BTZ horizon at $r=r_+$

\begin{figure}
\centering
\includegraphics[scale=0.4]{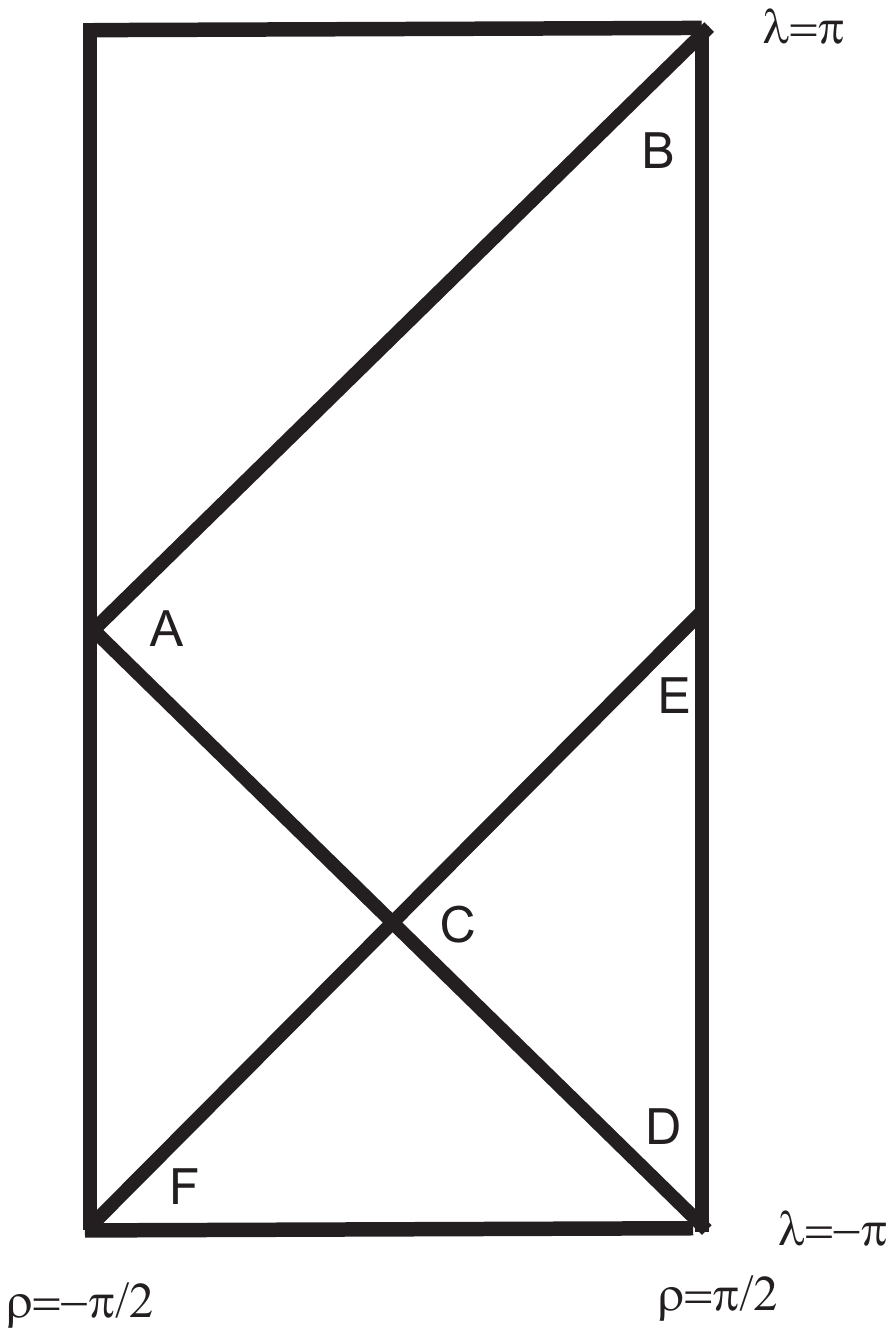}
\caption{\label{Fig:ads2} Regions in AdS spacetime in 1+1 dimensions covered by Poincar\'e and BTZ
coordinates. The region ABD is covered by Poincar\'e coordinates whereas the region CED (the `right exterior region' is covered by BTZ coordinates. Global coordinates cover all the manifold. The 1+1 dimensional BTZ spacetime consists of the `four-wedge' region AEDF bounded by the line AE (not drawn).}
\end{figure}

The charts for the three coordinate systems cover different regions of AdS spacetime as shown in 
Figure \ref{Fig:ads2}.

In order to draw this diagram we have conformally mapped AdS to the strip $-\pi/2<\rho<\pi/2$ in Minkowski spacetime
(see Note $<$\ref{Note:confboundary}$>$) and attached a (disconnected) boundary at $\rho=-\pi/2$ and $\rho=\pi/2$.  We also assume from now on (see Note $<$\ref{Note:wrap}$>$) that we are on the covering space i.e.\ we let $\lambda$ now range over $\mathbb{R}$.  Our BTZ coordinates range over $t\in (-\infty, \infty)$, $r\in (r_+, \infty)$ and cover the right-wedge region CDE which corresponds to the (right) exterior BTZ spacetime.  The full (1+1 dimensional) BTZ spacetime consists of the square (or `four-wedge') region AEDF.

In the massless case and for the two-dimensional AdS metric (\ref{E:a14}), our (classical, real) Klein Gordon equation (\ref{KG}) amounts to the wave equation
\begin{equation}
\label{2dwave}
\frac{\partial^2\phi}{\partial\lambda^2}-\frac{\partial^2\phi}{\partial\rho^2}=0
\end{equation}
and, in consequence $<$\ref{Note:cnfmlwt}$>$, the massless quantum two-point function $<$\ref{Note:2ptfn}$>$,
\[
{\cal G}_{\mathrm{global}}(\lambda_1,\rho_1; \lambda_2,\rho_2)=\omega_{\mathrm{globalground}}(\phi(\lambda_1,\rho_1)\phi(\lambda_2,\rho_2)),
\]
in the global AdS ground state is the same as that for the ground state of (\ref{2dwave}) with respect to Minkowski-time on our strip of 1+1 dimensional Minkowski space with vanishing boundary conditions at $\rho=\pm \pi/2$, which, working for convenience with  $y$, defined by $y=\pi/2-\rho$, is easily calculated to be
\begin{equation}
\label{2dmasslessG}
{\cal G}_{\mathrm{global}}(\lambda_1,y_1; \lambda_2,y_2)=\frac{1}{\pi}\sum_{n=1}^\infty \frac{e^{-in(\lambda_1-\lambda_2)}}{n}
\sin n y_1\sin n y_2.
\end{equation}
From this, one finds that the direct boundary limit $G_{\mathrm{global}}$ -- obtained from ${\cal G}_{\mathrm{global}}$, according to (\ref{Blim}) applied to (\ref{E:a14}), by taking the limit as $\rho_1$ and $\rho_2$ tend towards say, the right boundary, $\rho=\pi/2$, i.e as $y_1$ and $y_2$ tend to $y=0$ -- of $(\sin y_1\sin y_2)^{-1}{\cal G}_{\mathrm{global}}(\lambda_1,y_1; \lambda_2,y_2)$
is easily seen to be                                                                
\begin{equation}
\label{E:71}
G_{\mathrm{global}}\left(\lambda_{1},\lambda_{2}\right)=\frac{1}{2\pi}\frac{1}{\cos\left(\lambda_{1}-\lambda_{2}-i\epsilon\right)-1}.
\end{equation}
(See \cite{pLark07} for details but note that, there, the boundary limit is taken at the left boundary!)

We can also compute the boundary limit of the same two point function (i.e.\ for the global AdS ground state) -- this time coordinatizing the boundary by Poincar\'e time $T$ (which is the `Minkowski' time coordinate in the interpretation of the relevant region, i.e.\ the line DB in Figure \ref{Fig:ads2}, of the boundary as 1+0-dimensional Minkowski space).  One way to do this is to rely on the proof, \cite{Spradlin:1999bn}, by Spradlin and Strominger that, in 1+1 dimensions, the AdS ground state and the Poincar\'e ground state coincide.  The two-point function in the AdS ground state, restricted to a Poincar\'e chart, can then (see again $<$\ref{Note:cnfmlwt}$>$) in view of (\ref{E:79}), be identified with the 
two point function 
\begin{equation}
\label{Poincaretwopointfn}
{\cal G}_{\mathrm{Poincare}}(T_1,z_1; T_2,z_2)=\frac{1}{\pi}\int_0^\infty \frac{e^{-ik(T_1-T_2)}}{k}
\sin kz_1\sin kz_2\,dk
\end{equation}
for the ground state on the right half of 1+1 dimensional Minkowski space with vanishing boundary conditions at $z=0$.
Taking the direct boundary limit -- obtained, according to (\ref{Blim}) applied to (\ref{E:79}) -- by taking the limit as
$z_1$ and $z_2$ tend towards the right-boundary at $z=0$ -- of $(z_1z_2)^{-1}{\cal G}_{\mathrm{Poincare}}(T_1, z_1; T_2, z_2)$, one obtains
\begin{equation}
\label{E:92}
G_{\mathrm{Poincare}}(T_{1},T_{2})=-\frac{1}{\pi}\frac{1}{\left(T_{1}-T_{2}-i\epsilon\right)^{2}}
\end{equation}
which agrees, as it of course must, with (\ref{Wanom}) for the case $d=1$, $\Delta=1$.

One may alternatively obtain (\ref{E:92}) from (\ref{E:71}) by noticing first that (by comparing $u/v$ in the limit as
$x$ tends to $\infty$ in (\ref{E:a13}) with $u/v$ in the limit as $x$ tends to $\infty$ in (\ref{E:77})) $T$ coordinates are related to $\lambda$ coordinates on the boundary by
\begin{equation}
\label{E:81}
\frac{T}{\ell}=\tan\left(\frac{\lambda}{2}\right).
\end{equation}
From this expression, it follows that, if we define the metrics on the boundary, $g=d\lambda^2$ and $\tilde g=dT^2$, then $g$ and $\tilde g$ are related by
\begin{equation}
\label{E:84}
\tilde g=\Omega^{2}g
\end{equation}
where
\begin{equation}
\label{gpOmega}
\Omega=\frac{\ell^{2}+T^{2}}{2\ell},
\end{equation}
whereupon, by the formalism of Note $<$\ref{Note:cnfmlwt}$>$, we must have
\[
G_{\mathrm{Poincare}}(T_1, T_2)=\Omega(T_1)^{-\Delta}\Omega(T_2)^{-\Delta}G_{\mathrm{global}}(\lambda_1(T_1),\lambda_2(T_2))
\]
where (by (\ref{Delta}) for $d=1$ and $m=0$) $\Delta=1$.  With $\Omega$ as in (\ref{gpOmega}) and the relationship (\ref{E:81}) between $T$ and $\lambda$, this easily reproduces our formula (\ref{E:92}) for $G_{\mathrm{Poincare}}$ given the formula, (\ref{E:71}) for $G_{\mathrm{global}}$.

It is easy to see that the two-point function, $G_{\mathrm{Poincare}}(T_1, T_2)$ of (\ref{E:92}) satisfies the 
ground-state condition with respect to $T$-translations.  In fact, apart from a 1/4 factor (see Note $<$\ref{Note:1plus1Mink}$>$) it is equal to the twice-differentiated ground state two point function on a null line for the massless Klein Gordon equation in 1+1 dimensional Minkowski space -- where we identify $T$ here with, say the null-coordinate $U=T+X$ on the latter.

Next, we wish to find the boundary limit, $G_{\mathrm{BTZ}}(t_1, t_2)$, of the bulk two-point function for the global ground state of AdS, when expressed in terms of  BTZ time,  where we now think of the global ground state (see Section
\ref{Sect:intro}) as the appropriate version of the HHI state when the 4-wedge region of AdS is interpreted as 1+1 dimensional BTZ.  To find this, notice first that
(by comparing $v/x$ in the limit as
$\rho$ tends to $\pi/2$ in (\ref{E:a13}) with $v/x$ in the limit as $z$ tends to $0$ in (\ref{E:bb2})) $t$ coordinates are related to $\lambda$ coordinates on the boundary by
\begin{equation}\label{E:93a}
\tanh\kappa t=\cos\lambda.
\end{equation}
From this expression, it follows that, if we define the metrics on the boundary, $g=d\lambda^2$ and $\tilde g=dt^2$, then $g$ and $\tilde g$ are related by an equation of form (\ref{E:84}) where now
\begin{equation}
\label{E:98}
\Omega(t)=\frac{\cosh\kappa t}{\kappa},
\end{equation}
whereupon, again by the formalism of Note $<$\ref{Note:cnfmlwt}$>$, we must have
\[
G_{\mathrm{BTZ}}(t_1, t_2)=\Omega(t_1)^{-\Delta}\Omega(t_2)^{-\Delta}G_{\mathrm{global}}(\lambda_1(t_1),\lambda_2(t_2))
\]
where (again by (\ref{Delta}) for $d=1$ and $m=0$) $\Delta=1$.  With $\Omega$ as in (\ref{E:98}) and the relationship (\ref{E:93a}) between $t$ and $\lambda$, this easily gives
\begin{equation}\label{E:100}
G_{\mathrm{BTZ}}(t_{1},t_{2})=-\frac{1}{4\pi}\frac{\kappa^{2}}{\sinh^{2}\left(\kappa\frac{t_{1}-t_{2}-i\epsilon}{2}\right)},
\end{equation}
which, as one may easily check, is a KMS state at inverse temperature $2\pi/\kappa$.  In fact (again up to a factor of $1/4$) the two point function (\ref{E:100}) is equal to the twice-differentiated thermal two point function at the same temperature on a null line for the massless Klein Gordon equation in 1+1 dimensional Minkowski space -- where we identify $t$ here with, say, the null-coordinate $u=t+x$ on the latter.  (Below we shall think of $(t, x)$ as `Rindler' coordinates on a right wedge of a different copy of Minkowski space and $u$ as its restriction to a null half-line.)

It is interesting to notice, in fact, that the relation between the (Poincar\'e) $T$ coordinate and the (BTZ) $t$ coordinate on the boundary (DE in Figure \ref{Fig:ads2}) is (e.g.\ from (\ref{E:81}) and (\ref{E:93a}))
\begin{equation}
\label{Ttrelation}
-T=\ell e^{-\kappa t}.
\end{equation}
and we could, of course, alternatively, have obtained (\ref{E:100})
directly from (\ref{E:92}) and (\ref{Ttrelation}) $<$\ref{Note:TtGstory}$>$.

In fact, the situation for the two-point functions (\ref{E:92}) and (\ref{E:100}) on the (timelike) boundary line DEB in Figure \ref{Fig:ads2} is closely similar to the situation on a full null-line -- say the line coordinatized by $U=T+X$ in 1+1 dimensional Minkowski space $<$\ref{Note:1plus1Mink}$>$,  the only difference being the factor of $1/4$ and the need to twice-differentiate as we noted above in the case of each two-point function.  Aside from this difference, this close similarity, or analogy, is realized by identifying our `Poincar\'e' coordinate, $T$, with $U$ and (cf. (\ref{Ttrelation})) our `BTZ' coordinate, $t$ (which coordinatizes the boundary half-line DE)  with, $u$, related to $U$ by (cf. (\ref{Ttrelation})) $-U=\ell e^{\kappa u}$ which coordinatizes, say, the `left' half of the chosen null line in Minkowski space.  On the Minkowski side of this analogy, $u$ translations correspond to (the restriction to the null line of) Lorentz boosts and the thermality of $G_{\mathrm{BTZ}}$ amounts to the Unruh effect.  Here we mean, by the Unruh effect (see \cite{Unruh:1976db} and also \cite{Kay:1985zs}) the thermal nature of the usual ground state in Minkowski space when restricted to, say, the right Minkowski-wedge and regarded with respect to the `time-evolution' given by the one-parameter family of wedge-preserving Lorentz boosts (scaled by $\kappa$ to make our analogy closer) -- except we are considering here, instead of the full wedge algebra, the algebra for one of the two null half-lines which bounds the right wedge (say the one to the future) -- this being, by the way, for the massless scalar field in 1+1 dimensions, equivalent in an obvious way to the restriction to left-moving modes.  Moreover, (cf. \cite{Kay:1985zs}) just as the Minkowski ground state is a `double KMS state' in the sense of \cite{Kay:1985yw} (see also Note $<$\ref{Note:equiv}$>$) on the union of the right and left wedge of Minkowski space -- and, more relevantly, by restriction, on the full null-line we are considering, it is equally a `double KMS state' with respect to the union of the right half-line ($U>0$) and the left half line ($U<0$) (where `time-evolution' is given by $u$-translations where $u$ is defined by $U=\ell e^\kappa u$ on the right half line and by $-U=\ell e^{-\kappa u}$ on the left half-line).  In particular, and adopting `heuristic' language, it is thermal on each half-line ($U>0$ and $U<0$) but `entangled' in just such a way as to be a pure state (in fact a ground state for $U$-translations) on the full $U$ line.   And all this translates, pursuing our analogy in reverse, to the fact that the AdS ground state on our bulk AdS, restricted to the right (now timelike) boundary line DEB is a ground state with respect to $T$-translations on the full line DEB in Figure \ref{Fig:ads2} but it is a `double KMS state' with respect to $t$-translations (i.e.\ `BTZ time-translations') on the half-line $T>0$ which coordinatizes EB (with $t$ now defined by $T=\ell e^{\kappa t}$) and with respect to our BTZ $t$-translations on the half-line $T<0$ which coordinatizes DE (with $t$ as in (\ref{Ttrelation})).  Finally, we remark that, because of the way solutions to our bulk field theory (i.e.\ Equation (\ref{2dwave})) propagate, the situation on the half line EB (see again Figure \ref{Fig:ads2}) is essentially the same as the situation on the half-line FA and so we may alternatively think in terms of the boundary restriction (now to the left boundary-half-line, FA, and the right boundary-half-line, DE) of the global AdS ground state as being a double KMS state with respect to our $t$-translations on DE and the appropriate notion of $t$-translations on FA.  This latter situation is, as we shall see in Section \ref{Sect:1+2}, analogous to what happens in the case of bulk 1+2 dimensional BTZ, for the two boundary cylinders there.

\section{The boundary limit of the Klein Gordon QFT on 1+2 dimensional AdS and BTZ}
\label{Sect:1+2}

1+2 dimensional AdS with radius $\ell$, related to the cosmological constant, $\Lambda$, by $\Lambda=-1/\ell^2$ (we'll just call it AdS from now on in the rest of this section) can be defined to be the surface
\begin{equation}
\label{E:104}
u^{2}+v^{2}-x^{2}-y^{2}=\ell^{2}
\end{equation}
embedded in $\mathbb{R}^{4}$ with metric
\begin{equation}\label{E:1}
ds^{2}=du^{2}+dv^{2}-dx^{2}-dy^{2}.
\end{equation}

We are interested again in three parameterizations of
(\ref{E:104}) leading to global, Poincar\'{e} and BTZ coordinates. Global
coordinates $(\lambda,\rho,\theta)$ can be defined by
\begin{eqnarray}\label{E:105}
v=\ell\sec\rho\cos\lambda\hspace{1cm}u=\ell\sec\rho\sin\lambda\nonumber\\
x=\ell\tan\rho\cos\theta\hspace{1cm}y=\ell\tan\rho\sin\theta,
\end{eqnarray}
where
$(\lambda,\rho,\theta)\in[-\pi,\pi)\times[0,\pi/2)\times[-\pi,\pi)$.
$-\pi<\lambda\leq\pi$, $0\leq\rho<\pi/2$ and $-\pi\leq\theta<\pi$.
In these coordinates the metric is
\begin{equation}
\label{E:106aa}
ds^{2}=\ell^{2}\sec^{2}\rho\left(d\lambda^{2}-d\rho^{2}-\sin^{2}\rho
d\theta^{2}\right).
\end{equation}
Poincar\'{e} coordinates $(T,k,z)$ are given by
\begin{eqnarray}\label{E:107}
v=\frac{1}{2z}\left(z^{2}+\ell^{2}+k^{2}-T^{2}\right)\hspace{1cm}u=\frac{\ell T}{z}\nonumber\\
x=\frac{1}{2z}\left(\ell^{2}-z^{2}+T^{2}-k^{2}\right)\hspace{1cm}y=\frac{\ell k}{z}.
\end{eqnarray}
In these coordinates the metric is
\begin{equation}
\label{E:108}
ds^{2}=\frac{\ell^{2}}{z^{2}}\left(dT^{2}-dk^{2}-dz^{2}\right)
\end{equation}
where
$(T,k,z)\in(-\infty,\infty)\times(-\infty,\infty)\times(0,\infty)$.
From (\ref{E:108}) we see that the boundary, at $z=0$, of the region of AdS 
covered by Poincar\'{e} coordinates is conformal to 1+1 dimensional
Minkowski space.

\begin{figure}
\centering
\includegraphics[trim=0cm 4cm 0cm 0cm, clip=true, width= 8.2cm, height=9cm]{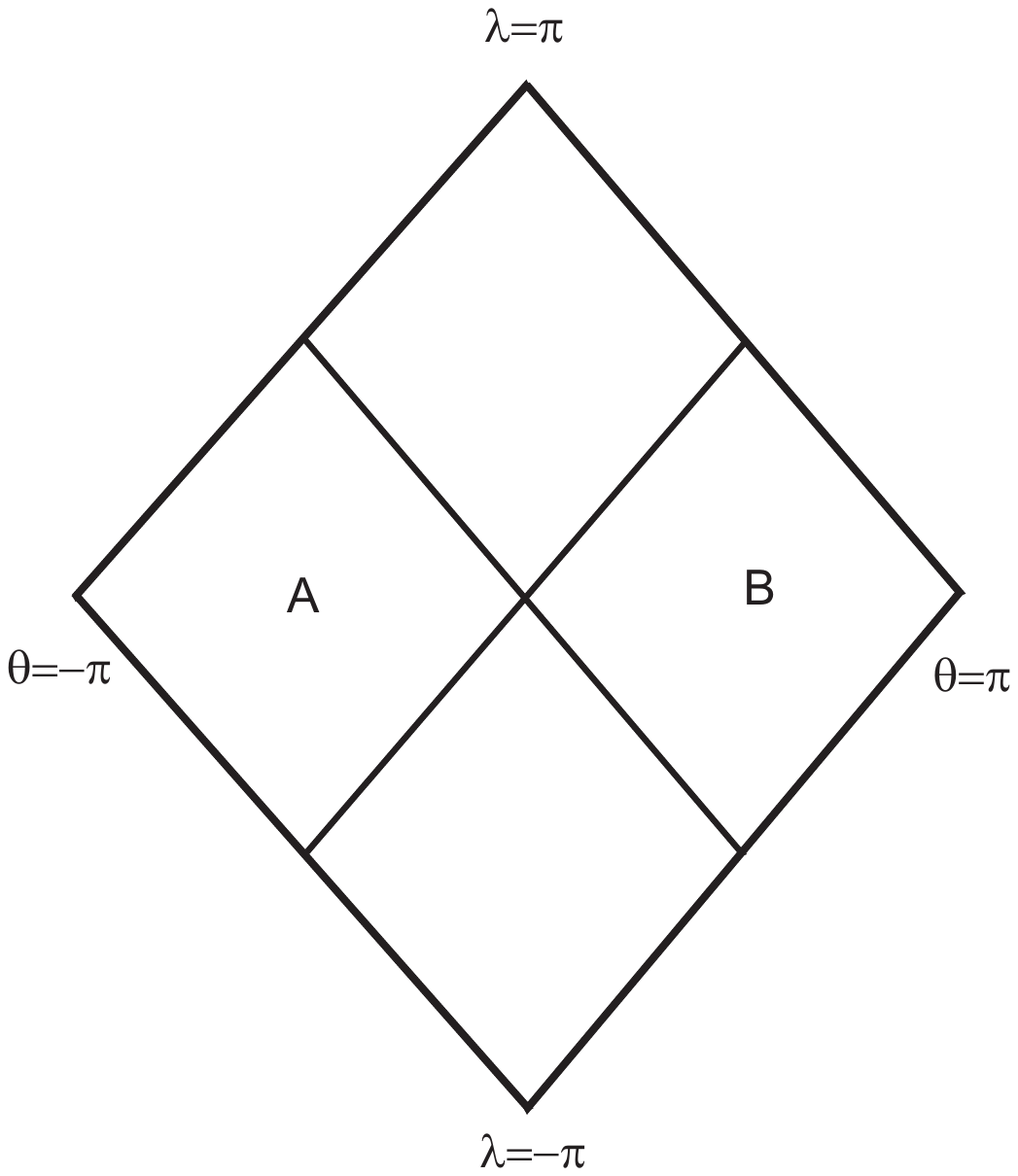}
\caption{\label{Fig:btzpoincare}A Penrose diagram showing how the intersection of the covering space of BTZ with the AdS boundary is related to the intersection of our Poincar\'{e} chart with the AdS boundary. The big diamond is the intersection of our Poincar\'{e} chart with the AdS boundary, whereas the small diamonds, A and B, are, respectively, the intersections with the boundary of the covering spaces of the two exterior BTZ regions.}
\end{figure}

The third parametrization, again, for a given choice of $M\in \mathbb{R}^+$ (which will again become the `BTZ mass') is given by
\begin{eqnarray}\label{E:112}
u=\ell\sqrt{\frac{r^{2}-r_{+}^{2}}{r_{+}^{2}}}\sinh\kappa t\hspace{1cm}v=\ell\frac{r}{r_{+}}\cosh \ell\kappa\varphi\nonumber\\
y=\ell\sqrt{\frac{r^{2}-r_{+}^{2}}{r_{+}^{2}}}\cosh\kappa
t\hspace{1cm}x=\ell\frac{r}{r_{+}}\sinh \ell\kappa\varphi,
\end{eqnarray}
where $r_+=\ell\sqrt{M}$ and $\kappa=r_{+}/\ell^{2}$ ($=\sqrt{M}/\ell$) and
$(t,r,\varphi)\in(-\infty,\infty)\times(r_+,\infty)\times(-\infty,\infty)$.
The metric in this case is
\begin{equation}
\label{E:109}
ds^{2}=f(r)dt^{2}-f(r)^{-1}dr^{2}-r^{2}d\varphi^{2},
\end{equation}
where
\begin{equation}
\label{E:110}
f(r)=\frac{r^{2}}{\ell^{2}}-M.
\end{equation}
The resulting coordinate patch, together with the metric (\ref{E:109}) becomes one of the exterior (i.e.\ $r > r_+$) regions of the BTZ black hole if we make $\varphi$ periodic (i.e.\ if we quotient by the equivalence relation $\varphi\sim \varphi+2\pi$) \cite{Banados:1992gq}.  The maximally extended BTZ spacetime has two exterior regions and it turns out \cite{Ortiz:2011mi, lOrt11}
that the intersection with the conformal boundary of AdS of the corresponding regions {\it before} we make $\varphi$ periodic are the subregions of the intersection of the Poincar\'e chart with the boundary as depicted in Figure \ref{Fig:btzpoincare}.  One of us (LO, see \cite{lOrt11, Ortiz:2011mi}) has shown that if one quantizes the Klein Gordon equation, (\ref{KG}), for an arbitrary mass on AdS with vanishing boundary conditions on the conformal boundary, then the boundary limit of the two point function in either the global ground state or the Poincar\'e ground state $<$\ref{Note:PoincareGlobalEquiv}$>$  is, when expressed in Poincar\'e coordinates, given by  

\begin{equation}
\label{E:140ll}
G_{\mathrm{Poincare}}(T_1, k_1; T_2, k_2)=\frac{1}{2\pi}\frac{1}{((k_1-k_2)^2-(T_1-T_2-i\epsilon)^2)^\Delta}
\end{equation}
where $\Delta$ is given by (\ref{Delta}) with $d=2$.  

From (\ref{E:107}) and (\ref{E:112}) we find that, say in the right diamond region labelled B in Figure (\ref{Fig:btzpoincare}), 
the Poincar\'e $(T,k)$ coordinates are related to our BTZ $(t,\varphi)$ coordinates by
\begin{equation}\label{E:116}
T=\ell e^{-\kappa\ell\varphi}\sinh\left(\kappa
t\right)\hspace{1cm} k=\ell e^{-\kappa\ell\varphi}\cosh\left(\kappa t\right).
\end{equation}
Defining the metrics, $g=dt^2-\ell^2d\varphi^2$ and $\tilde g=dT^2-dk^2$, we obtain
\begin{equation}\label{E:117}
\tilde g=\Omega(\varphi)^{2}g,
\end{equation}
where
\begin{equation}\label{E:140rr}
\Omega(\varphi)=\frac{e^{\kappa\ell\varphi}}{\ell\kappa}.
\end{equation}
By a similar analysis to that in the previous section, we have, by (\ref{confstate}) in Note $<$\ref{Note:cnfmlwt}$>$,  
\begin{equation}
\label{E:140tt}
\fl G_{\mathrm{preBTZ}}(t_1, \varphi_1'; t_2, \varphi_2')=\frac{1}{2\pi
2^\Delta}\frac{\kappa^{2\Delta}}{(\cosh\kappa\ell(\varphi_1' - \varphi_2')-\cosh\kappa(t_1-t_2-i\epsilon))^\Delta}.
\end{equation}
We have called this two point function, for the global ground state of AdS in BTZ coordinates (before making $\varphi$ periodic) `$G_{\mathrm{preBTZ}}$'.  

\begin{figure}
\includegraphics[scale = 1.2, trim = 0 0 580 650]{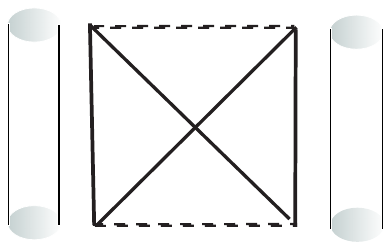}
\caption{\label{Fig:BTZcylinders}\footnotesize{Schematic picture of BTZ spacetime with its two boundary cylinders}}
\end{figure}

Now, as discussed in the introduction,  we expect the HHI state on BTZ to be the natural state induced on BTZ by the AdS ground state on AdS when one quotients by the equivalence relation $\varphi\sim \varphi+2\pi$.  Moreover it is clear that the quotient of the boundary by $\varphi\sim \varphi+2\pi$, consists of two disconnected cylinders, each of radius $\ell$ (see Figure (\ref{Fig:BTZcylinders})), one being the boundary of the right exterior BTZ region and one the boundary of the left exterior BTZ region.   In consequence, the two-point function in the HHI state on, say the right exterior region of true BTZ will be defined on the right cylinder and will be given by  the image sum $<$\ref{Note:image}$>$

\begin{equation}
\label{E:140vv}
\fl G_{\mathrm{BTZ}}(t_1, \varphi_1'; t_2, \varphi_2')=\sum_{n\in{\mathbb Z}}\frac{1}{2\pi
2^\Delta}\frac{\kappa^{2\Delta}}{(\cosh\kappa\ell(\varphi_1 - \varphi_2+2\pi n)-\cosh\kappa(t_1-t_2-i\epsilon))^\Delta}.
\end{equation}

\medskip

\noindent
(\ref{E:140vv}) has been obtained before \cite{KeskiVakkuri:1998nw, Maldacena:2001kr}.  (See Section \ref{Sect:intro}.) 

$G_{\mathrm{BTZ}}$ clearly inherits the KMS property from $G_{\mathrm{preBTZ}}$ -- being, in fact, the two-point function on our cylinder of radius $\ell$ at inverse temperature $2\pi/\kappa$, i.e.\ the inverse Hawking temperature of the BTZ black hole.  In fact, as already discussed by Maldacena in  \cite{ Maldacena:2001kr},  the boundary limit of the HHI state on the union of the left cylinder and the right cylinder is expected to be, what in our language (see  Note $<$\ref{Note:equiv}$>$ and  \cite{Kay:1985yw, Kay:1985zs}) we would call a double-KMS state -- i.e.\ KMS with respect to the appropriate $t$ (going to the future on the right cylinder and the past on the left cylinder) when restricted to either of the cylinders but entangled in just such a way as to be pure.   (See \cite{ Maldacena:2001kr} for the two point function with one point in the right cylinder and one in the left and cf. similar formulae in \cite{Kay:1985yw} and \cite{Kay:1985zs})
(A similar remark could of course have been made about $G_{\mathrm{preBTZ}}$.)

\section{The brick wall modification of the HHI state for the Klein Gordon equation on BTZ and its boundary limit in $d=1$ and $d=2$}
\label{Sect:brick}

In Sections \ref{Sect:1+1} and \ref{Sect:1+2} we have seen that, in 1+1 dimensional and 1+2 dimensional BTZ, the appropriate HHI state  -- i.e.\ the global ($=$ Poincar\'e)  AdS ground state in the 1+1 dimensional case and the state (with two point function defined by the appropriate image sum) inherited from the global AdS ground state (see Note $<$\ref{Note:PoincareGlobalEquiv}$>$) under the BTZ quotient construction in the 1+2 dimensional case --  goes over to a state on the (say, right) boundary line (in 1+1) or cylinder (in 1+2) which is a KMS state at the appropriate Hawking temperature.   Moreover, although we have not derived this explicitly here, it is straightforward to check that each of these bulk HHI states is a KMS state at the Hawking temperature with respect to BTZ time translations in, say, the right BTZ wedge (and is a `double KMS state' on the bulk double wedge -- cf. e.g.\ the discussion after (\ref{Ttrelation}) and Note $<$\ref{Note:equiv}$>$).  So it is fair to say that, in boundary-limit holography for BTZ, the thermal nature of the HHI state in the bulk is faithfully reflected by the thermal nature of its boundary limit on the (say) right conformal boundary.   However, in line with our discussion in the introduction, in both cases, both bulk and boundary state have infinite entropy -- corresponding to the fact that, if we attempt, formally, to express these states as Gibbs states, with, say, in the case of 
1+1 dimensional bulk BTZ,  two-point function defined by
\[
{\cal G}_{\mathrm{BTZ}}(t_1,r_1; t_2,r_2)=\tr(e^{-\beta H}\phi(t_1,r_1)\phi(t_2,r_2))/Z
\]
etc., where the fields are represented in the appropriate ground-state (or `Boulware') representation, $Z=\tr(e^{-\beta H})$, $\beta=2\pi/\kappa$, and $H$ is the appropriate Hamiltonian, then $H$ will have continuous spectrum and so the trace in the standard formula $S=-\tr\rho\ln\rho$ where $\rho=e^{-\beta H}/Z$ (as well as the trace in the definition, $Z=\tr(e^{-\beta H})$, of $Z$ itself) will not really exist -- or be `infinite'.  (And similarly on the boundary and similarly for bulk and boundary in the 1+2 dimensional case).

To remedy this, we will adopt the brick-wall approach of 't Hooft \cite{'tHooft:1984re} and of Mukohyama and Israel
\cite{Mukohyama:1998rf} and others, as discussed in the introduction, and introduce a suitable `brick wall' in the bulk so as to make the bulk entropy finite.  We shall then take the boundary limit of the brick-wall modified thermal two-point function at the Hawking temperature and, in the case of our 1+1 dimensional BTZ with our zero-mass Klein Gordon equation, find the explicit form of the boundary-limit two-point function and show that this is both thermal at the Hawking temperature (for an appropriately, brick-wall modified boundary Hamiltonian) and also has the same finite entropy as the bulk brick-wall modified state.   We shall calculate the bulk entropy in the 1+1 dimensional brick wall model by the Mukohyama-Israel approximation method \cite{Mukohyama:1998rf} and show that, for this case, this method reproduces the exact result.

In the 1+2 dimensional BTZ case, we shall re-obtain, by the same Mukohyama-Israel approximation method, the result of  Kim et al.\ \cite{Kim:1996eg} which is that (at least to a good approximation) the brick-wall modified bulk entropy is proportional to the `area' (i.e circumference) of the event horizon.  We shall also obtain in-principle exact sums (albeit difficult to perform explicitly) for the bulk brick-wall modified two-point function and its boundary limit and show, without summing explicitly, that the bulk and boundary entropies must be exactly equal.

\subsection{The brick wall model and its boundary limit for 1+1 dimensional BTZ}

We locate our brick wall, in BTZ coordinates, at $r=r_+ + \epsilon$ where $\epsilon$ is suitably small.  This is a natural choice since it respects BTZ time-translational invariance.   In Figure \ref{Fig:ads2}, the brick wall is a smooth curve joining D to E and passing just to the right of the piecewise-linear curve DCE.   Our aim is to quantize our field to the right of this brick wall, with vanishing boundary conditions on the brick wall as well as on the conformal boundary.  The Hamiltonian will, as expected, turn out to have discrete spectrum.  So we will be able to construct a thermal state (with inverse Hawking temperature, $\beta=2\pi/\kappa$) as a Gibbs state in the ground-state representation.  

To do all this, it is convenient to work in $(t,r^*)$ coordinates, where the tortoise-like coordinate (cf. (\ref{Schwtortoise}))  $r^*$, is related to the BTZ $r$ coordinate of (\ref{E:bb3}) and (\ref{E:16}) by 
\begin{equation}
\label{defrstar}
\frac{dr^*}{dr}=f(r)^{-1}=\left(\frac{r^2}{\ell^2}-M\right)^{-1}=\frac{\ell^2}{(r+r_+)(r-r_+)},
\end{equation}
with solution
\begin{equation}
\label{BTZtortoise}
r^{*}=\frac{\ell^{2}}{2r_{+}}\textrm{ln}\frac{r-r_{+}}{r+r_{+}}. 
\end{equation}

The metric is then given by
\begin{equation}
\label{E:141b}
ds^{2}=f(r(r^*))\left(dt^2-{dr^*}^2\right).
\end{equation}

On the full exterior BTZ region, $r^*$ ranges from $-\infty$ at the BTZ horizon to $0$ at the AdS boundary (where $r=\infty$).  However, the brick wall will lie at the finite value $r^*=-B$, where $B=(\ell^2/(2r_{+}))\ln(2r_+/\epsilon+1)$.  So, the classical theory which we wish to quantize is the massless covariant Klein-Gordon equation, which, in $(t,r^*)$ coordinates, becomes
\begin{equation}
\label{E:142b}
\left(\frac{\partial^{2}}{\partial
t^{2}}-\frac{\partial^{2}}{\partial
{r^{*}}^{2}}\right)\phi(t,r^{*})=0,
\end{equation}
with the boundary conditions
\begin{equation}
\label{E:143b}
\phi(t,-B)=0=\phi(t,0), \ \ \forall t\in {\mathbb R}.
\end{equation}
By essentially the same arguments which we gave for the form of  ${\cal G}_{\mathrm{global}}$ in Equation (\ref{2dmasslessG}), the two-point function for the ground state for this theory then takes the form 
\[
{\cal G}_{\mathrm{BTZ1+1brick}}(t_1,r^*_1; t_2, r^*_2) = \hbox{`}\langle 0|\hat{\phi}(t_1,r^*_1)\hat{\phi}(t_2, r^*_2)|0\rangle\hbox{'}
\]
\begin{equation}
\label{E:148b}
=\sum_n\frac{1}{B\omega_n}e^{-i\omega_n(t_1-t_2)}\sin(\omega_n
r^*_1)\sin(\omega_n r^*_2)
\end{equation}
where $\omega_n$ stands for $n\pi/B$, $n=1,2,3,\dots$.  In other words, it is the same as the two-point function of a (real) massless Klein Gordon field in a strip, $t \in {\mathbb R}, r^*\in (-B, 0)$ of 1+1 dimensional Minkowski space with vanishing boundary conditions at its edges.  In fact, as may readily be checked, (\ref{E:148b}) may be derived by the usual creation and annihilation operator method, for mode functions adapted to these boundary conditions.

To get a suitable boundary limit of ${\cal G}_{\mathrm{BTZ1+1brick}}$,  we expect, in accordance with (\ref{Blim}), that the right thing to do will be to multiply (\ref{E:148b}) by $\ell^{-2}f(r(r_1^*))^{1/2}f(r(r_2^*))^{1/2}$ and then to take the limit as $r_1^*$ and $r_2^*$ both tend to zero.  To this end, we notice first that $f(r)^{1/2}\sim r/\ell$.  Also, recalling that $r(r^*)$ is given by Equation (\ref{BTZtortoise}) one easily sees that the inverse function, $r(r^*)$ is 
\begin{equation}
\label{BTZesiotrot}
r(r^*)=-r_+\coth\left(\frac{r_+r^*}{\ell^2}\right)
\end{equation}
from which we see that when $r$ is large (i.e.\ $r^*$ is close to $0$) we have $r\sim -\ell^2/r^*$ (we can actually also see this directly from (\ref{BTZtortoise})) and hence
$f(r(r^*))\sim r(r^*)/\ell\sim -\ell/r^*$.    In fact, one easily sees, with these considerations, that the above `right thing to do' amounts to taking the limit as $r_1^*$ and $r_2^*$ tend to zero of $1/(r_1^*r_2^*)$ times ${\cal G}_{\mathrm{BTZ1+1brick}}(t_1,r^*_1; t_2, r^*_2)$ which is easily seen to be
\begin{equation}\label{E:149b}
G_{\mathrm{BTZ1+1brick}}(t,t')=\sum_n\frac{\omega_n}{B}e^{-i\omega_n(t-t')}.
\end{equation}

As one can see from the discreteness of the sum in (\ref{E:148b}) (or from the discreteness of the modes in its derivation with creation and annihilation operators) the `one-particle Hamiltonian', $h$, whose second quantization $d\Gamma(h)=
0 \oplus h \oplus (h\otimes 1 + 1\otimes h) \oplus \dots$ (see e.g.\ \cite{RSVol2}) generates bulk BTZ time-evolution in the region to the right of our brick wall, has discrete spectrum, with, in fact, eigenvalues
$\omega_n=n\pi/B$, $n=1,2,3\dots$ (and multiplicity 1).  Thus $d\Gamma(h)$ will have discrete spectrum too.  And, from (\ref{E:149b}), we may conclude that, also, the one-particle Hamiltonian generating time-evolution on the BTZ boundary for the (conformal, generalized-free, with scaling dimension $\Delta=1$)) boundary restriction of this bulk theory has the same spectrum, as will also {\it its} second quantization.  In fact it will clearly be unitarily equivalent to $h$, so we will also call it $h$ and its second quantization $d\Gamma(h)$.

Because the spectra are discrete -- more precisely, because
\begin{equation}
\label{partition}
Z=\tr\left(e^{-\beta d\Gamma(h)}\right)
\end{equation}
is finite -- we may thus compute the bulk and boundary thermal two-point functions
\[
{\cal G}_{\mathrm{BTZ1+1brick}}^\beta(t_1,r^*_1; t_2, r^*_2) = Z^{-1}\tr(e^{-d\Gamma(h)}\hat{\phi}(t_1,r^*_1)\hat{\phi}(t_2, r^*_2))
\]
and
\[
G_{\mathrm{BTZ1+1brick}}^\beta(t_1, t_2) = Z^{-1}\tr(e^{-d\Gamma(h)}\hat{\phi}(t_1)\hat{\phi}(t_2))
\]
at the appropriate inverse Hawking temperature $\beta=2\pi/\kappa$.  E.g.\ by using creation and annihilation operator methods again, these are easily calculated to be
\begin{equation}
\label{E:150b}
\fl {\cal G}_{\mathrm{BTZ1+1brick}}^\beta(t_1,r^*_1; t_2, r^*_2)=\sum_n\frac{\sin(\omega_n
r_1^{*})\sin(\omega_n {r_2^{*}})}{B\omega_n}
\left(\frac{e^{-i\omega_n(t-t_1)}+e^{i\omega_n(t-t_1)}e^{-\beta\omega_n}}{1-e^{-\beta\omega_n}}\right)
\end{equation}
and 
\begin{equation}
\label{E:152b}
G_{\mathrm{BTZ1+1brick}}^\beta(t_1, t_2) = 
\sum_{n}\frac{\omega_n}{B}\left(\frac{e^{-i\omega_n(t_1-t_2)}+e^{i\omega_n(t_1-t_2)}e^{-\beta\omega_n}}{1-e^{-\beta\omega_n}}\right)
\end{equation}
both of which are easily seen to satisfy the KMS condition (see Section \ref{Sect:prelim}) for this $\beta$.  We also notice that, just as we obtained
$G_{\mathrm{BTZ1+1brick}}$ (\ref{E:149b}) from ${\cal G}_{\mathrm{BTZ1+1brick}}$ (\ref{E:148b}), we could have obtained $G_{\mathrm{BTZ1+1brick}}^\beta$ (\ref{E:152b}) as the limit as $r_1^*$ and $r_2^*$ tend to zero of $1/(r_1^*r_2^*)$ times ${\cal G}_{\mathrm{BTZ1+1brick}}^\beta$ (\ref{E:150b}). 

Recalling that $\omega_n$ stands for $n\pi/B$, we notice that (\ref{E:152b}) can be rewritten as
\begin{equation}
\label{Gbrickbetan}
\fl G_{\mathrm{BTZ1+1brick}}^\beta(t_1, t_2) = 
\sum_{n=1}^\infty\frac{\pi n}{B^2}\left(\frac{e^{-i(n\pi/B)(t_1-t_2)}+e^{i(n\pi/B)(t_1-t_2)}e^{-\beta n\pi/B}}{1-e^{-\beta n\pi/B}}\right).
\end{equation}
In the limit as the brick wall is removed, $B$ tends to infinity and we expect this to be well-approximated by replacing the sum over $n$ by an integral over, say, $y$.  After making the substitution $x=y/B$ we will thus have
\begin{equation}
\label{Gbrickbetax}
\int_0^\infty \pi x\left(\frac{e^{-i(\pi x)(t_1-t_2)}+e^{i(\pi x)(t_1-t_2)}e^{-\beta\pi x}}{1-e^{-\beta\pi x}}\right)dx
\end{equation}
which $<$\ref{Note:Gradshteyn}$>$  is equal to 
\[
-\frac{\pi}{\beta^2}\frac{1}{\sinh^2\left(\frac{\pi(t-t'-i\epsilon)}{\beta}\right)}
\]
which, recalling that $\kappa=2\pi/\beta$, agrees with (\ref{E:100}).  This confirms that the brick wall does what we anticipated in Section \ref{Sect:intro} that it would do in this case:  

\emph{The boundary limit of the thermal state at the Hawking temperature built on the modification, due to the presence of the brick wall, to the boundary limit of the BTZ ground state (i.e.\ the `Boulware state') tends, as the brick wall is removed, to the boundary limit of the BTZ HHI state.}  

Although we have not shown it expicitly here, the corresponding (and more straightforward) statement also holds for the bulk theory:  The bulk thermal state at the Hawking temperature built on the brick-wall modified BTZ ground state tends, as the brick wall is removed, to the BTZ HHI state (restricted to the right BTZ wedge).

Moreover, we saw above that (again in line with what we anticipated in Section \ref{Sect:intro} -- see Note $<$\ref{Note:Ham}$>$) in the presence of the brick wall, the Hamiltonians for the bulk and boundary theory are unitarily equivalent mathematical operators.   But now, since they have discrete spectrum (and a finite partition function (\ref{partition})) they will, in particular, have finite (and equal!) von Neumann entropies, given by $S=-\tr(\rho\ln\rho)$ for $\rho=e^{-\beta H}/Z$, $Z=\tr(e^{-\beta H})$, $H$ being what we called above $d\Gamma(h)$.  We next calculate the value of this entropy.  

In this simple massless 1+1 dimensional BTZ case, we can in fact calculate the entropy exactly:  In fact, 
thinking of it as bulk entropy, in view of Equations (\ref{E:142b}) and (\ref{E:143b}) etc., for given $\beta=1/{\cal T}$, it is just the usual entropy for a special relativistic real free scalar field in a `one-dimensional box' (i.e.\ line) of length $B$ in thermal equilibrium at temperature, $\cal T$, which is given (to a very good approximation $<$\ref{Note:scalarentropy}$>$) by the (standard) formula: 
\begin{equation}
\label{1dentropy}
S=\frac{\pi B{\cal T}}{3}.
\end{equation}
Taking $\cal T$ here to be the appropriate Hawking temperature, which we take to be $\kappa/(2\pi) = r_+/(2\pi \ell^2)$ ($= M^{1/2}/(2\pi \ell)$), this is
\begin{equation}
\label{1dentropywithTH}
S=\frac{r_+ B}{6\ell^2}.
\end{equation}

It is also instructive to apply the approximation method of Mukohyama and Israel \cite{Mukohyama:1998rf} to this case.
For any metric with the spherically symmetric `Schwarzschild-like' form (\ref{generalsphmetric}) of any spacetime dimension this method amounts to assuming that the total entropy, $S$, of a thermal state at temperature $\cal T$ in a region, $r_+ + \epsilon < r < L$, say  ($r_+$ the value of $r$ at the relevant horizon) should be well approximated by the integral
\begin{equation}
\label{Sapprox}
S=\int_{r_+ +\epsilon}^L  s({\cal T}_{\mathrm{loc}}(r)) A(r)\frac{dr}{f(r)^{1/2}}
\end{equation}
where, for any temperature, $\cal T$,  $s({\cal T})$ is the entropy density of a free real scalar field in Minkowski space of the same dimension and the local temperature, ${\cal T}_{\mathrm{loc}}(r)$ is given by 
\begin{equation}
\label{Tlocal}
{\cal T}_{\mathrm{loc}}(r)={\cal T}f^{-1/2}(r).
\end{equation}
$A(r)$ in (\ref{Sapprox}) denotes the area of the $d$-1 sphere of radius $r$ ($=2\pi r$ in 1+2 dimensions, $4\pi r^2 $ in 1+3 dimensions, etc.; in 1+1 dimensions, we replace $A(r)$ by 1) so that $A(r)dr/f(r)^{1/2}$ represents the proper volume element.

Applying the formula (\ref{Sapprox}) to find the total entropy outside the brick wall in 1+1 dimensional BTZ, we take $L=\infty$.  Also we have, by (\ref{E:16}), $f(r)=(r^2/\ell^2-M)^{-1}$ while the entropy density $s({\cal T})$ (by Note $<$\ref{Note:scalarentropy}$>$) is $\pi {\cal T}/3$.   So (\ref{Sapprox}) becomes
\[
S=\frac{\pi {\cal T}}{3}\int_\epsilon^\infty \left(\frac{r^2}{\ell^2}-M\right)^{-1}dr
\]
which, in view of (\ref{defrstar}), is just $(\pi {\cal T}/3) \int_{-B}^0 dr^* = \pi B{\cal T}/3$ which agrees with (\ref{1dentropy}) exactly!  (And hence will also lead to the formula (\ref{1dentropywithTH})).  Our main reason for deriving (\ref{1dentropy}), (\ref{1dentropywithTH}) is as a preliminary to deriving the corresponding formulae for 1+2 dimensional BTZ to which we next turn.  See also Note $<$\ref{Note:300}$>$.

\subsection{The brick wall model and its boundary limit for 1+2 dimensional BTZ}

All of the above generalizes to 1+2 dimensional BTZ.  The (zero-mass) covariant Klein-Gordon equation (\ref{KG}) 
becomes
\begin{equation}
\label{BTZ1+2KG}
\frac{\partial^2\phi}{\partial t^2}-\frac{1}{r}\left(\frac{r^2}{l^2}-M\right)\frac{\partial}{\partial r}\left(r\left(\frac{r^2}{l^2}-M\right)\frac{\partial\phi}{\partial r}\right)-\left(\frac{r^2}{l^2}-M\right)\frac{\partial^2\phi}{\partial\varphi^2}=0
\end{equation}
and may be solved, subject to our vanishing boundary conditions at $r=r_+ +\epsilon$ and $r=\infty$, by the usual method of separation of variables.  We seek (complex-valued) solutions of form
\[
\phi(t,r,\varphi)=e^{-i\omega t}e^{in\varphi}f_n(r)
\]
where $f_n(r)$ satisfies the appropriate radial equation
\begin{equation}
\label{radialeq}
\fl -\frac{1}{r}\left(\frac{r^2}{l^2}-M\right)\frac{d}{dr}\left(r\left(\frac{r^2}{l^2}-M\right)\frac{df_n(r)}{dr}\right)
+n\left(\frac{r^2}{l^2}-M\right)f_n(r)=\omega^2f_n(r).
\end{equation}
We remark that (\ref{radialeq}) may be written, alternatively, 
in the form of the Schr\"odinger-like equation with positive potential:
\[
\left(-\frac{d^2}{{dr^*}^2}+\left(\frac{r^2}{l^2}-M\right)\left(n^2+\frac{3}{4l^2}+\frac{M}{4r^2}\right)\right)r^{1/2}f_n(r^*)=\omega^2r^{1/2}f_n(r^*)
\]
where $r^*$ is related to $r$ in the same way as in (\ref{BTZtortoise}) and (\ref{BTZesiotrot}).
For each $n$, this will clearly have solutions which satisfy our boundary conditions only for certain discrete values of $\omega$, which we shall call $\omega_n^m$.   We also recall \cite{lOrt11} that, for any $\omega$, Equation (\ref{radialeq}) has (up to a multiplicative constant) the unique solution which vanishes at $r=\infty$, $f_n(r)=R_{\omega,n}(r)$, where

\begin{equation}
\label{hyper}
\fl R_{\omega,n}(r)=M^{\alpha+\beta}\left(\frac{r^2}{r_+^2}-1\right)^\alpha \left(\frac{r^2}{r_+^2}\right)^{\beta-\alpha}
F\left(\alpha+\beta+1,\alpha-\beta+1;2;\frac{r_+^2}{r^2}\right),
\end{equation}
where $\alpha=ir_+\omega/(2M)$ and $\beta=ir_+n/(2M\ell)$.
(cf. \cite{Kenmoku:2008qx} who give an apparently different expression).

The discrete allowed values, $\omega_n^m$, of $\omega$ may then, in principle, be found by imposing the boundary condition $R_{\omega,n}(r)=0$ at $r=r_+ +\epsilon$ and we shall, from now on, let $f_{nm}(r)$ stand for
$R_{\omega_n^m,n}(r)$ (defined, as in (\ref{hyper}), for $\alpha=ir_+\omega_n^m/(2M)$ and $\beta$ as before).

We will then have (see again \cite{lOrt11}) that the functions
\begin{equation}
\label{E:157b}
F_{nm}(t,\varphi, r)=\left(\frac{A}{\omega_n^m}\right)^{1/2}e^{-i\omega t}e^{in\phi}f_{n\omega}(r)
\end{equation}
will be a complete set of positive frequency modes for a suitable constant, $A$.

By creation and annihilation operator methods, one may then show (in analogy with equation (\ref{E:148b})), that the ground state two-point function will take the form
\[
{\cal G}_{\mathrm{BTZ1+2brick}}(t_1, \varphi_1, r_1; t_2, \varphi_2, r_2) = \hbox{`}\langle 0|\hat{\phi}(t_1, \varphi_1, r_1)\hat{\phi}(t_2, \varphi_2, r_2)|0\rangle\hbox{'}
\] 
\begin{equation}
\label{E:160b}
=\sum_{nm}\frac{A}{\omega_n^m}
e^{-i\omega_n^m(t_1-t_2)}e^{in(\varphi_1-\varphi_2)}f_{nm}(r_1)f_{nm}(r_2)^*.
\end{equation}
(Here $\sum_{nm}$ is shorthand for the sum over $n$ and $m$.)
The boundary limit of (\ref{E:160b}) will be defined (combining (\ref{Blim}) with Equation (\ref{1+2conf}) in Note $<$\ref{Note:confboundary}$>$  -- cf. the passage from (\ref{E:148b}) to (\ref{E:149b})) 
as the limit as $r_1^*$ and $r_2^*$ tend to zero of
$1/(r_1^*r_2^*)$ times -- equivalently the limit as $r_1$ and $r_2$ tend to infinity of $(r_1/r_+)^2(r_2/r_+)^2$  times -- ${\cal G}_{\mathrm{BTZ1+2brick}}(t_1,\varphi_1, r_1; t_2, \varphi_2, r_2)$, which is easily seen to be
\begin{equation}
\label{E:161b}
G_{\mathrm{BTZ1+2brick}}(t_1, \varphi_1; t_2, \varphi_2)=\sum_{nm}\frac{A}{\omega_n^m}e^{-i\omega_n^m(t_1-t_2)}e^{in(\varphi_1-\varphi_2)}
\end{equation}
(in analogy with equation (\ref{E:148b})).

Similarly,  in analogy with equations (\ref{E:150b}) and (\ref{E:152b}), we'll have
\[
{\cal G}_{\mathrm{BTZ1+2brick}}^\beta(t_1, \varphi_1, r_1; t_2, \varphi_2, r_2)=\sum_{nm}\frac{A}{\omega_n^m}
e^{in(\varphi_1-\varphi_2)}f_{nm}(r_1)f_{nm}(r_2)^*  \ \ \times
\]
\begin{equation}
\label{1+2betacal}
\left(\frac{e^{-i\omega_n^m(t_1-t_2)}+e^{i\omega_n^m(t_1-t_2)}e^{-\beta\omega_n^m}}{1-e^{-\beta\omega_n^m}}\right)
\end{equation}
and 
\begin{equation}
\label{1+2beta}
\fl G_{\mathrm{BTZ1+2brick}}^\beta(t_1, \varphi_1; t_2, \varphi_2) = 
\sum_{nm}\frac{A}{\omega_n^m}e^{in(\varphi_1-\varphi_2)}\left(\frac{e^{-i\omega_n^m(t_1-t_2)}+e^{i\omega_n^m(t_1-t_2)}e^{-\beta\omega_n^m}}{1-e^{-\beta\omega_n^m}}\right).
\end{equation}
both of which are, again, easily seen to satisfy the KMS condition (see Section \ref{Sect:prelim}).  Further, 
just as we showed above that the limit in which the brick wall is removed of $G_{\mathrm{BTZ1+1brick}}^\beta$
(\ref{E:152b}), for $\beta$ equal to the inverse Hawking temperature, equals the boundary limit, (\ref{E:100}), of the two-point function of the HHI state on 1+1 dimensional BTZ, so we expect that the limit in which 
the brick wall is removed of $G_{\mathrm{BTZ1+2brick}}^\beta$ will be equal to the boundary limit, (\ref{E:140vv}), of the two-point function of the HHI state on 1+2 dimensional BTZ.  However we have not shown this explicitly and, to do so, one would need to have more mathematical control on the values of the $\omega_n^m$.

What is important for us is that it is clear from the above that (even in the absence of any mathematical control on the values of the $\omega_n^m$) similarly to the 1+1 dimensional case, the one particle Hamiltonians
generating time evolution in both bulk and boundary in the ground-state representations corresponding to each of the brick-wall-modified two-point functions, ${\cal G}_{\mathrm{BTZ1+2brick}}$ (\ref{E:160b}) and $G_{\mathrm{BTZ1+2brick}}$ (\ref{E:161b}) are both unitarily equivalent to one another (both being the second quantizations, $d\Gamma(h)$, of the one-particle Hamiltonian, $h$, with non-degenerate spectrum now consisting of the $\omega_n^m$ (taken with appropriate multiplicity if/when any of these turn out to be degenerate).

So, just as in the 1+1 dimensional case, we expect that, in the presence of the brick wall, both bulk and boundary will have finite $<$\ref{Note:omeganmcondition}$>$ (and equal!) von Neumann entropies, given by $S=-\tr(\rho\ln\rho)$ for $\rho=e^{-\beta H}/Z$, $Z=\tr(e^{-\beta H})$, $H$ now meaning $d\Gamma(h)$ with $h$ (with spectrum consisting of the $\omega_n^m$) as in the previous paragraph.  

It turns out to be difficult to compute this entropy exactly, because of the difficulty of obtaining sufficient mathematical control over the $\omega_n^m$ although an attempt to do so was made by Ichinose and Satoh \cite{Ichinose:1994rg}.  Instead, we have applied the approximation method of Mukohyama and Israel, which we also used above in the 1+1 dimensional case, to obtain  Equations (\ref{1plus2brickfirstresult}) and (\ref{1plus2bricksecondresult}) below.  Essentially the same results as these (together with generalizations to rotating BTZ) have been obtained before, by Kim et al.\ \cite{Kim:1996eg} who also comment on the work of Ichinose and Satoh and explain how it can be reconciled with their ($=$ our) result.  We give the details below since our method (i.e.\ the application of the method of Mukohyama and Israel \cite{Mukohyama:1998rf}), is particularly short and direct and also since, by basing our discussion on the approach of Mukohyama and Israel \cite{Mukohyama:1998rf} we are able to draw a conclusion (about the effective number of Klein Gordon fields needed in a consistent quantum theory of gravity) which does not appear to have been drawn before in a 1+2 dimensional (BTZ) context.

Our starting point is again Equation (\ref{Sapprox}). Applying this formula to find the total entropy outside the brick wall in 1+2 dimensional BTZ, we again take $L=\infty$ and again have, by (\ref{Tlocal}), ${\cal T}_{\mathrm{loc}}(r)={\cal T}f^{-1/2}(r)$ where, by (\ref{E:110}), $f(r)$ is again equal to $(r^2/\ell^2-M)^{-1}$.  The entropy density, $s({\cal T})$ (by Note $<$\ref{Note:scalarentropy}$>$) will now be given by 
\begin{equation}
\label{2dimentropydens}
s({\cal T})=3\zeta(3){\cal T}^2/(2\pi)
\end{equation}
and $A(r)$ is $2\pi r$.   So (\ref{Sapprox}) becomes
\[
S=3\zeta(3){\cal T}^2\int_{r_+ + \epsilon}^\infty r\left(\frac{r^2}{\ell^2}-M\right)^{-3/2}dr
\]
\[
=-\frac{3\zeta(3){\cal T}^2\ell^3}{2}\left((r_+ +\epsilon)^2-r_+^2\right)^{-1/2}
\] 
where, for the last line, we recall that  $M=r_+^2/\ell^2$.  
For small $\epsilon$, this will be well-approximated by
\begin{equation}
\label{1plus2brickfirstresult}
\frac{3\zeta(3)}{2}{\cal T}^2\ell^3(2\epsilon r_+)^{-1/2}.
\end{equation}
We next set $\cal T$ equal to the Hawking temperature,
\begin{equation}
\label{THawking}
{\cal T}=\kappa/(2\pi) = r_+/(2\pi \ell^2), 
\end{equation}
and, following Mukohyama and Israel, express our result in terms of the metrical distance, $\alpha$, of the brick wall from the horizon, related to $\epsilon$ by
\[
\alpha=\int_{r_+}^{r_++\epsilon}f(r)^{-1/2}dr=\ell\int_{r_+}^{r_++\epsilon}\frac{dr}{(r^2-r_+^2)^{1/2}}\simeq 2\ell\left(\frac{\epsilon}{2r_+}\right)^{1/2},
\]
where for the last approximate equality, we have approximated $(r^2-r_+^2)$ by $2r_+(r-r_+)$.  Equivalently
\begin{equation}
\label{epsilonalpharelation}
\epsilon\simeq\frac{r_+}{2\ell^2}\alpha^2.
\end{equation}
With (\ref{THawking}) and (\ref{epsilonalpharelation}), (\ref{1plus2brickfirstresult}) becomes $3\zeta(3)r_+/(8\pi^2 \alpha)$.  The entropy, $S_{\mathsf{N}}$, of a collection of matter fields equivalent (see Section \ref{Sect:intro}) to $\mathsf{N}$ effective Klein Gordon fields may then be written in the suggestive form $<$\ref{Note:300}$>$  

\begin{equation}
\label{1plus2bricksecondresult}
S_{\mathsf{N}}={\mathsf{N}}\left(\frac{3\zeta(3)}{4\pi^3}\right)\frac{1}{\alpha}\left(\frac{2\pi r_+}{4}\right).
\end{equation}
We see from (\ref{1plus2bricksecondresult}) that the entropy is proportional to the appropriate notion of `horizon area' for 1+2 dimensions -- i.e.\ circumference, $2\pi r_+$.  Moreover, following the line of argument of Mukohyama and Israel \cite{Mukohyama:1998rf}, if the metrical distance, $\alpha$, from the brick wall to the horizon is taken to be $\approx$ 1 in natural units (see Section \ref{Sect:prelim}) then, for $\mathsf{N}$ around $4\pi^3/(3\zeta(3)) \approx 34$, the total matter entropy will take the Hawking value of $2\pi r_+/4$.  This suggests that a consistent theory of quantum gravity (with cosmological constant) in 1+2 dimensions will require a collection of matter fields equivalent to a number, $\mathsf{N}$, within an order of magnitude of 34 or so of (massless, real) effective Klein Gordon fields.  

We remark that, with a brick wall in the (1+3 dimensional) Schwarzschild metric, as discussed in \cite{'tHooft:1984re} and
\cite{Mukohyama:1998rf}, and for finite $L$ in (\ref{Sapprox}), the entropy arises as a sum of a (large) `area piece', due to a small region near the brick wall and proportional to horizon area, and a (typically smaller) `volume piece' associated with the region far away from the horizon.  In contrast, in the case of BTZ, even though $L$ is now taken to be $\infty$, we see from Equation (\ref{1plus2bricksecondresult}) that -- at least in our ($=$ Kim et al.'s \cite{Kim:1996eg}) approximation -- our entire entropy consists of an `area piece' and the `volume piece'  vanishes!

\section{Discussion}
\label{Sect:discuss}

\subsection{Summary of Results}
\label{subsect:summary}

We have provided general evidence that, at least for our simple massless Klein Gordon equation model, there seem to be well-defined versions of fixed-background holography, not only for plain AdS, but also for certain asymptotically AdS spacetimes, such as, in $1+ d$ dimensions, the Schwarzschild-AdS black hole spacetime.  In particular, specializing to 1+2 dimensions (BTZ) we have confirmed that the boundary theory is (locally) the same (i.e.\ that for the  conformal generalized free field with anomalous dimension $\Delta=1$) as that of plain 1+2 dimensional AdS and also that, as expected, the BTZ counterpart to the HHI state has, as its boundary limit (on, say, the right boundary cylinder) a thermal state of that boundary theory, at the appropriate Hawking temperature.  Moreover, and this was our main new result at a technical level, we have shown that, while these states of this theory have infinite entropy, in both bulk and boundary, if one imposes a suitable brick-wall cutoff in the bulk, labelled by the metrical distance, $\alpha$, from the brick wall to the horizon, then both bulk and boundary theories will have equal, finite, entropies, proportional to the `area' (i.e.\ circumference) of the event horizon.   Moreover, by the way, we have pointed out, following 't Hooft and Mukohyama and Israel, that, if one assumes that a suitable number, $\mathsf{N}$, of our Klein Gordon fields will serve to model the thermodynamic behaviour of the basic (matter and gravity) fields which make up a consistent theory of quantum gravity (in 1+2 dimensions, with cosmological constant) then, in order for the entropy of both bulk and boundary to equal the, presumed physically correct, Hawking value of one quarter of the `area' of the event horizon (i.e.\ $\pi r_+/2$), ${\mathsf{N}}/\alpha$ needs to be taken to be approximately 34.  (So, in particular, if $\alpha$ is around 1, then $\mathsf{N}$ needs to be around 34.)
 
Below, we shall attempt to draw some tentative conclusions from our results regarding the relationship between 
fixed-background holography (in particular, Rehren's algebraic holography) and mainstream Maldacena holography.  
In the course of our discussion, we shall recall the `matter-gravity entanglement hypothesis' of one of us (BSK) and the way in which this claims to offer a resolution to the Mukohyama-Israel `complementarity' puzzle.   Some (but not all) of our tentative conclusions assume this hypothesis and this resolution to be correct.  In particular, we point out that this resolution seems to suggest a radically different from usual understanding of the nature of Maldacena holography (namely as a one-way map from bulk to boundary rather than a bijection).  In the sequel, we shall find it convenient (despite our earlier terminological conventions in Section \ref{Sect:intro} and Note $<$\ref{Note:unfortunate}$>$) to adopt Arnsdorf and Smolin's terminology `Rehren dual'  to denote the bulk theory which corresponds to a given boundary CFT according to Rehren's algebraic holography.  One of our tentative conclusions which doesn't depend on our `matter-gravity entanglement hypothesis' (although which survives in an interestingly modified form with it) concerns the nature of the Rehren dual of the boundary CFT which arise in Maldacena's AdS/CFT correspondence. Finally, we shall also comment further on the Arnsdorf-Smolin puzzle in the light of our tentative conclusions $<$\ref{Note:Smolinprovisos}$>$.  

We begin, in the next subsection, by gathering together some, we think reasonable, extrapolations from what we have shown:

\subsection{Tentative generalizations and working assumptions}
\label{subsect:working}

We adopt a number of working assumptions throughout the sequel: First: that the result which we found in the 1+2 dimensional case holds for Schwarzschild AdS (with arbitrary mass $M$ and radius $\ell$) in arbitrary dimension 1+$d$ and for other quantum theories (not just the Klein-Gordon field) defined in terms of nets of local *-algebras on the fixed Schwarzschild-AdS background.  In particular, we shall assume that, for such quantum theories on such bulk spacetimes, the boundary theory, defined by a suitable generalization of (\ref{Blim}) on the appropriate 1+($d$-1) dimensional (say right) boundary cylinder will be locally the same boundary conformal $<$\ref{Note:same}$>$ quantum theory that one obtains on the boundary for the same bulk quantum theory on AdS.  Moreover, we shall assume that the HHI state will go over to a thermal state at the appropriate Hawking temperature on the boundary theory.  We shall assume further that, if necessary with a suitable brick wall in the bulk to make it finite, the entropy of the induced state on the boundary will always equal the bulk entropy.

Turning to consider the reverse direction of boundary to bulk:  Second: It also seems reasonable to expect (see $<$\ref{Note:expect}$>$)  that, given our conformal generalized free field with anomalous dimension given by (\ref{Delta}) for some choice of $d$ and $m$, defined on a 1+($d$-1) dimensional cylinder of radius $\ell$, then the bulk quantum theory corresponding to the covariant Klein Gordon equation (\ref{KG}) for mass $M$ on the right Schwarzschild wedge of the 1+$d$ dimensional Schwarzschild-AdS spacetime of mass $M$ and radius $\ell$ will be related to it by boundary limit holography.  (For the prospects for it also being related according to a suitable extension of Rehren's algebraic holography see Note $<$\ref{Note:bhalgholo}$>$).  Furthermore, we would expect a thermal state for any temperature, $\cal T$, on the former CFT to go over to a thermal state at the same temperature on each of the latter, bulk right Schwarzschild wedges (as well, of course, as on all of plain AdS).  However, in view of the uniqueness results in \cite{Kay:1988mu} and \cite{Kay:1992gr}, for given $M$ and $\ell$, we would only expect one of these bulk states to extend in a non-singular way from the right Schwarzschild wedge to the full *-algebra for the full (maximally extended) Schwarzschild-AdS bulk -- namely  the one for which $\cal T$ coincides with the Hawking temperature.  (And when it does extend, the extended state will be the HHI state.)

We shall also assume that this second working assumption generalizes to an arbitrary boundary CFT.  i.e.\ we shall assume that, given an arbitrary CFT whose Rehren dual is a given quantum theory in the bulk of plain AdS, then, if defined on a 1+ ($d$-1) dimensional cylinder of radius $\ell$, it will be related by boundary limit holography to the \textit{same} $<$\ref{Note:same}$>$ bulk theory defined on the right Schwarzschild wedge of the 1+$d$ dimensional Schwarzschild AdS spacetime of arbitrary mass, $M$, and radius $\ell$. A thermal state at temperature, $\cal T$, of the CFT on the boundary cylinder could then be seen as arising from the boundary limit of a thermal state at the same temperature of the same bulk theory.  But we would only expect such a state to arise from a non-singular state on the *-algebra when the same theory is defined on the maximally extended Schwarzschild AdS with mass, $M$, such that $\cal T$ is the Hawking temperature for that particular $M$ and $\ell$. (And when it does extend, the extended state would again be the HHI state.)

With these working assumptions, we next remark that, in fixed background holography, the very fact that the same boundary theory (say our conformal generalized free field with anomalous dimension $\Delta$ on a 1+($d$-1) dimensional cylinder of radius $\ell$) is related to more than one bulk theory (i.e.\ the Klein Gordon equation on 1+$d$ dimensional plain AdS and also on the right Schwarzschild wedge of Schwarzschild AdS for each value of $M$)  is a feature that is not shared with (presumably background independent!) mainstream Maldacena AdS/CFT.  In the latter theory, we expect, of course, that the different bulk geometries themselves arise from classical approximations to different states of the same bulk theory (in particular a ground state and a thermal state at the relevant Hawking temperature $\sqrt{M}/(2\pi\ell)$ for each $M$).  This contrast is well-illustrated by the Maldacena scenario involving an AdS$_5\times \mathbb{S}^5$ bulk, where \cite{Witten:1998zw} at low temperatures, a thermal state on a cylinder of the boundary CFT is believed to correspond to a state of the bulk theory $<$\ref{Note:double}$>$ which, in the large $N$/weak coupling limit consists of the relevant (Kaluza Klein reduced) supergravity theory on a bulk spacetime which is locally plain AdS$_5$, while at high temperatures (i.e.\ above the relevant phase transition \cite{Witten:1998zw}) and in the same limit, it consists of the same supergravity theory, but now on a five dimensional Schwarzschild-AdS background with an appropriate mass, $M$; whereas, on the other hand, one can, presumably, ask about the fixed-background holography counterpart to thermal states of the boundary CFT on the given radius-$\ell$ boundary cylinder at any temperature for bulk Schwarzschild-AdS backgrounds of any radius $\ell$ and any mass $M$ (or for plain AdS) -- and these will constitue a one-parameter family (i.e.\ parametrized by $M$) of \textit{different}  fixed-background holographic relationships.

\subsection{The connection between the Mukohyama-Israel puzzle and the Arnsdorf-Smolin puzzle}
\label{subsect:MI/ASconnection}

For definiteness, and also to make closer contact with \cite{Arnsdorf:2001qb},  let us take as our examplar of a quantum gravity theory the bulk (string) theory involved in the Maldacena AdS$_5$ correspondence (see Note $<$\ref{Note:AdSCFT}$>$).  It is reasonable and usual to suppose that a state of this theory which physically represents a (stable) quantum black hole in equilibrium with its thermal atmosphere will have an approximate description in terms of a classical 5-dimensional Schwarzschild AdS spacetime together with a thermal atmosphere consisting of the fields which result from a Kaluza Klein reduction of the relevant supergravity theory on the 5 dimensional Schwarzschild AdS times $\mathbb{S}^5$ -- defined on (say) the right Schwarzschild wedge $<$\ref{Note:double}$>$. We recall from Section \ref{Sect:intro} that the Mukohyama-Israel puzzle is the puzzle as to how it comes about that the entropy of this state can be calculated -- see again Note $<$\ref{Note:AdSCFT}$>$ -- in two seemingly quite different ways: On the one hand, it can be calculated as the gravitational entropy, i.e.\ one quarter of the area of the event horizon (divided by the appropriate string theory formula for Newton's constant) of the classical black hole background, while, on the other hand, it can presumably also be calculated as the entropy of the thermal atmosphere -- the latter calculation presumably requiring a suitable brick wall cutoff in order to compensate for a presumed flaw in the description of the thermal atmosphere  in terms of quantum fields on a fixed curved spacetime (which will, again presumably, get repaired in a more correct quantum gravitational description -- [but] see Subsection \ref{subsect:dual} below).

The connection with the Arnsdorf-Smolin puzzle (see \cite{Arnsdorf:2001qb} and Section \ref{Sect:intro})  is that there are now two seemingly quite different ways of arguing that the entropy of the quantum black hole state is equal to the entropy of the boundary CFT which arises in the AdS$_5$ version of AdS/CFT.  Firstly, thinking of the entropy of the state as given by the former gravitational entropy calculation, we can argue that this must be equal to the entropy of the boundary CFT using Maldacena holography.  On the other hand, if we assume that the (brick wall modified) thermal atmosphere is related to the same boundary CFT by fixed background holography, then, with our working assumptions, thinking of the entropy of our state as the entropy of this thermal atmosphere, we can argue that this must be equal to the entropy of the boundary CFT but, this time, the argument for the equality is based on fixed background holography.

Admittedly, the above remarks, as they stand, amount to just an interesting connection between our two puzzles rather than a resolution of either of them.  However, this connection would seem to suggest one interesting conclusion:  Namely, that the bulk theory which is related to our boundary CFT by boundary limit holography must at least be `something like' our thermal atmosphere -- i.e.\ the (Kaluza Klein reduction of) our supergravity theory on the right Schwarzschild wedge.  And, with our (generalization of our) second working assumption, this then implies that, in {\it plain} AdS,  the Rehren dual to our boundary CFT must be `something like' the same supergravity theory.  This would seem to support one of the tentative possible resolutions of their paradox raised by  Arnsdorf and Smolin themselves (see the paragraph  labelled `M1' on page 7 of \cite{Arnsdorf:2001qb} where they write that, ``in the [large $N$ and] $g_s\rightarrow 0$ limit, Maldacena holography and the Rehren duality could coincide'').  See however Subsection \ref{subsect:dual} below where we will offer a clarification of this and, in particular, argue for a certain clarification of the phrase `something like' here.

\subsection{The matter-gravity entanglement hypothesis and its implications for the Mukohyama-Israel puzzle and for Maldacena holography}
\label{subsect:matgrav}

In a series of articles \cite{Kay:1998vv, Kay:1998cj, Abyaneh:2005tc, Kay:2007rx} dating back to 1998 (see especially Endnotes (i)-(v), in \cite{Kay:2007rx}) one of us (BSK) proposed a physical interpretation of quantum gravity which, while seemingly consistent with conservative principles -- it assumes quantum gravity is a conventional quantum theory with a unitary time evolution -- entails a radically different from usual picture of quantum black holes. The main idea is to posit that any closed quantum gravitational system (be it one that contains a black hole or otherwise) is described by an (ever pure and unitarily evolving) pure state (here, we take `pure state' to be synonymous with `vector state'), $\Psi_{\mathrm{total}}$, on a total Hilbert space, ${\cal H}_{\mathrm{total}}$, which is a tensor product of a gravity Hilbert space, ${\cal H}_{\mathrm{gravity}}$, and a matter Hilbert space ${\cal H}_{\mathrm{matter}}$.   However, rather than identifying its physical entropy with the von Neumann entropy of the total state (which would of course be ever zero for a total pure state) it is proposed to identify it with the total state's matter-gravity entanglement entropy.  This can perfectly well be non-zero, and increasing with time, even though the total state is assumed to be pure and to evolve unitarily.  It thereby seems to offer a simple and attractive solution to the information loss puzzle \cite{Hawking:1976ra}.

As explained in Endnotes (i) and (iii) in \cite{Kay:2007rx} this `matter-gravity entanglement hypothesis' also seems to offer a natural resolution to, what we have called in the present paper, the Mukohyama-Israel puzzle.  For, in a vector state, the entanglement entropy of a bipartite system must, by an easy
theorem, be both the von Neumann entropy of the reduced density operator on the first system and equally the von Neumann entropy of the reduced density operator on the second system.  But, applying this theorem to our pure entangled state of matter-gravity, our physical entropy must be the von Neumann entropy of the reduced density operator of the gravitational field, which is plausibly the same thing as the gravitational entropy (i.e.\ one quarter of the area of the event horizon); but it must also be the von Neumann entropy of the reduced density operator of the matter field which is plausibly (close to -- see below) the same thing as the state of the (perhaps brick wall modified) thermal atmosphere in a quantum field theory in curved spacetime description.

As far as the operators of the quantum gravity theory are concerned, these presumably are described by a total *-algebra which is a tensor product of a gravity *-algebra, ${\cal A}_{\mathrm{gravity}}$, and a matter *-algebra,
${\cal A}_{\mathrm{matter}}$ (the full tensor product being represented on ${\cal H}_{\mathrm{total}}$).  Combining all this with the conclusions of the previous subsection, we seem to be led to the following view as to the nature of Maldacena holography in which, rather than being a bijection between the total bulk theory and the boundary CFT: 

\medskip

\noindent
\textit{Maldacena holography (say in its AdS$_5$ version) is a one-way mapping from bulk to boundary in which the boundary CFT is identified with (and arises as a suitable boundary limit of) bulk operators of form $I \otimes {\cal A}_{\mathrm{matter}}$ ($I$ here denoting the identity operator on the gravity Hilbert space).}  

\smallskip

\noindent
\textit{Moreover, the proper understanding of the well-established  (see Section \ref{Sect:intro} and Note $<$\ref{Note:AdSCFT}$>$)  equality between the bulk gravitational entropy and the boundary CFT entropy in the case of a thermal state on the 1+3 dimensional boundary cylinder (at a temperature above the Witten phase transition temperature) is that the total bulk state (as usual, approximately described by the relevant classical Schwarzschild-AdS background with the relevant Kaluza Klein-reduced supergravity fields/strings propagating on it) is a {\it pure} total state, $\Psi_{\mathrm{total}}$, of matter-gravity, entangled in just such a way that the reduced density operators (i.e.\ of 
$|\Psi_{\mathrm{total}}\rangle\langle\Psi_{\mathrm{total}}|$) on each of ${\cal H}_{\mathrm{matter}}$ and 
${\cal H}_{\mathrm{gravity}}$ are each separately (approximately) thermal at the relevant Hawking temperature.   When we say that the entropies of bulk and boundary are equal, what is really happening is that the von Neumann entropy of the (mixed) boundary thermal density operator (equals the von-Neumann entropy of the reduced density operator on ${\cal H}_{\mathrm{matter}}$ which) equals the matter-gravity entanglement entropy of the (pure!) bulk state.}

\medskip

Before discussing this further, we should clarify/emphasize again one important point (this is the same point we make in Note $<$\ref{Note:double}$>$ -- see $<$\ref{Note:double}$>$  and other places in the text where this note is referred to):  When we discuss quantum field theory on a fixed Schwarzschild AdS background, it is appropriate to consider the background to be full maximally extended Schwarzschild AdS spacetime which contains a left wedge and a right wedge as well as future and past wedges, and has two disconnected ${\mathbb R}\times {\mathbb S^3}$ cylinders as its conformal boundary.  This is indeed the 1+4-dimensional counterpart to the 1+2 dimensional situation we discussed here in Section \ref{Sect:1+2} (see Figure \ref{Fig:BTZcylinders}) where, indeed we found that the thermal nature of the HHI state on the right wedge region resulted from entanglement in the overall-pure HHI state between the right wedge and the left wedge.  In our above discussion of full quantum gravity and, in particular, in our above italicised statement, on the other hand, we have assumed that (to the extent that it is still possible to talk in terms of a classical spacetime background) all the physics takes place in the right wedge and the thermal nature of the matter in the right wedge is explained in terms of entanglement with the gravitational field (again in the right wedge).   (Note that, by \textit{monogamy} $<$\ref{Note:monogamy}$>$ it could not be simultaneously entangled with the matter in a left wedge and with the gravitational field in the right wedge.)

Thus we have made the assumption that, in passing from quantum field theory in curved spacetime to quantum gravity, the left wedge and the future and past wedges disappear from the picture.    Since the first version of this paper was written, one of us \cite{Kay:enclosed} has given new evidence that this must happen based on an argument  that the horizons of Schwarzschild AdS spacetime (and other \textit{enclosed} horizons with a similar causal structure) are \textit{unstable} against the switching on of the coupling to gravity.

The above italicised statements seem quite heretical although we are unaware of any decisive evidence against them.  We note, in this regard, that Arnsdorf and Smolin have argued in \cite{Arnsdorf:2001qb} that it may well be consistent with what is known (was known as of mid 2001) about Maldacena holography that it might just be a one-way map from bulk to boundary, rather than a bijection (see also a similar point by Giddings, reiterated in his recent paper \cite{arXiv:1105.6359}); and such a result, would, as they themselves point out, go at least part of the way towards resolving their paradox.   In any case, even if there may turn out to be evidence against it, such evidence should, we feel, be weighed against the virtue of the above view that it would seem to fit better with the hope that string theory can be part of the solution to the information loss puzzle than the conventional view about the interpretation of string theory (where one adopts the conventional picture of black hole equilibrium states as non-pure states -- a picture which seems to be at odds with any such resolution).     

Of course, for these italicised statements to be less than vague, we would need to specify what might be meant by the `matter-gravity split' in a string-theory context.  We have to admit that we presently do not have a clear answer to this but would make the following remarks:  First (cf. especially Note (ii) in \cite{Kay:2007rx}) the splitting between matter and gravity is only expected to be an approximate notion which only makes sense (well) below the Planck energy.  Secondly, in a more physically realistic string theory (where, say, there are compact `extra' dimensions which are small) then we can, of course, identify, at low energies, in a field-theoretic approximation involving a Kaluza-Klein reduced supergravity theory etc., what are photons, electrons etc. (i.e.\ matter) and what are gravitons (gravity) etc.  Thirdly, as Susskind \cite{Susskind:1994vu} has argued, we expect a (far from extremal) black hole to correspond in the weak string-coupling limit to a very long string so it seems plausible that such states of very long strings should get lumped together as part of the gravitational field.  (\textit{Added note:} Since the first version of this paper was written, one of us has recently pursued the implications of the matter-gravity entanglement hypothesis adopting a view about the nature of the matter-gravity split consistent with this expectation, i.e. that in a black hole equilibrium state there is, in an appropriate weak-string coupling limit, a rough correspondence: \textit{gravity} $\leftrightarrow$ \textit{long string}; \textit{matter} $\leftrightarrow$ \textit{atmosphere of small strings} \cite{Kay:thermality, Kay:stringy, Kay:morestring} with promising results.)  Finally, it is tempting to wonder if one can, naively, make the identification:  \textit{open strings} $\leftrightarrow$ \textit{matter}; \textit{closed strings} $\leftrightarrow$ \textit{gravity}.  But it is then not clear how this would fit with our above remark about long strings etc.  We should also remark that there may be more than one way of realizing a splitting between closed and open strings in the bulk.   In fact, quite aside from the fact that such a splitting can of course anyway only make sense in an approximate way at weak string coupling, there may also possibly be alternative dual (now really in the sense of equivalent!) descriptions of the bulk theory in which the parts of the theory which are identified as closed-string and open-string sectors are different.  For example, IY Park  \cite{Park:1999xz} has argued for a dual (i.e. equivalent) description of the bulk theory in which, rather than a Type IIB theory,  it is a theory with open strings (some with fully Neumann boundary condistions) as well as closed strings and D-branes.

As a final remark on this issue: it may be that what we need to do is turn things around and simply {\it define} ${\cal A}_{\mathrm{matter}}$ to be the Rehren dual (see Subsection \ref{subsect:dual}) of the *-algebra of the Maldacena boundary CFT.

We next make a couple of further comments on the scenario proposed in our italicised statements above.  First, another way of expressing things is that the pure vector state, $\Psi_{\mathrm{total}}$, in the total Hilbert space ${\cal H}_{\mathrm{total}}$ which describes the bulk matter-gravity equilibrium {\it purifies} the thermal density operator on ${\cal H}_{\mathrm{matter}}$ (cf. \cite{Kay:1985yw,Kay:1985yx,Kay:1985zs, Kay:1988mu}).  Secondly, on our scenario, an arbitrary element of ${\cal H}_{\mathrm{total}}$ will be approximated as closely as one likes by acting on $\Psi_{\mathrm{total}}$ with operators of the form $I \otimes {\cal A}_{\mathrm{matter}}$ -- which is identified with the *-algebra of operators of the boundary CFT.  So in this sense, it would still be correct  to say (cf. \cite{Maldacena:1997re}) that \textit{the bulk (string theory) Hilbert space is the same as the boundary CFT Hilbert space}.  Cf. the analogy with the `right-wedge Reeh Schlieder property' which we point out in Note $<$\ref{Note:double}$>$.  Another way of saying this is that the total Hilbert space can be equated with the GNS Hilbert space (see Section \ref{Sect:prelim}) for the (algebraic) state ${\rm trace}(\rho(\cdot))$ on ${\cal A}_{\mathrm{matter}}$.  In other terminology, it is the Hilbert space in the sense of thermofield dynamics \cite{TakUme, Takahashi:1996zn}.   However, at zero temperature (i.e.\ for the ground state on the boundary CFT -- now on Minkowski space) when the bulk is plain AdS this will no longer be the case.  In fact, in line with our `matter-gravity entanglement hypothesis' and the expectations above, we would expect the total ground state to be (approximately) a tensor product (i.e.\ an unentangled state) of a gravity ground state and a matter ground state and therefore its entropy (as we define it) will be zero -- which is of course in line with the conventional expectation.

Lastly, in the paper \cite{arXiv:1105.6359} by Giddings which we cited in a parenthetical remark above, it is speculated that the bulk-to-boundary mapping may be such that the boundary theory only captures an ``appropriately coarse-grained'' description of bulk physics.  We remark that our proposal in our italicised statement above might perhaps be regarded as a concrete implementation of this idea.  After all, the process of taking a partial trace (in our proposal, over gravity so as to obtain the reduced density operator of matter in the bulk which will then be equivalent to the thermal state of the boundary CFT) can be regarded as a particular sort of quantum counterpart to classical coarse-graining -- a sort which in our context (in particular, because we suppose the bulk to be in a pure state) is, as we indicated above, consistent with the resolution to the information loss puzzle proposed in \cite{Kay:1998vv, Kay:1998cj, Abyaneh:2005tc, Kay:2007rx}.

\subsection{More about the Rehren dual of the boundary CFT of Maldacena AdS$_5$ holography}
\label{subsect:dual}

Finally, we return to our tentative conclusion, in Subsection \ref{subsect:MI/ASconnection}, that the Rehren dual of the Maldacena boundary CFT must at least be `something like' our thermal atmosphere -- i.e.\ our (Kaluza Klein reduction of) our supergravity theory on our right Schwarzschild wedge.   First of all, we recall that Rehren pointed out in \cite{Rehren:2004yu} that the large $N$ limit of the Maldacena CFT is not itself a quantum field theory -- and it is of course strictly only this large $N$ limit which (at fixed 't Hooft coupling) is believed to correspond to classical supergravity.  So the question we ask ourselves is:  What is the Rehren dual of the Maldacena CFT at large but finite $N$?  What we would seem to learn from our results (with the extrapolations and working assumptions we made in Subsection \ref{subsect:working} ) is that \textit{this cannot be an ordinary quantum field theory}!  After all, if it were an ordinary quantum field theory, it would presumably have a formally infinite entropy requiring a brick wall to be made finite.  And yet the Maldacena CFT (unlike e.g.\ the conformal generalized free field whose Rehren dual is the Klein Gordon equation) \textit{is} presumably an ordinary quantum field theory and therefore is expected to have a finite entropy density and hence a finite total entropy when defined on a cylinder, {\it without} the need for any brick wall in the bulk!   As to what its Rehren dual could be, it would seem reasonable to suggest that the phrase `something like our thermal atmosphere' should, in the light of this, be (partially) clarified in the following way:

\medskip

\noindent
\textit{Whatever the Rehren dual of the Maldacena CFT (i.e.\ of ${\cal N}=4$ supersymmetric Yang Mills theory) may be, for fixed $N$ and at finite Yang-Mills coupling constant $g_{\mathrm{YM}}$, in the limit $g_{\mathrm{YM}}\rightarrow 0$, it is something like the (Kaluza Klein reduction of) the bulk string theory of AdS$_5 \times {\mathbb S}^5$ in the limit of zero string coupling $g_s$ (but at finite string length $\ell_s$).}

\medskip

Part of our rationale for saying this is that, at zero $g_{\mathrm{YM}}$ (which corresponds -- see Note $<$\ref{Note:AdSCFT}$>$ -- to zero
$g_{s}$) the string theory will be a theory which makes sense on a fixed background.  Another part is that we suppose that (such a limit of) a string theory will have a better chance than a quantum field theory of not needing a brick wall in order to have a finite entropy.

This (still partly vague) conclusion did not depend on the `matter-gravity entanglement hypothesis' of \cite{Kay:1998vv, Kay:1998cj, Abyaneh:2005tc, Kay:2007rx}.  However, on that hypothesis, it would be natural to add the further clarification that it is in fact `something like the {\it matter sector} of the (Kaluza Klein reduction of) the bulk string theory'.   (If/when that can be given a clear meaning.)

In support of either of these tentative conclusions (i.e.\ with or without the above further clarification) we recall first that Rehren himself has suggested (in \cite{Rehren:2000tp}) that the Rehren dual of the Maldacena CFT may well involve strings.  Also there is work (e.g.\ by Dimock \cite{Dimock:2000yv, Dimock:2001dy}) suggesting that it will be possible to incorporate (at least free, open) string field theory into the framework of nets of local *-algebras.

All these tentative conclusions are, of course, more or less speculative.  But, assuming they are on the right track, they would seem to go at least some of the way towards a possible resolution of the Arnsdorf-Smolin puzzle.  The puzzle will probably only be fully resolved when we have a satisfactory background-independent formulation of string theory, and if and when we have a clearer understanding  of the nature of the Rehren dual of the Maldacena boundary CFT.  Such a clearer understanding of the Rehren dual would, on our matter-gravity entanglement hypothesis, as we remarked above, also presumably clarify how the matter-gravity split should be defined in the bulk string theory.

\section{Notes}
\label{Sect:notes}

\begin{enumerate}

\item
\label{Note:AdSCFT} 
The principal examples of AdS/CFT involve AdS$_5\times {\mathbb S}^5$ and AdS$_3\times {\mathbb S}^3\times X$ bulks where $X$ denotes either $\mathbb{T}^4$ or $K3$.  See \cite{Aharony:1999ti} or \cite{Ross:2005sc}.  The latter review by Ross provides a useful bridge between work on quantum field theory in curved spacetime and on Euclidean quantum gravity on the one hand and work on AdS/CFT on the other.  

\smallskip

We briefly recall, in the remainder of this note, more details about the  AdS$_5\times {\mathbb S}^5$ example since we shall refer to this extensively in Section \ref{Sect:discuss}.  We shall also include an outline of the calculation of the entropy of a thermal state of its boundary CFT, when defined on a cylinder, in the regime in which the bulk is well described by a classical Schwarzschild AdS gravitational field, since this result is central to much of our discussion and appears difficult to find in a single place.  (The account below is pieced together  from  \cite{Maldacena:1997re, Gubser:1996de, Witten:1998zw, Klebanov:2000me, Ross:2005sc}.  See also e.g.\ \cite{Ortin} for the definition of Newton's constant in string theory. ) 

\smallskip

In the AdS$_5\times {\mathbb S}^5$ example, the bulk theory is Type IIB string theory (with string length $\ell_s$ and string coupling constant $g_s$) with 5 units of Ramond Ramond flux on AdS$_5\times {\mathbb S}^5$ .  Both the AdS$_5$ and the ${\mathbb S}^5$ have the same radius $\ell$.  The boundary theory is ${\cal N}=4$ supersymmetric Yang Mills theory for the gauge group $SU(N)$ with coupling constant $g_{YM}$.  The `dictionary' that relates the two theories includes the entries (now only setting $\hbar$ and $c$ equal to 1): $g_{\mathrm{YM}}^2=4\pi g_s$; $g_{\mathrm{YM}}^2N$ ($=$ the `'t Hooft coupling') $=(\ell/\ell_s)^4$.  Also,   Newton's constant, $G_{10}$, in 10 dimensions is related to $g_s$ and $\ell_s$ by $G_{10}=8\pi^6g_s^2\ell_s^8$ while, we remark that, in a Kaluza Klein reduced picture, the 5-dimensional Newton's constant, $G_5$, is equal to $G_{10}$ divided by the `area', 
$\pi^3\ell^5$, of the 5-sphere.  (Below, we will also need that the `area' of a 3-sphere of radius $R$ is $2\pi^2R^3$.)
When $N$ is very large, $g_{\mathrm{YM}}$ is small, but the 't Hooft coupling, $g_{\mathrm{YM}}^2N$, is large, then the bulk theory is believed to be well-approximated by ${\cal N}=4$ classical supergravity on
AdS$_5\times {\mathbb S}^5$. 

\smallskip

The entropy of the supersymmetric Yang Mills theory at temperature $\cal T$ on a 1+3 dimensional cylinder of radius $\ell$ (and hence spatial volume, say $V$, equal to $2\pi^2\ell^3$) in this regime is calculated, according to AdS/CFT   (assuming, for the given $\ell$, that $\cal T$ is above the Hawking-Page-like \cite{Hawking:1982dh} phase transition discussed in \cite{Witten:1998zw})  by simply equating it with the Hawking entropy of the 10-dimensional black hole, consisting of a 5-dimensional classical Schwarzschild AdS of asymptotic radius $\ell$ times the 5-sphere of the same radius $\ell$, for which the Hawking temperature is $\cal T$.  There are actually two such classical metrics, as one can see as follows.  First, we notice that the metric takes \cite{Witten:1998zw} the form (\ref{generalsphmetric}) for $d=4$ with $f(r)= 1-\mu/r^2+r^2/\ell^2$ (where, by the way, $\mu$ is related \cite{Witten:1998zw} to the black hole mass, $M$, by $\mu=8\pi G_5/(3\pi^2)$) and one easily sees that, for each $\mu$, there is a unique value of $r$ for which $f(r)=0$, namely $r_+$ ($\simeq \sqrt\mu$).  Secondly, the surface gravity, $\kappa$ (see Section \ref{Sect:prelim}) is $f'(r_+)/2$, which (using $f(r)=0$ again) is easily seen to be $2r_+/\ell^2 + 1/r_+$.   Equating $\cal T$ with the Hawking temperature, $\kappa/(2\pi)$, we find that $r_+$ must be a root of the quadratic equation $2r_+^2-(2\pi\ell^2{\cal T})r_++\ell^2=0$.  Just as in the $1+3$ dimensional case \cite{Hawking:1982dh}, there are two of these: one `small', representing an unstable equilibrium, the other `large', with $r_+\simeq \pi\ell^2{\cal T}$ representing a stable equilibrium.  Choosing the latter, the entropy is given by $S=$ `Area$/(4G)$' $=(2\pi^2r_+^3)(\pi^3\ell^5)/(4G_{10})$ (or alternatively, but equivalently, by $2\pi^2r_+^3/(4G_5)$)  which we can write as $(\pi^3/4)(V\ell^3{\cal T}^3/G_{10})$ which, using the dictionary entries and the formula for $G_{10}$ and the definition of $V$ above, is equal to $\pi^2N^2V{\cal T}^3/2$.  This is to be compared with the value $2\pi^2N^2V{\cal T}^3/3$ for the boundary CFT at fixed $N$ and weak Yang Mills coupling -- see \cite{Klebanov:2000me} both for the derivation of this latter formula and for a discussion of the physical significance of the factor of $3/4$ discrepancy.   As explained there, one takes these calculations as (part of the) evidence that, for any $N$, and any value of the Yang Mills coupling, the boundary entropy equals the bulk entropy.

\item
\label{Note:thinloc} The *-algebra isomorphism in \cite{Kay:2007rf} involves a certain `thinning-out' of test functions -- see \cite{Kay:2007rf} and Note  $<$\ref{Note:vNvsCstar}$>$.  There is also a significant caveat about the nature of the localization of the smeared boundary fields in the isomorphism, explained in Note [14] of \cite{Kay:2007rf} 

\item 
\label{Note:vNvsCstar}  
Rehren had in mind von Neumann algebras.  Kay and Larkin's construction \cite{Kay:2007rf} referred to below will (see the place in \cite{Kay:2007rf} where reference is made to \cite{Kay:2006jn}) be in terms, say, of $C^*$ algebras. However one can easily see that it can be extended, once one chooses to represent fields in the global AdS vacuum representation, to a von Neumann algebra correspondence and, indeed, it seems conceivable that the nicety Kay and Larkin found in \cite{Kay:2007rf} about the need to `thin-out' the test functions on the boundary in order to obtain a bulk-boundary isomorphism will go away when such an extension is made.

\item
\label{Note:wrap} 
We should point out that the algebraic holography work of Rehren concerns the ${\mathbb Z}_2$ quotient of true AdS (i.e.\ with closed timelike curves) while the Kay-Larkin work is done on the covering space and no ${\mathbb Z}_2$ quotient is taken.  We shall assume throughout this paper that, unless it is clear otherwise from the context, when we refer to AdS, we mean the covering space (what is sometimes called CAdS) of true AdS.

Related to this, we should point out that, in Rehren's paper, the usual notion of commutativity at spacelike separation gets replaced by the commutativity of local algebras for regions which are not connectable by timelike \textit{geodesics}.   On CAdS, however, one may adopt the usual notion of spacelike commutativity and also one may regard the ${\mathbb Z}_2$ quotient of AdS as a single Poincar\'e chart (see e.g. \cite{Kay:2007rf}) of CAdS together with the addition of certain boundary points with certain identifications.  With this point of view,  Rehren's map restricts to a map from wedges in a Poincar\'e chart of CAdS to boundary double cones in the conformal boundary of the Poincar\'e chart which, in turn, is of course conformal to Minkowski space.  On the other hand, the conformal boundary of the ${\mathbb Z}_2$ quotient of AdS which Rehren considers may be identified with the usual
(cf.\ e.g.\ \cite{Haag}) conformal compactification of Minkowski space.

\item
\label{Note:Wightman} 
We remark that to have something which deserves to be called a Wightman theory in the bulk is roughly equivalent to a specification of a bulk field algebra together with the choice of the ground state for the time translations of global coordinates.  The Wightman functions will then be interpretable as expectation values of products of fields in that state.   Also, when we refer to a Wightman theory on the conformal boundary, what we really mean is that the theory will be a Wightman theory on $d$ dimensional Minkowski space when that is identified in the usual way (see \cite{Kay:2007rf}) with the boundary of a Poincar\'e chart.

\item
\label{Note:confboundary}
In Equation (\ref{Blim}), we assume $(q,x^i)$, $x^i\in {\mathbb R}^d$ denotes any system of coordinates such that the 
metric of AdS takes the form
\[
ds_{\mathrm{AdS}}^2=\ell^2\Xi(q)^2(dq^2+ds_{\mathrm{boundary}}(x^i)^2)
\]
where $\ell$ is the AdS radius (see (\ref{E:a12}) and (\ref{E:104})) and $\Xi(q)$ is positive and tends to infinity as $q$ tends to $q_0$, which is the location of the conformal boundary.

\smallskip

In Equation (\ref{E:a14}), which relates to 1+1 dimensional AdS in global coordinates, $q$ is identified with $\rho$, $x^i$ with $\lambda$, $\Xi(q)$ with $\sec(\rho)$, and $q_0$ with $\rho=\pm \pi/2$ etc.  Similarly for Equations (\ref{E:79}) (Poincar\'e coordinates), (\ref{E:106aa}) (global coordinates on 1+2 dimensional BTZ) and (\ref{E:108}) (Poincar\'e coordinates).  In the case of BTZ coordinates in 1+1 and 1+2 dimensions, we remark that Equation (\ref{E:bb3}) can be rewritten, using our tortoise-like coordinate (see (\ref{BTZtortoise}) and (\ref{BTZesiotrot})) in the above form as (\ref{E:141b})
($ds^{2}=f(r(r^*))\left(dt^2-{dr^*}^2\right)$) while equation (\ref{E:108}) can be rewritten as
\begin{equation}
\label{1+2conf}
ds^2=f(r(r^*))\left(dt^2-{dr^*}^2-\frac{r(r^*)^2}{f(r(r^*))}d\varphi^2\right)
\end{equation}
where $f(r)$ is given by (\ref{E:16}).

\item
\label{Note:Rehgeo}
We note, by the way, that the notion of commutativity at spacelike separation in \cite{Rehren:1999jn} is different from the usual one and corresponds to the non-existence of timelike geodesics connecting the two regions in question rather than the, more usual, non-existence of timelike curves.   However, if one reformulates Rehren's algebraic holography in terms of the covering space (CAdS) of AdS (as is done implicitly in \cite{Kay:2007rf} -- see Note $<$\ref{Note:wrap}$>$ and/but Note $<$\ref{Note:vNvsCstar}$>$) then the relevant notion can be taken to be non-existence of timelike curves connecting the two regions.  

\item
\label{Note:blog}
The objections to algebraic holography referred in the main text include a number of articles in the blogosphere; see in particular \hfil\break
http://golem.ph.utexas.edu/$\sim$distler/blog/archives/000987.html

\item
\label{Note:slice}
The time-slice condition on a quantum field theory on a given spacetime with a choice of time-coordinate is that the 
*-algebra for the region of the spacetime between two times is equal to the *-algebra of the full spacetime.  

\item
\label{Note:unfortunate}
The terminology issue is possibly confused further because, in a number of references (including \cite{Arnsdorf:2001qb} as well as the paper, \cite{Kay:2007rf}, co-authored by one of us) the Rehren algebraic holography correspondence between quantum theories in the bulk and on the boundary is also referred to as a `duality'  -- the word being used, in these references, as if synonmyous with `isomorphism'.  We shall also use the word `dual' ourselves in this sense in Section \ref{Sect:discuss} when we talk about the `Rehren dual' of the Maldacena boundary CFT.

\item
\label{Note:bhalgholo} 
In addition to direct boundary-limit holography, we also expect the algebraic holography of Rehren to extend from plain AdS to Schwarzschild AdS (and in particular, BTZ)  in a suitable way:

\smallskip

If, for a given double cone on, (say) the right Schwarzschild-AdS boundary cylinder, we define (cf. the equivalent definition given by \cite{Arnsdorf:2001qb} to Rehren's definition of `wedge' for plain AdS) the bulk wedge which corresponds to it to be the `causal completion' in the bulk of the given boundary double cone (i.e.\ the set of all points in the bulk which can be connected by both past and future directed causal curves to points in the given double cone) then it is clear that spacelike related boundary double cones will correspond to spacelike related bulk wedges.   So at least an important part of the main `geometric Lemma' of \cite{Rehren:1999jn} will generalize to the case of Schwarzschild AdS and still give us a bijection between the net of local *-algebras for the right cylinder boundary and the net of local *-algebras for the exterior right wedge region of the BTZ bulk which maps the sub*-algebras for double cones to the sub*-algebras for wedges.  However, some of the properties of the bijection in the case of plain AdS may not generalize because e.g.\ of the lower degree of symmetry.  In particular, it seems that all the bulk wedges which correspond to boundary cones `point in the same direction' so one cannot take intersections of them to obtain bulk double cones.

\smallskip

We would also expect there to be a suitable generalization of the pre-holography work of \cite{Kay:2007rf} showing that, subject to the caveats mentioned in Note $<$\ref{Note:thinloc}$>$, 
in the case of the bulk (real) scalar Klein Gordon field, this bijection will be between the net of local *-algebras for the 
bulk covariant Klein Gordon equation on the bulk right wedge of Schwarzschild AdS and the net of local *-algebras for the 
conformally invariant generalized free field of (\ref{Wanom}) on the right cylinder.  

\smallskip

Further we expect this bijection to extend (both abstractly and in the concrete case of the Klein Gordon bulk field) in a suitable way to a bijection from the net of local *-algebras for the full bulk Schwarzschild AdS and the net of local *-algebras for the union of the right and left boundary cylinder.

\item
\label{Note:onequarter}
Given that we raise a number of puzzles and apparent paradoxes in our introduction which relate to black hole entropy, it is worth saying explicitly that one thing we shall not question is the belief that the entropy of a black hole has (at least approximately) the Hawking value of $1/4$ of the area of its event horizon. The strongest reason for believing this for a physical black hole is  the original thermodynamic argument from the Hawking radiation formula as given in \cite{Hawking:1974sw}.  But it is also believed to hold for AdS black holes (see e.g.\ \cite{Hawking:1982dh}) and in dimensions other than 4 and we shall not question these beliefs either.  For the 1+2 dimensional BTZ black hole, it also holds (see e.g.\ \cite{Ross:2005sc}) with `area', of course, understoood to mean circumference -- i.e.\ (in the notation of Section \ref{Sect:1+1} and \ref{Sect:1+2}) $2\pi r_+$.  (But we do not have anything new to say here as to why the relevant coefficient of area takes the value $1/4$.)

\item
\label{Note:expect}
Our remarks which contain the words `expectations' and `arguments' in the main text and in the notes (especially Notes $<$\ref{Note:bhalgholo}$>$, $<$\ref{Note:equiv}$>$, and $<$\ref{Note:Ham}$>$) should be interpreted not as mathematical proofs but rather as conjectures or sketches which still need to be filled in; they may turn out to require provisos and extra conditions etc. For example, in view of the results in \cite{Kay:2007rf}, it seems possible that
(and it would be interesting to investigate whether) the statement in the main text to the effect that the direct boundary limit on bulk fields would be expected to inherit the entanglement and thermal properties of the bulk theory might (in the case in which the bulk field theory is (\ref{KG})) turn out to hold only when the mass, $m$ is such that $\Delta$ in (\ref{Delta}) is an integer or half-integer.  

\smallskip

The results that we will obtain in Sections (\ref{Sect:1+1}) and (\ref{Sect:1+2}) will show that our expectations are all actually fulfilled at least for the zero mass Klein Gordon fields which we study there.

\item
\label{Note:equiv} We expect that the *-algebra ${\cal A}_{\mathrm{DW}}$ (which will, up to possible small technicalities, be the same as the *-algebra for full BTZ) will arise in the form of a tensor product of  *-algebras, ${\cal A}_{\mathrm{LW}}$ for the left wedge (i.e.\ the triangle ACF in Figure \ref{Fig:ads2}) and ${\cal A}_{\mathrm{RW}}$ for the right wedge (the triangle ECD in Figure \ref{Fig:ads2}) in such a way that the pair $({\cal A}_{\mathrm{DW}}, \alpha_{\mathrm{DW}}(t))$, together with the appropriate wedge-reflection involutary antiautomorphism, $\iota_{\mathrm{W}}$ (see \cite{Kay:1985yw} and \cite{Kay:1985zs}) will be what is called in \cite{Kay:1985yw} a `double dynamical system' and similarly we expect that the *-algebra ${\cal A}_{\mathrm{DC}}$ will arise in the form of the tensor product of  *-algebras, ${\cal A}_{\mathrm{LC}}$ for the left cylinder (see Figure \ref{Fig:BTZcylinders}) and ${\cal A}_{\mathrm{RC}}$ for the right cylinder, in such a way that the pair $({\cal A}_{\mathrm{DC}}, \alpha_{\mathrm{DC}}(t))$, together with an appropriate involutary antiautomorphism, $\iota_{\mathrm{C}}$, which maps between ${\cal A}_{\mathrm{LC}}$ and ${\cal A}_{\mathrm{RC}}$ and reverses the sense of time will also be a `double dynamical system' (and we expect, further, that these, in turn, will each arise by second quantization from appropriate `double linear dynamical systems' for the appropriate underlying classical theories as in \cite{Kay:1985yx}).  With this terminology, the expectation, stated in the main text, that ``the direct boundary limit on the pair of cylinders of a bulk theory defined on BTZ would be expected to inherit the entanglement and thermal properties of the bulk theory'' can then be formulated and argued for more precisely by saying, first, that we expect that the direct boundary limit will induce a natural isomorphism between the double dynamical systems $({\cal A}_{\mathrm{DW}}, \alpha_{\mathrm{DW}}(t), \iota_{\mathrm{W}})$ and $({\cal A}_{\mathrm{DC}}, \alpha_{\mathrm{DC}}(t), \iota_{\mathrm{C}})$ and, second, that the HHI state on the bulk spacetime will be a `double KMS state',  in the sense of \cite{Kay:1985yw}, at the Hawking temperature, on $({\cal A}_{\mathrm{DW}}, \alpha_{\mathrm{DW}}(t), \iota_{\mathrm{W}})$.  Hence, by the natural isomorphism just mentioned, the state induced on the boundary by taking the direct boundary limit of fields, will be a double KMS state at the Hawking temperature on the (isomorphic) double dynamical system $({\cal A}_{\mathrm{DW}}, \alpha_{\mathrm{DW}}(t), \iota_{\mathrm{W}})$.  (We note that in the case of 1+1 dimensional BTZ, the right and left cylinders of course reduce to lines.)

\item
\label{Note:Ham} 
The quantum dynamical system (see Section \ref{Sect:prelim}) consisting of the bulk (total) *-algebra, ${\cal A}_{\mathrm{RW}}$, for, say, the bulk quantum field theory of (\ref{KG}), on, say the right BTZ wedge, together with the automorphism group, $\alpha_{\mathrm{RW}}(t)$, corresponding to the one-parameter subgroup, of right-BTZ-wedge preserving AdS isometries (see the next remark) is expected to be equivalent to the quantum dynamical system consisting of the *-algebra, ${\cal A}_{\mathrm{RC}}$, for the conformal generalized free field with anomalous scaling dimension $\Delta$ on the  right cylinder together with the one-parameter group of automorphisms, $\alpha_{\mathrm{RW}}(t)$, which time-translates towards the future in the right cylinder.

\smallskip

We remark that $\alpha_{\mathrm{RW}}(t)$ will, of course, just be the restriction from ${\cal A}_{\mathrm{DW}}$
to ${\cal A}_{\mathrm{RW}}$ of the ${\cal A}_{\mathrm{DW}}$ mentioned in Note $<$\ref{Note:equiv}$>$
and in the paragraph in the main text to which that note refers (and similarly for ${\cal A}_{\mathrm{DC}}$
and ${\cal A}_{\mathrm{RC}}$).

\smallskip

It follows (see Section \ref{Sect:prelim}) that the Hamiltonian generating dynamics in the GNS representation of the ground state on $({\cal A}_{\mathrm{RW}}, \alpha_{\mathrm{RW}}(t))$ (i.e.\ of the BTZ analogue of the Boulware state \cite{Boulware:1974dm}) will be unitarily equivalent to the Hamiltonian generating dynamics in the GNS representation of the ground state for $({\cal A}_{\mathrm{RC}}, \alpha_{\mathrm{RW}}(t))$.

\item
\label{Note:metrical}

When we refer, in the main text, to the metrical distance from the horizon, we mean the metrical distance from the horizon within a surface of constant Schwarzschild time.

\item
\label{Note:MIcorrection}
Our symbol $\mathsf{N}$ for what might be termed the `effective number of (real) Klein Gordon fields in nature' 
corresponds to Mukohyama and Israel's $\cal N$ \cite{Mukohyama:1998rf} .  It plays a similar role to the quantity denoted by $Z$ in 
\cite{'tHooft:1984re}.   Actually our $\mathsf{N}$ is defined so as to be $90/\pi^4$   ($\approx 0.92$) times the $\cal N$ defined in \cite{Mukohyama:1998rf}.   This (unimportant) discrepancy is slightly 
complicated by the fact that there appears to be an (unimportant) error in equation (3.14) of \cite{Mukohyama:1998rf}
which seemingly only holds if one takes $\cal N$ there to be as we have defined $\mathsf{N}$ here (and not as $\cal N$ was defined earlier in \cite{Mukohyama:1998rf}).

\item 
\label{Note:complementarity}
The Mukohyama-Israel complementarity argument is attractive insofar as the alternatives seem unattractive: After all, if one were to take $\alpha$ to lead to a different fraction (say $f$) than one quarter of the area of the event horizon, then there would only seem to be two reasonable possibilities: Either this is to be equated with the total entropy of the black hole, or it should be added to the entropy arising from the Gibbons-Hawking Euclidean classical gravitational action.  In either case, though, this would result in a fraction of the area of the event horizon different from one quarter -- namely $f$ on the first view and $1/4 + f$ on the second view.

\smallskip

We note that there is a more sophisticated variant of the view that the area part of the thermal atmosphere entropy and the entropy derived from the classical gravitational action should be added together, according to which 
(see \cite{Susskind:1994sm}) the overall  coefficient of ``one quarter area'' (divergent as $\alpha\rightarrow 0$ in our formulae (\ref{1plus2bricksecondresult}) and (\ref{originalbrick})) is interpreted as an inverse renormalized Newton's constant.  (See e.g. \cite{hep-th/0011176}.)  However  there seem to be difficulties with this view too.  In particular (cf. \cite{Barbon:1995im}) it seems to rely on a viewpoint, according to which the distance from the brick wall to the horizon (called $\alpha$ here and $\epsilon$ in \cite{Susskind:1994sm} and \cite{Barbon:1995im}) is regarded as a comparable quantity to the invariant-geodesic distance between a pair of close-by events -- used, in point-splitting regularization, to regulate the ultra-violet divergences in products of matter fields at a single point.  However, it is not clear that these are comparable quantities and, in fact, as pointed out in \cite{Barbon:1995im}, an attempted comparison would lead to frame-dependent results.
In any case, we note that this point of view involving the renormalization of Newton's constant  would seem to be incompatible with the view taken by 't Hooft and by Mukohyama and Israel and to adopt it would involve abandoning what Mukohyama and Israel call their complementarity principle and also abandoning the prediction of a value for the effective number, $\mathsf{N}$, of particle species in a consistent theory of quantum gravity.

\item
\label{Note:cnfmlwt}
 In general (see e.g.\ \cite{Wald} as well as \cite{pdFranpMatdSene97}), a conformally invariant scalar field, $\phi$, on a manifold, $M$, equipped with the spacetime metric, $g$, transforms to the field, $\tilde\phi$, on the same manifold equipped with the metric
$\tilde g=\Omega^2 g$ according to
\[
\tilde\phi=\Omega^{-\Delta}\phi
\]
where $\Delta$ is the scaling dimension of $\phi$.  Related to this, if $G(x_1, x_2)=\omega(\phi(x_1)\phi(x_2))$ is the expectation value of a product of fields (i.e.\ a `two-point function') in some state, $\omega$, on $(M, g_{ab})$, then 
\begin{equation}
\label{confstate}
\tilde G(x_1,x_2)=\Omega(x_1)^{-\Delta}\Omega(x_2)^{-\Delta}G(x_1,x_2)
\end{equation}
will define the corresponding two-point function -- i.e.\ the expectation value, $\tilde\omega(\tilde\phi(x_1)\tilde\phi(x_2))$, of the product of the conformally transformed fields in the appropriately `conformally transformed' state $\tilde\omega$.  We note that equation (\ref{confstate}) (and its counterparts for higher $n$-point functions) actually tells us what we mean here by the `conformally transformed state'.

\smallskip

For the massless Klein Gordon equation on the bulk of AdS$_2$, we rely on the fact that that (just in 1+1 dimensions -- in any other spacetime dimension, $n$, one would need, \cite{Wald}, to include a conformal coupling term $\xi R\phi$ on the left hand side of
(\ref{KG}) where $\xi=(n-2)/(4(n-1))$) the massless Klein Gordon equation is conformally invariant with (non-anomalous) scaling dimension $\Delta=0$.  (For the $n$ dimensional conformally coupled version of (\ref{KG}) indicated above, $\Delta$ is related to $n$ by the `non-anomalous' relation $\Delta=n/2-1$.)  It is thanks to all this that the global two-point function on 1+1 dimensional AdS is, in the massless case, equal to the two-point function (\ref{2dmasslessG}) for the ground state on our strip of Minkowski space, and similarly that the Poincar\'e two-point function is equal to the two point function (\ref{Poincaretwopointfn}) for the ground state on the right half of Minkowski space.

\item
\label{Note:2ptfn}
Here, we use the terminology and notation of algebraic quantum field theory \cite{Haag, Kay:2006jn} where our ground state $\omega_{\mathrm{globalground}}$ is a state in the sense of a `positive normalized linear functional' (say, on the algebra of smeared fields).  We might write our two-point function in `physicist's notation' as
$\langle O_{\mathrm{globalground}}|(\phi(\lambda_1,\rho_1)\phi(\lambda_2,\rho_2)|O_{\mathrm{globalground}}\rangle$.

\item
\label{Note:TtGstory}

To get (\ref{E:100}) directly from (\ref{E:92}), we can define $g=dT^2$ and $\tilde g=dt^2$.  Then we easily have, by (\ref{Ttrelation}), $\tilde g=\Omega^2g$ with $\Omega=1/(\kappa T)=e^{\kappa t}/(\kappa \ell)$, whereupon by (\ref{confstate}) in Note $<$\ref{Note:cnfmlwt}$>$, $G_{\mathrm{BTZ}}$ is easily seen to be given by (\ref{E:100}).

\item
\label{Note:1plus1Mink}
As is well-known, or easy to calculate, the two point function, ${\cal G}_{\mathrm{Minkowski}}$, for the ground state with respect to Minkowski time translations of the massless Klein Gordon equation on 1+1 dimensional Minkowski space can be written 
\[
\fl {\cal G}_{\mathrm{Minkowski}}(T_1, T_2; X_1, X_2)=-\frac{1}{4\pi}\ln((T_1-T_T)^2-(X_1-X_2)^2-2i\epsilon(T_1-T_2)) + C
\]
where $C$ is an `ill-defined' constant which doesn't matter for us because it will go away when we take derivatives.
Thus, if we introduce the double-null coordinates, $U=T+X$ and $V=T-X$, it can be written (now ignoring the constant) as the sum of 
\begin{equation}
\label{left}
G_{\mathrm{MinkowskiLeft}}(U_1, U_2)=-\frac{1}{4\pi}\ln(U_1-U_2-i\epsilon) 
\end{equation}
and
\[
G_{\mathrm{MinkowskiRight}}(V_1, V_2)=-\frac{1}{4\pi}\ln(V_1-V_2-i\epsilon),
\]
which can be thought of as the restrictions of ${\cal G}_{\mathrm{Minkowski}}$ to the null lines $T=X$ and $T=-X$
(or alternatively as the restrictions to the sectors consisting of left-moving and right-moving modes).

\smallskip

In the main text, we point out that, if we replace $T$ by $U$, the two-point function (\ref{E:92}) $G_{\mathrm{Poincare}}$ can be identified with the the double derivative, ${\partial^2/\partial U_1\partial U_2}$ of $G_{\mathrm{MinkowskiLeft}}$ up to a factor of 4.  Moreover, if we introduce Rindler coordinates on the right Rindler-wedge of our 1+1 dimensional Minkowski space via $U=\ell e^{\kappa u}$, $-V=\ell e^{-\kappa v}$ (and then introduce $t$ and $x$ according to $u=t+x$ and $v=t-x$) for $U$ positive and $V$ negative (and similarly we introduce Rindler coordinates on the left wedge by  $-U=\ell e^{-\kappa u}$, $V=\ell e^{\kappa v}$ for $U$ negative and $V$ positive and then again define $t$ and $x$ according to $u=t+x$ and $v=t-x$) then we notice that the analogy extends to relate the fact that the 1+1 massless Minkowski ground state (say restricted to the null line $T=X$) is a `double KMS state' in the sense of \cite{Kay:1985yw, Kay:1985zs} and Note $<$\ref{Note:equiv}$>$ with respect to Lorentz boosts to the fact that the global 1+1 dimensional AdS ground state, restricted to the right boundary line, is a double KMS state (now with respect to the restriction to the right AdS boundary line of BTZ time translations -- i.e.\ what we call $t$-translations in the main text -- on the region $T<0$ and the corresponding $t$-translations on the region $T>0$).  In particular, it relates the fact that $G_{\mathrm{MinkowskiLeft}}$ is a KMS state (at inverse temperature $2\pi/\kappa$) with respect to Lorentz boosts (which act as $u$-translations).  In fact, we easily have (cf. (\ref{E:100})) that
${\partial^2/\partial u_1\partial u_2}$ of $G_{\mathrm{MinkowskiLeft}}(e^{\kappa u_1}, e^{\kappa u_2})$ is
\[
=-\frac{1}{16\pi}\frac{\kappa^{2}}{\sinh^{2}\left(\kappa\frac{u_{1}-u_{2}-i\epsilon}{2}\right)},
\]
which is easily seen to satisfy the KMS condition at inverse temperature $2\pi/\kappa$.
(Above, we included factors of $\ell$ and $\kappa$ to make the analogy closer where $\ell$ is identified with the AdS radius and $\kappa$ is identified with the BTZ surface gravity of the main text.) 

\item
\label{Note:PoincareGlobalEquiv}
Unlike in the 1+1 dimensional case, which was settled by Spradlin and Strominger in \cite{Spradlin:1999bn}
it appears an open question as to whether the global and Poincar\'e ground states coincide in 1+2 dimensional AdS.  (It seems very likely that they do though!)  In \cite{Ortiz:2011mi, lOrt11}, one of us (LO) shows, in any case, that both ground states have the same boundary limit for their bulk two-point functions (as given by (\ref{E:140ll})). 

\item
\label{Note:image} 
As pointed out in \cite{Kay:2006jn} (see the Section there entitled `Warnings') in general, a two-point function defined by an image sum does not necessarily satisfy the necessary positivity properties to be the expectation value of a product of quantum fields.  However, one can see that this is not a problem in the case of our $G_{\mathrm{BTZ}}$  as defined by (\ref{E:140vv}).  

\item
\label{Note:Gradshteyn}
The passage from (\ref{Gbrickbetan}) to (\ref{Gbrickbetax}) can be made on noticing that (see e.g.\ formula {\bf 3.41} 31, page 327 of Gradshteyn \cite{Gradshteyn})
\[
\int_0^\infty\frac{e^{-qx}+e^{q-p}x}{1-e^{-px}}xdx=\left(\frac{\pi}{p}{\rm cosec}{\frac{q \pi}{p}}\right)^2 \quad 0<q<p
\]
with some care over `epsilons'.

\item
\label{Note:scalarentropy}
To derive equations (\ref{1dentropy}) and (\ref{2dimentropydens}), we start with $S=-\tr(\rho\ln\rho)$, which, since $\rho=e^{-\beta H}/Z$, is given by the (standard, statistical mechanical) formula
\begin{equation}
\label{SfromZ}
S=\left(1-\beta\frac{\partial}{\partial\beta}\right)\ln Z 
\end{equation}
where $Z=\tr(e^{-\beta d\Gamma(h)})$.  To calculate $Z$ for a one-dimensional box of side $L$ say, we may think of our real scalar quantum field as a collection of quantum harmonic oscillators  with angular frequencies $\omega_m=\pi m/L$.
Each such single oscillator will obviously have the partition function (ignoring zero-point energy), $Z_m= \sum_{n=0}^\infty e^{-\beta n\omega_m}=(1-e^{-\beta\omega_m})^{-1}$.  So the collection of oscillators will have partition function given by 
\begin{equation}
\label{1dlogZ}
\ln Z=\sum_{m=1}^\infty \ln\frac{1}{1-e^{1-\beta\pi m/L}}.
\end{equation}
We approximate the sum by the integral
\[
\frac{L{\cal T}}{\pi}\int_0^\infty\ln\left(\frac{1}{1-e^{-x}}\right)dx
\]
(${\cal T}=1/\beta$) 
\[
=\frac{L{\cal T}}{\pi}\frac{\pi^2}{6}=\frac{\pi L{\cal T}}{6}.
\]
By (\ref{SfromZ}) we then have $S$ $(=Z-\beta\partial Z/\partial\beta)=Z+{\cal T}dZ/d{\cal T}=\pi L{\cal T}/3$.

\smallskip

For a two-dimensional, say square, box of side $L$, the calculation is similar, except we now have modes labelled by two integers, say $m$ and $p$ with angular frequencies $\omega_{mp}=\pi (m^2+p^2)^{1/2}/L$.  In consequence,
(\ref{1dlogZ}) gets replaced by
\begin{equation}
\label{2dlogZ}
\ln Z=\sum_{m=1}^\infty\sum_{p=1}^\infty \ln\frac{1}{1-e^{-\beta(\pi/L)(m^2+p^2)^{1/2}}}
\end{equation}
which we can approximate as an integral over the positive quadrant of ${\mathbb R}^2$, which, when we convert to polar coordinates, becomes
\[
\frac{L^2{\cal T}^2}{\pi^2}\frac{2\pi}{4}\int_0^\infty x\ln\left(\frac{1}{1-e^{-x}}\right)dx
\]
\[
=\frac{\zeta(3)}{2\pi}L^2{\cal T}^2
\]
whereupon $S=Z+{\cal T}dZ/d{\cal T}=3\zeta(3)L^2{\cal T}^2/(2\pi)$ where $\zeta(3)$ is the Riemann zeta function of 3 (approximately $1.202$).

\smallskip

The obvious 3-dimensional counterpart to these calculations gives the well-known results, for the partition function and entropy of a real scalar field:  $\ln Z=L^3{\cal T}^3\pi^2/90$, $S=2\pi^2L^3{\cal T}^3/45$.

\item
\label{Note:omeganmcondition}
In order for the brick-wall modified $\rho$ and $Z$ to exist in the 1+2 dimensional case, we require, of course, that 
$Z=\tr(e^{-\beta H})$ where $H$ is the second quantization, $d\Gamma(h)$ of the one-particle Hamiltonian, $h$, whose spectrum consists of the $\omega_n^m$.  We haven't actually checked this since we haven't obtained mathematical control on the $\omega_n^m$.  However, we expect it to be finite.

\item
\label{Note:300}
The counterpart to our Equation (\ref{1plus2bricksecondresult}) for a collection of $\mathsf{N}$ Klein-Gordon fields in a finite box, outside a suitable brick wall in the (1+3 dimensional) Schwarzschild spacetime is Mukohyama and Israel's \cite{Mukohyama:1998rf} Equation (3.14) (see also Note $<$\ref{Note:MIcorrection}$>$)  for the area piece of their entropy, which can be written in the form
\begin{equation}
\label{originalbrick}
S_{\mathsf{N}} = \left(\frac{\mathsf{N}}{90\pi}\right)\frac{1}{\alpha^2}\left(\frac{4\pi r^2}{4}\right)
\end{equation}
where $r$ is the Schwarzschild radius.  In \cite{Mukohyama:1998rf} it was concluded from this that ``$\alpha$ is very near the Planck length if the effective number, $\mathsf{N}$, of basic quantum fields in nature is on the order of 300'' ($\approx 90\pi$).  One might equally conclude, though, that, if quantum gravitational effects are correctly taken into account by setting $\alpha$ (the metrical distance from the brick wall to the horizon) approximately equal to the Planck length, then the number of basic quantum fields in nature (or perhaps we should say the number of basic quantum fields in a consistent theory of quantum gravity) must be of the order of 300.  We draw a corresponding conclusion in the main text from our Equation (\ref{1plus2bricksecondresult}).

\smallskip

There doesn't appear to be a physically meaningful corresponding conclusion for the case of 1+1 dimensional BTZ.  In this case, it is easy to see that the metrical distance, $\alpha$, from the brick wall to the horizon is related to $B$ (see after Equation (\ref{E:141b})) by the approximate formula 
\begin{equation}
\label{1plus1brickdist}
\ln\left(\frac{\alpha}{2\ell}\right)\simeq -\frac{r_+B}{\ell}.
\end{equation}
So for $\mathsf{N}$ (real, massless) Klein Gordon fields, we would have, by (\ref{1dentropywithTH}) and (\ref{1plus1brickdist}), the 1+1 dimensional counterpart to Equation (\ref{1plus2bricksecondresult})
\[
S_{\mathsf{N}}=\left(\frac{2{\mathsf{N}}}{3}\ln\left(\frac{2\ell}{\alpha}\right)\right)\left(\frac{1}{4}\right).
\]
If one considers the Hawking value of the entropy to be $1/4$  in this case, then we observe that, for this to take the Hawking value, we would need to set ${\mathsf{N}}=(3/2)/(\ln(2\ell/\alpha))$ which, for $\alpha=1$ and $\ell$ much larger than 1 in natural units, would be less than 1!

\item
\label{Note:Smolinprovisos}
In their paper, \cite{Arnsdorf:2001qb}, Arnsdorf and Smolin consider the possibilities that the resolution to their puzzle might be either that string theory is not a theory of quantum gravity or that the Maldacena dual (i.e. ${\cal N}=4$ supersymmetric Yang Mills theory) in the AdS$_5$ version of holography does not exist as a quantum field theory in Minkowski space satisfying the basic axioms of nets of local *-algebras.  While we of course can not discount either of these possibilities, we shall aim in Section \ref{Sect:discuss} to arrive at at least a partial resolution to the Arnsdorf Smolin puzzle without contemplating either of these possibilities.  

\item
\label{Note:same} 
For the subtleties involved in determining what is meant by the `same quantum field theory on two different curved spacetime backgrounds' see \cite{Fewster:2011pe, Fewster:2011pn}.

\item
\label{Note:double}
(See also the two paragraphs immediately following the italicized statement in Section \ref{subsect:matgrav}.)
When we ask about the state of quantum gravity (as described by string theory) corresponding to a thermal state defined on a boundary cylinder, then the picture (see Figure \ref{Fig:BTZcylinders}) of the full maximally extended Schwarzschild AdS spacetime with its two boundary cylinders is probably misleading.  This picture is meaningful in the context of quantum field theory on a fixed curved spacetime.  But,  already in  Euclidean quantum gravity (see \cite{Gibbons:1976ue}, \cite{Hawking:1980gf}), it seems not to be meaningful, for a given cylinder, to regard it as a `right cylinder' and to assume the existence of a left wedge and a left cylinder.  (Note that this point is unrelated to the `single exterior black holes' considered by Louko and Marolf \cite{Louko:1998hc} and by Maldacena \cite{Maldacena:2001kr}.)  With our `matter-gravity entanglement hypothesis' (discussed in Subsection \ref{subsect:matgrav}) one might say, indeed, that the role which would be played by the left wedge is, in a sense, taken over by the quantum gravitational field. (Since the first version of this paper was written, new evidence for this was given by one of us in \cite{Kay:enclosed}.)   In particular, our expectation (see Subsection \ref{subsect:matgrav}) that the vector state, $\Psi_{\mathrm{total}}$ in ${\cal H}_{\mathrm{total}}$ which represents the equilibrium involving our AdS black hole and its thermal atmosphere `purifies' our thermal density operator on  ${\cal H}_{\mathrm{matter}}$ is analogous to the way in which the double-wedge Hilbert space purifies the thermal density operator for the right wedge in a quantum field theory in curved spacetime context.  Similarly our expectation, in Subsection \ref{subsect:matgrav}, that an arbitrary vector in ${\cal H}_{\mathrm{total}}$ can be arbitrarily closely approximated by vectors obtained by acting with elements of $I\otimes {\cal A}_{\mathrm{matter}}$ on $\Psi_{\mathrm{total}}$  is analogous to the `Reeh Schlieder property' of the right wedge algebra in the fixed background context.  (Cf. \cite{Kay:1985zs}).  

\item
\label{Note:monogamy}  For the notion of \textit{monogamy}, see e.g. \cite{DongYang}.  We remark that (apparently before the term `monogamy' was coined) the notion of monogamy and the use of monogamy arguments were already implicit in the intuition behind Theorem 6.5 in \cite{Kay:1988mu} (about the impossibility of certain `triple-wedge' situations) as discussed in the fourth paragraph of Section 6.1 of \cite{Kay:1988mu}.

\end{enumerate}

\ack 

LO thanks the Mexican National Council for Science and Technology (CONACYT) for funding his research studentship in York.  BSK is grateful to Michael Kay for helpful comments and suggestions.


\section*{References}
\begin{thebibliography}{99}

\bibitem{Aharony:1999ti}
Aharony O, Gubser SS, Maldacena J, Ooguri J and Oz Y 2000
Large N field theories, string theory and gravity
\textit{Phys. Rept.}  {\bf 323} 183-386
(arXiv:hep-th/9905111)

\bibitem{Maldacena:1997re}
  Maldacena JM 1998 The large N limit of superconformal field theories and supergravity
  \textit{Adv.\ Theor.\ Math.\ Phys.}  {\bf 2} 231 \hfil\break
  [also published in 2000 as \textit{Int.\ J.\ Theor.\ Phys.}  {\bf 38}, 1113 (1999)]
  (arXiv:hep-th/9711200)

\bibitem{Rehren:1999jn}
      Rehren KH 2000 Algebraic Holography 
      \textit{Annales\ Henri\ Poincar\'e} {\bf 1} 607-623
(arXiv:hep-th/9905179)

\bibitem{Rehren:2000tp}
Rehren KH 2000 Local quantum observables in the anti-de-Sitter-conformal QFT
correspondence \textit{Phys.\ Lett.\ B}  \textbf{493} 383-388
(arXiv:hep-th/9905179)

\bibitem{Haag} Haag R 1996 \textit{Local Quantum Physics (2nd ed.)} 
(Berlin: Springer)

\bibitem{Kay:2007rf}
  Kay BS and Larkin P 2008 Pre-Holography 
  \textit{Phys.\ Rev.\  D} {\bf 77} 121501R
  (arXiv:0708.1283) 

\bibitem{Greenberg:1961mr}
  Greenberg OW 1961
  Generalized free fields and models of local field theory
  \textit{Annals\ Phys.}\  {\bf 16} 158-176

\bibitem{Duetsch:2002hc}
Duetsch M and Rehren KH 2003
Generalized free fields and the AdS-CFT correspondence
  \textit{Annales\ Henri\ Poincar\'e} {\bf 4} 613-635
  (arXiv:math-ph/0209035)

\bibitem{Bertola:2000pp}
      M. Bertola, J. Bros, U. Moschella and R. Schaeffer 2000
      A general construction of conformal field theories from scalar anti-de Sitter quantum field
      theories
      \textit{Nucl.\ Phys.\ B} {\bf 587}, 619 (arXiv:hep-th/9908140)

\bibitem{Arnsdorf:2001qb}
  Arnsdorf M and Smolin L 2001
  The Maldacena conjecture and Rehren duality
  (arXiv:hep-th/0106073)

\bibitem{Rehren:2004yu}
  Rehren KH 2004
  QFT lectures on AdS-CFT
  arXiv:hep-th/0411086

\bibitem{Witten:1998qj} 
  Witten E 1998
  Anti-de Sitter space and holography
  \textit{Adv.\ Theor.\ Math.\ Phys.}  {\bf 2} 253 
  (arXiv:hep-th/9802150)

\bibitem{Gubser:1998bc}
  Gubser SS, Klebanov IR and Polyakov AM 1998
  Gauge theory correlators from non-critical string theory
  \textit{Phys.\ Lett.\  B} {\bf 428} 105 
  (arXiv:hep-th/9802109)

\bibitem{Duetsch:2002wy}
  Duetsch M and Rehren KH 2002
 A Comment on the dual field in the AdS-CFT correspondence  \textit{Lett.\ Math.\ Phys.}  {\bf 62} 171 
  (arXiv:hep-th/0204123)

\bibitem{BuchJung86} 
Buchholz D and Junglas P 1986
Local properties of equilibrium states and the particle spectrum in
quantum field theory \textit{Lett.\ Math.\ Phys.} {\bf 11} 51 

\bibitem{Banados:1992gq}
  Ba\~nados M, Henneaux M, Teitelboim C and Zanelli J 1993
 Geometry of the (2+1) black hole
  \textit{Phys.\ Rev.\  D} {\bf 48} 1506 
  (arXiv:gr-qc/9302012)

\bibitem{Lifschytz:1993eb}
Lifschytz G and Ortiz M 1994
 Scalar field quantization on the (2+1)-dimensional black hole background
 \textit{Phys.\ Rev.\  D} {\bf 49} 1929 
 (arXiv:gr-qc/9310008)

\bibitem{Avis:1977yn}
Avis SJ, Isham CJ and Storey D 1978
Quantum field theory in anti-de Sitter space-time
  \textit{Phys.\ Rev.\  D} {\bf 18} 3565

\bibitem{Hartle:1976tp}
  Hartle JB and Hawking SW 1976
 Path integral derivation of black hole radiance
 \textit{Phys.\ Rev.\  D} {\bf 13} 2188

\bibitem{Israel:1976ur}
 Israel W 1976
 Thermo field dynamics of black holes
  \textit{Phys.\ Lett.\  A} {\bf 57} 107

\bibitem{Kay:1988mu}
  Kay BS and Wald RM 1991
 Theorems on the uniqueness and thermal properties of stationary,
 nonsingular, quasifree states on space-times with a bifurcate Killing
  horizon
  \textit{Phys.\ Rept.}  {\bf 207}  49-136

\bibitem{Kay:1992gr}
  Kay BS 1993
 Sufficient conditions for quasifree states and an improved uniqueness
 theorem for quantum fields on space-times with horizons
 \textit{J.\ Math.\ Phys.}  {\bf 34}  4519

\bibitem{Unruh:1976db}
  Unruh WG 1976
 Notes on black hole evaporation
 \textit{Phys.\ Rev.\  D} {\bf 14} 870

\bibitem{Maldacena:2001kr}
 Maldacena JM 2003
 Eternal black holes in anti-de Sitter  JHEP {\bf 0304} 021 
 (arXiv:hep-th/0106112)

\bibitem{Ross:2005sc}
 Ross SF 2005
 Black hole thermodynamics
  arXiv:hep-th/0502195

\bibitem{Gibbons:1976ue}
  Gibbons GW and Hawking SW 1977
  Action integrals and partition functions in quantum gravity
  \textit{Phys.\ Rev.\ D} {\bf 15} 2752

\bibitem{Hawking:1982dh}
  Hawking SW and Page DN 1983
 Thermodynamics of black holes in anti-de Sitter space
  \textit{Commun.\ Math.\ Phys.}  {\bf 87} 577

\bibitem{pLark07}
      Larkin, P 2007
     \textit{Pre-Holography} PhD Thesis, University of York

\bibitem{Ortiz:2011mi}
 Ort\'iz L 2013
 Hawking effect in the eternal BTZ black hole: an example of Holography in AdS spacetime J. Gen. Rel. Grav. {\bf 45}, 427 
 (arXiv:1110.4451)

\bibitem{lOrt11}
 Ort\'iz L 2011
 \textit{Quantum fields on BTZ black holes} PhD Thesis, University of York

\bibitem{KeskiVakkuri:1998nw}
  E.~Keski-Vakkuri 1999
  Bulk and boundary dynamics in BTZ black holes
  \textit{Phys.\ Rev.\  D} {\bf 59}, 104001
  (arXiv:hep-th/9808037)

\bibitem{Boulware:1974dm}
  Boulware DG 1975
  Quantum field theory in Schwarzschild and Rindler spaces
  \textit{Phys.\ Rev.\  D} {\bf 11} 1404

\bibitem{'tHooft:1984re}
  't Hooft G 1985
 On the quantum structure of a black hole
  \textit{Nucl.\ Phys.\  B} {\bf 256} 727

\bibitem{Mukohyama:1998rf}
  Mukohyama S and Israel W 1998
 Black holes, brick walls and the Boulware state
  \textit{Phys.\ Rev.\  D} {\bf 58} 104005
  (arXiv:gr-qc/9806012)

\bibitem{Hawking:1980gf}
  Hawking SW 1979
  The path-integral approach to quantum gravity
{\it In} Hawking SW and Israel W (editors)  \textit{General Relativity: An Einstein Centenary Survey} 
(Cambridge: Cambridge University Press)

\bibitem{hep-th/0011176}
  E.~Winstanley 2001
 Renormalized black hole entropy in anti de Sitter space via the `brick wall' method
  \textit{Phys.\ Rev.\ D} {\bf 63}  084013
 (arXiv:hep-th/0011176)


\bibitem{Kim:1996eg}
Kim S-W, Kim WT, Park Y-J and Shin H 1997
Entropy of the BTZ black hole in 2+1 dimensions
\textit{Phys.\ Lett.\ B} {\bf 392} 311-318
(arXiv:hep-th/9603043)

\bibitem{Spradlin:1999bn}
  Spradlin M and Strominger A 1999
 Vacuum states for AdS(2) black holes
  JHEP {\bf 9911} 021
  (arXiv:hep-th/9904143)

\bibitem{Kay:2006jn}
Kay BS 2006
Quantum field theory in curved spacetime.  In 
\textit{Encyclopedia of Mathematical Physics} edited by J.-P. Fran\c coise, G. Naber and S.T. Tsou (Amsterdam, New York: Academic [Elsevier]) Vol. 4, p. 202 (arXiv:gr-qc/0601008)

\bibitem{Hawking:1974sw}
 Hawking SW 1975
 Particle creation by black holes 
  Commun.\ Math.\ Phys.\  {\bf 43} 199
  [Erratum-ibid.\ 1976 {\bf 46} 206]

\bibitem{pdFranpMatdSene97}
      Francesco PD, Mathieu P, S\'{e}n\'{e}chal D 1997
      \textit{Conformal Field Theory}
      (New York: Springer)

\bibitem{saFullsnRui87}
      Fulling SA and Ruijsenaars SNM 1987
      Temperature, periodicity and horizons
      \textit{Physics Reports} {\bf 152} 135-176

\bibitem{Gradshteyn}
      I. S. Gradshteyn and I. M. Ryzhik 1980
      \textit{Table of integrals, series and products}
      (New York:Academic)

\bibitem{Haag:1967sg}
Haag R, Hugenholtz NM and Winnink M 1967 
On the equilibrium states in quantum statistical mechanics
Commun. math. Phys. {\bf 5} 215-236

\bibitem{Kay:1985yw}
  Kay BS 1985
  Purification of KMS States
  \textit{Helv.\ Phys.\ Acta} {\bf 58}, 1030-1040

\bibitem{Kay:1985yx}
Kay BS 1985
A uniqueness result for quasifree KMS states
\textit{Helv.\ Phys.\ Acta} {\bf 58} 1017-1029

\bibitem{Kay:1985zs}
  Kay BS 1985
  The double wedge algebra for quantum fields on Schwarzschild and Minkowski spacetimes
  \textit{Commun.\ Math.\ Phys.}  {\bf 100} 57-81

\bibitem{Ichinose:1994rg}
  Ichinose I and Satoh Y 1995
  Entropies of scalar fields on three-dimensional black holes
  \textit{Nucl.\ Phys.\  B} {\bf 447} 340-370
  (arXiv:hep-th/9412144)

\bibitem{RSVol2}
Reed, M and Simon, B 1975
\textit{Fourier Analysis, Self-Adjointness (Methods of Modern Mathematical Physics, Vol. 2)}
(New York: Academic)

\bibitem{Wald} Wald, RM 1984
\textit{General Relativity} (Chicago and London: University of Chicago Press)

\bibitem{Kenmoku:2008qx}
  Kenmoku M, Kuwata M and Shigemoto K 2008
 Normal modes and no zero mode theorem of scalar fields in BTZ black hole spacetime
  \textit{Class.\ Quant.\ Grav.}  {\bf 25 } 145016
  (arXiv:0801.2044)

\bibitem{Susskind:1994sm}
  Susskind L and Uglum J 1994
  Black hole entropy in canonical quantum gravity and superstring theory
 \textit{Phys.\ Rev.\ D} {\bf 50 } 2700-2711
  (arXiv:hep-th/9401070)

\bibitem{Barbon:1995im}
Barb\'on JLF and Emparan R 1995
On quantum black hole entropy and Newton constant renormalization
\textit{Phys.\ Rev.\ D} {\bf 52 } 4527-4539
(arXiv:hep-th/9502155)

\bibitem{Witten:1998zw}
Witten E 1998
Anti-de Sitter space, thermal phase transition, and confinement in gauge
theories
\textit{Adv.\ Theor.\ Math.\ Phys.}  {\bf 2} 505
(arXiv:hep-th/9803131)

\bibitem{Kay:1998vv}
 Kay BS 1998
 Entropy defined, entropy increase and decoherence understood, and some black hole puzzles solved
 arXiv:hep-th/9802172

\bibitem{Kay:1998cj}
Kay BS 1998
Decoherence of macroscopic closed systems within Newtonian quantum gravity
\textit{Class.\ Quant.\ Grav.}  {\bf 15 } L89-L98
(arXiv:hep-th/9810077)

\bibitem{Abyaneh:2005tc}
Abyaneh V and Kay BS
The Robustness of a many-body decoherence formula of Kay under changes in graininess and shape of the bodies
arXiv:gr-qc/0506039

\bibitem{Kay:2007rx}
Kay BS and Abyaneh V
Expectation values, experimental predictions, events and entropy in quantum gravitationally decohered quantum mechanics
arXiv:0710.0992

\bibitem{Susskind:1994vu}
Susskind L 1995
The world as a hologram
\textit{J.\ Math.\ Phys.}  {\bf 36} 6377
(arXiv:hep-th/9409089)

\bibitem{Park:1999xz}
 Park IY 1999
 Fundamental versus solitonic description of D3-branes  
\textit{Phys.\ Lett.}  {\bf B468 } 213-218
(arXiv:hep-th/9907142)

\bibitem{Dimock:2000yv}
Dimock J 2000
Locality in free string field theory
\textit{J.\ Math.\ Phys.}  {\bf 41 } 40-61
  
\bibitem{Dimock:2001dy}
Dimock J 2002
Locality in free string field theory
\textit{Annales Henri Poincare} {\bf 3 } 613-634
(arXiv:math-ph/0102027)

\bibitem{Gubser:1996de}
Gubser SS,  Klebanov IR and Peet AW 1996
Entropy and temperature of black 3-branes  
\textit{Phys.\ Rev.}  {\bf D54 } 3915-3919
(arXiv:hep-th/9602135)

\bibitem{Klebanov:2000me}
Klebanov IR 2000
TASI lectures: Introduction to the AdS/CFT correspondence
arXiv:hep-th/0009139

\bibitem{Ortin}
Ort\'in T 2004
\textit{Gravity and Strings}
(Cambridge: Cambridge University Press)

\bibitem{Fewster:2011pe}
Fewster CJ and Verch R 2011
Dynamical locality and covariance: What makes a physical theory the same in all spacetimes?  
(arXiv:1106.4785)

\bibitem{Fewster:2011pn}
Fewster CJ and Verch R 2011
Dynamical locality of the free scalar field
(arXiv:1109.6732)

\bibitem{Louko:1998hc}
Louko J and Marolf D 1999
Single exterior black holes and the AdS/CFT conjecture
\textit{Phys.\ Rev.}  {\bf D59 } 066002
(arXiv:hep-th/9808081)

\bibitem{TakUme}
Takahashi Y and Umezawa H 1975
Thermo field dynamics
\textit{Collect. Phenom.}  {\bf 2} 55-80

\bibitem{Takahashi:1996zn}
Takahashi Y and Umezawa H 1996
Thermo field dynamics
\textit{Int.\ J.\ Mod.\ Phys.}  {\bf B10 } 1755-1805

\bibitem{Hawking:1976ra}
Hawking SW 1976
Breakdown of predictability in gravitational collapse
\textit{Phys.\ Rev.\  D} {\bf 14} 2460-2751

\bibitem{arXiv:1105.6359}
Giddings SB 2011
Is string theory a theory of quantum gravity? To appear in: G 't Hooft, E. Verlinde, D. Dieks, S. de Haro (eds.) 2012 \textit{Forty Years of String Theory: Reflecting on the Foundations}
\textit{Found. Phys. special issue} 
arXiv:1105.6359 

\bibitem{Kay:thermality} Kay BS 2012 On the origin of thermality arXiv:1209.5215

\bibitem{Kay:stringy} Kay BS 2012 Modern foundations for thermodynamics and the stringy limit of black hole equilibria
arXiv:1209.5085

\bibitem{Kay:morestring} Kay BS 2012 More about the stringy limit of black hole equilibria arXiv:1209.5110

\bibitem{Kay:enclosed} Kay BS 2013 Instability of enclosed horizons arXiv:1310.7395

\bibitem{DongYang} Yang D 2006 A simple proof of monogamy of entanglement \textit{Physics Letters A} {\bf 360} 249-250 (arXiv:quant-ph/0604168)
 
\end{thebibliography}
\end{document}